\documentclass[12pt]{article}
\usepackage[usenames]{color}
\usepackage{graphicx}
\usepackage{amsmath}
\usepackage{amssymb}
\usepackage{amsthm}

\usepackage{hyperref}
\hypersetup{
setpagesize=false,
 bookmarksnumbered=true,%
 bookmarksopen=true,%
 colorlinks=true,%
 linkcolor=blue,
 citecolor=blue,
}
\usepackage{bm}
\usepackage{cite}
\usepackage{framed,color}
\definecolor{shadecolor}{gray}{0.95}
\usepackage{fancybox}
\usepackage{ulem}
\definecolor{shadecolor}{gray}{0.95}
\usepackage{afterpage}
\usepackage{multirow}
\usepackage{slashed}
\numberwithin{equation}{section}

\newcommand{\vev}[1]{\left\langle #1 \right\rangle}
\newcommand{\der}{\partial}
\newcommand{\Tr}{\mbox{\rm Tr}}

\newcommand{\ad}[1]{\mathbf{ad}({#1})}

\begin{document}
\begin{flushright}
\end{flushright}
\begin{center}
{\LARGE\bf Vacuum structure of an eight-dimensional $SU(3)$ gauge theory on a magnetized torus}
\vskip 1.4cm
{\large  
Kentaro Kojima$^{a,}$\footnote{E-mail:~kojima@artsci.kyushu-u.ac.jp},
Yuri Okubo$^{b,}$\footnote{E-mail:~okubo.yuri226@gmail.com},
and 
Carolina Sayuri Takeda$^{b,}$\footnote{E-mail:~takeda.carolinasayuri.555@s.kyushu-u.ac.jp}
}\\ \vskip .5cm
{\it
$^a$ Faculty of Arts and Science, Kyushu University, Fukuoka 819-0395, Japan\\%
$^b$ Graduate School of Science, Kyushu University, Fukuoka 819-0395, Japan
}\\
\vskip 1.5cm
\begin{abstract}
    We analyze the vacuum structure of an eight-dimensional non-abelian gauge theory with a compactified four-dimensional torus as the extra dimensions. As a non-trivial background configuration of the gauge field of an $SU(n)$ gauge group, we suppose a magnetic flux in two extra dimensions, and continuous Wilson line phases are also involved. We introduce matter fields and calculate the mass spectrum of low-energy modes appearing in a four-dimensional effective theory in an $SU(3)$ model as an explicit example. As expected, potentially tachyonic states in four-dimensional modes appear from extra-dimensional gauge fields that couple to the flux background since the gauge group is simply connected. The Wilson line phases give a non-vanishing contribution to their masses, and we have a low-energy mass spectrum without tachyonic states, given that these phases take an appropriate value. To verify the validity of the values of the Wilson line phases, we examine the one-loop effective potential for these phases and explicitly show the contribution from each type of field present in our model. It is clarified that, although there seems to be no local minimum in the potential for the Wilson line phases in the pure Yang-Mills case, by including matter fields, we could find a vacuum configuration where tachyonic states disappear.
\end{abstract}
\end{center}

\newpage

\tableofcontents
\bigskip \bigskip

%
\section{Introduction}
\label{sec:intro}
%
The Standard Model (SM) has shown to be very successful, but there still remain many mysteries to be explored. The past decades have seen a significant increase in research on higher-dimensional theories as a potential framework for physics beyond the SM. For instance, identifying the Higgs as a scalar originating from an extra-dimensional gauge field, a model known as Gauge-Higgs Unification (GHU)~\cite{manton,fair,hosotani1,Hosotani:1988bm}, gives a new perspective on understanding its origin and solving the hierarchy problem~\cite{Hatanaka:1998yp}. Thus, exploring extra-dimensional gauge theories can provide insights into new physics beyond the limitations of our usual four dimensions. 

Non-trivial background configurations for extra-dimensional gauge fields can lead to interesting phenomena. A constant background value, which is the vacuum expectation value (VEV) of the extra-dimensional gauge fields, is closely related to the physical degrees of freedom of Wilson line (WL) phases. Since these phases parametrize physical vacua along flat directions of tree-level potentials for gauge fields, they are interesting candidates for the Higgs in GHU models. Consequently, the Higgs obtains a finite effective potential through quantum corrections and a finite mass even at higher-loop level~\cite{Maru:2006wa,Hosotani:2007kn}, characterized by the size of the extra dimensions, also clarifying the origin of the electroweak symmetry breaking~\cite{Hatanaka:1998yp,Hall:2001zb,Antoniadis:2001cv,Kubo:2001zc,Csaki:2002ur,Burdman:2002se,Gogoladze:2003bb,Scrucca:2003ra,Scrucca:2003ut,Haba:2004qf,Haba:2004jd}. In the context of Grand Unified Theories (GUTs)~\cite{GG}, WL phases can contribute to the spontaneous breaking of a more extensive gauge symmetry to the SM symmetry~\cite{Lim:2007jv,Kojima:2011ad,Yamashita:2011an,Hosotani:2015hoa,Yamatsu:2015oit,Furui:2016owe,Kojima:2016fvv,Kojima:2017qbt,Hosotani:2017edv,Maru:2019lit,Englert:2019xhz,Angelescu:2021nbp,Nakano:2022lyt,Angelescu:2022obm,Kojima:2023mew,Maru:2024ghd}. Moreover, introducing a constant magnetic flux in the background configuration brings extra phenomenologically desirable properties. First, having a flux background gives rise to chiral fermions in the effective theory~\cite{Generations,Cremades}, which is one fundamental feature of the SM. They exhibit a generation structure that can be used to explain the existence of multiple quark-lepton generations~\cite{Abe:2008sx,Kobayashi:2010an,Abe:2015yva,Abe:2015mua,Sakamoto:2020pev,Sakamoto:2020vdy,Kobayashi:2022tti,Imai:2022bke,Imai:2023yuf} and the flavor structure~\cite{Abe:2014vza,Fujimoto:2016zjs,Abe:2016eyh,Kobayashi:2016qag,Buchmuller:2017vut,Buchmuller:2017vho,Neutrino}, and the flux was shown to be a source for breaking supersymmetry (SUSY)~\cite{Bachas}.

There have been some recent studies considering the flux background in various setups. In models with more than six dimensions, massless scalars arising in the four-dimensional (4D) effective theory were identified as the Nambu-Goldstone (NG) bosons associated with the translational symmetry that is broken by the magnetic flux. For abelian gauge theories, quantum corrections of these scalars were shown to cancel in both SUSY and non-SUSY cases~\cite{Buchmuller,Ghilencea:2017jmh,B1,Honda:2019ema,Maru:2023esr,Hirose:2024vvx}. There are also studies focused on non-abelian cases~\cite{xi,Maru,Maru2,Akamatsu}.

Recently, we have investigated the mass spectra of a six-dimensional (6D) $SU(n)$ gauge theory with a magnetic flux background in the extra-dimensional torus~\cite{KYS}. We have also included WL phases in the background and could verify that their values along the flux direction have no physical contribution to the masses. In models with a simply-connected gauge group, mass spectrum and vacuum structure generally become complicated compared to the abelian case since tachyonic states appear in a simple setup~\cite{KYS}. A few attempts have been made to stabilize this type of system~\cite{Buchmuller:2019zhz,Franken:2022pug}, and tachyonic condensation was discussed in a SUSY model~\cite{Buchmuller:2020nnl}. However, there is less research on vacuum structure in non-abelian cases, including quantum corrections for potentials of the WL phases, which could lead to the development of phenomenologically interesting models.

In this work, we expand our setup to eight dimensions to address the above issue. The extra dimensions are compactified on a 4D torus with magnetic flux in only two directions. We examine dynamics of WL phases along the remaining two directions, whose values can now affect the masses of low-energy modes as a non-vanishing contribution. We can find a parameter region of the WL phases where tachyonic states disappear for a given flux background. However, as previously mentioned, these phases have no potential at tree level. Thus, it is essential to calculate quantum corrections for the potential to analyze the validity of the vacuum. By taking an $SU(3)$ gauge theory as a simple example, and also introducing matter fields, we show the mass spectrum of 4D modes and the one-loop effective potential for the WL phases. We find local minima of the potential where no tachyonic states appear in a low-energy mass spectrum in models with matter fields, whereas the pure Yang-Mills case has no local minimum. Using the flux and WL phases in the background configuration, we can generate many symmetry-breaking patterns and diverse low-energy effective theories. Further exploration in this field can lead to the development of new theories beyond the SM, such as GUT and GHU frameworks.  

The structure of this paper is as follows. In section~\ref{sec:setup}, we introduce definitions and basic concepts of an $SU(n)$ gauge theory on an eight-dimensional spacetime. Subsequently, we take the $SU(3)$ case as a simple example for further discussion. In section~\ref{sec:su3}, we elucidate the gauge fields and matter fields present in our model and show the masses of 4D modes appearing at low energy. As expected, some 4D modes can be tachyonic, which obtain positive mass squared with the help of non-vanihing WL phases. Thus, we discuss the conditions for the WL phases to stabilize potentially tachyonic states. In section~\ref{sec:effpot}, we compute the one-loop effective potential for the WL phases, indicating the different contributions from each type of field. Finally, in section~\ref{sec:vacuum}, we explore the vacuum structure, searching for local minima of the potential. We obtain qualitative insights from an analytical discussion of the potential. Then, we find local minima where tachyonic states disappear in the potential using numerical analysis. Section~\ref{sec:conclusions} concludes our work, and the appendixes contain details of derivations of the mass spectrum and the effective potential.

%
\section{Setup and notations}
\label{sec:setup}
%
We consider an eight-dimensional (8D) setup, which is an extension of the one discussed in Ref.~\cite{KYS}. It consists of ${\cal M}^4\times T^4$, where ${\cal M}^4$ is the Minkowski spacetime, and the extra dimensions are given by a 4D torus, $T^4$. The coordinates are denoted as usual, $x^M$ $(M=0,1,2,3,5,6,7,8)$ with $x^\mu$ $(\mu=0,1,2,3)$ on ${\cal M}^4$ and $x^m$ $(m=5,6,7,8)$ on $T^4$. As a simple case, we define that the torus coordinates satisfy the identification 
\begin{align}
  (x^5,x^6,x^7,x^8)\sim
  \left(x^5+Ln_5, \ x^6+Ln_6, \
  x^7+L'n_7, \ x^8+L'n_8  \right),
 \label{identT4}
\end{align}
where $n_5,n_6,n_7,n_8\in\mathbb Z$, and $L$ and $L'$ parametrize the size of the torus.  We set $L=1$ without loss of generality and define $L/L'=\hat M_w$, which is a free parameter and expresses the relative size of the torus in our theory. 

For an $SU(n)$ gauge theory, the gauge field $\bm A_M\in su(n)$ is expanded as $\bm A_M=A_M^at_a$ $(A_M^a\in \mathbb R,\ a=1,\dots,n^2-1)$, where $t_a\in su(n)$ are the generators that span the Lie algebra $su(n)$. Given the identification in eq.~\eqref{identT4} above, we have that the gauge fields $\bm A_M(x^\mu,x^m)$ must be physically equivalent to $\bm A_M(x^\mu,\tilde{x}^m)$, where $\tilde{x}^m$ is $x^m$ translated as in eq.~\eqref{identT4}. Therefore, it is sufficient that they are the same up to a gauge transformation. Let us define
  \begin{align}
    {\cal T}_n x^m=
    \begin{cases}
      (x^5+L,x^6,x^7,x^8), \quad &{\rm for}\quad n=5,\\
      (x^5,x^6+L,x^7,x^8),\quad &{\rm for}\quad n=6,\\
      (x^5,x^6,x^7+L',x^8),\quad &{\rm for}\quad n=7,\\
      (x^5,x^6,x^7,x^8+L'),\quad &{\rm for}\quad n=8.
    \end{cases}
  \end{align}
  Then, we have 
\begin{align}\label{AmBC}
  \bm A_M({\cal T}_n{x}^m) = T_n\bm A_M(x^m)T_n^\dag+\frac{i}{g} T_n\der_MT_n^\dag,
\end{align}
which are the boundary conditions for the gauge fields in the torus. The matrices $T_m\in SU(n)$ are called the twist matrices, $g$ is the gauge coupling constant, and $x^\mu$ was suppressed to simplify the notation. From now on, we will keep this notation for all functions of $x^\mu$.

We start by discussing the pure Yang-Mills theory, which has the following Lagrangian
\begin{align}\label{Lagrangian}
{\cal L}=-{1\over 2}\Tr[\bm F_{MN}\bm F^{MN}],
\end{align}
where we have used the definitions
\begin{align}\label{defFmn}
  \bm F_{MN}={i\over g}[\bm D_M,\bm D_N]=\der_M\bm A_N-\der_N\bm A_M -ig[\bm A_M,\bm A_N], \quad \text{and} \quad \bm D_M=\der_M-ig \bm A_M,
\end{align}
for the field strength tensor and covariant derivative.  Later, we introduce matter fields.

Since we are considering non-trivial background configurations for the extra-dimensional gauge fields, we make the following replacement
\begin{align}\label{repl1}
  \bm A_M(x^m)\to \bm B_M(x^m) + {\bm A}_M(x^m), 
\end{align}
where $\bm B_M$ denotes the background configuration and ${\bm A}_M$ on the right-hand side represents the fluctuations around $\bm B_M$. By imposing 4D Lorentz invariance at the vacuum, we hereafter set $\bm B_\mu=0$.  We define the background field strength tensor and covariant derivative as
\begin{align}\label{bgdfdef1}
  {\cal F}_{mn}=\der_m\bm B_n-\der_n\bm B_m-ig[\bm B_m,\bm B_n], \qquad
  {\cal D}_m=\der_m-ig\ad{\bm B_m},
\end{align}
where $\ad{X}Y=[X,Y]$. Using these definitions, we perform the standard $R_\xi$ gauge fixing by adding the term, 
\begin{align}\label{gaugefixing}
  {\cal L}_{\rm GF}
=-\frac{1}{\xi}\Tr\left[(\partial_\mu \bm A^\mu+\xi {\cal D}_m\bm A^m)^2\right],
\end{align}
to the Lagrangian given by eq.~\eqref{Lagrangian}. In the above, $\xi$ is a real parameter called a gauge parameter.

This background has to satisfy the equation of motion for consistency. Accordingly, we obtain the background equation of motion, which is given by
\begin{align}\label{eom}
  {\cal D}^m  {\cal F}_{mn} =0.
\end{align}
A solution is 
\begin{gather}\label{bggensol1}
  \bm B_5(x^m)=\bm v_5-\left(1+\gamma_1\right) \bm f_1 x^6/2, \qquad
  \bm B_6(x^m)=\bm v_6+\left(1-\gamma_1\right) \bm f_1 x^5/2, \\\label{bggensol2}
  \bm B_7(x^m)=\bm v_7-\left(1+\gamma_2\right) \bm f_2 x^8/2, \qquad
  \bm B_8(x^m)=\bm v_8+\left(1-\gamma_2\right) \bm f_2 x^7/2, 
\end{gather}
where
\begin{align}
    [\bm v_m,\bm v_n]=[\bm v_m,\bm f_1]=[\bm v_m,\bm f_2]=[\bm f_1,\bm f_2]=0.
\end{align}
Here, $\bm v_m,\bm f_p\in su(n)$ $(p=1,2)$ and $\gamma_p\in \mathbb R$ are constants. The constants $\bm v_m$ are called continuous WL phases, and $\bm f_p$ parametrize the constant magnetic flux present in the background of the extra dimensions. In the above, $\gamma_p$ has no effect on physical results and labels different choices of gauge. For instance, $\gamma_p=\pm 1$ and $\gamma_p=0$ are often called the Landau and symmetric gauge, respectively.  

Now, let us discuss our choice of basis of $su(n)$. It is convenient to choose the Cartan-Weyl basis, where we write the $su(n)$ generators $\{t_a\}$ $(a=1,\dots,n^2-1)$ as $\{t_a\}=\{H_k\}\cup \{E_{\bm \alpha}\}$. The Cartan generators $\{H_k\}$ $(k=1,\dots,n-1)$ are Hermitian, and the step operators $E_{\bm \alpha}$ associated to a root vector $\bm \alpha$ satisfy $E_{\bm \alpha}^\dag=E_{-\bm \alpha}$. Their commutation relations are given by
\begin{align}\label{CWcom1}
  [H_k,H_\ell]=0, \qquad
  [H_k,E_{\bm \alpha}]=\alpha_{k}E_{\bm \alpha}, 
\end{align}
where $\alpha_{k}\in \mathbb R$ is the $k$-th component of the root vector ${\bm \alpha}$.  

We also choose the basis of the generators to be in the fundamental representation space of $su(n)$ for simplicity. Consequently, we write the Cartan generators as
\begin{align}
H_1 =
  \begin{pmatrix}
      1&0&0&\cdots&0\\
      0&-1&0&\cdots&0\\
      0&0&0&\cdots&0\\
      &&&\ddots
      &\\
      0&0&0&\cdots&0
  \end{pmatrix}, \ 
  H_2=
  \begin{pmatrix}
      0&0&0&\cdots&0\\
      0&1&0&\cdots&0\\
      0&0&-1&\cdots&0\\
      &&&\ddots
      &\\
      0&0&0&\cdots&0
  \end{pmatrix}, \
  \dots,\
  H_{n-1}=
  \begin{pmatrix}
      0&0&\cdots&0&0\\
      0&0&\cdots&0&0\\
      &&\ddots
      &&\\
      0&0&\cdots&1&0\\
      0&0&\cdots&0&-1
  \end{pmatrix},
\end{align}
and the $n(n-1)$ step operators as
\begin{align}\label{Eij}
  E_{ij}^{(+)}={\hat e}_{ij}, \qquad 
  E_{ij}^{(-)}={\hat e}_{ji}, \qquad 1\leq i<j\leq n,
\end{align}
where we have defined the basis matrices ${\hat e}_{ij}$ to have the $(i',j')$ element given by $(\hat e_{ij})_{i'j'}=\delta_{ii'}\delta_{jj'}$, and $\delta_{ii'}$ is the Kronecker delta. 

The magnetic flux $\bm f_p$ and $\bm v_m$ can be simultaneously diagonalized and therefore can be expanded by $su(n)$ Cartan generators:
\begin{align}
  \bm f_p=f^k_pH_k, \qquad \bm v_m=v_m^kH_k, \qquad f^k_p,v_m^k\in \mathbb R, 
\end{align}
where summations over $k$ are taken. In addition, the flux background was found to be quantized, such as
\begin{align}
  f^k_1=\frac{2\pi}{g L^2} N_1^k
  =\hat f_1 N^k_1,
  \qquad
  f^k_2=\frac{2\pi}{g L'{}^2} N_2^k
  =\hat f_2 N^k_2,  
  \qquad N_1^k,N_2^k\in \mathbb Z,
  \label{qcondhatb1}
\end{align}
where we have introduced the unit of flux $\hat f_1=2\pi /(gL^2)$ and $\hat f_2=2\pi /(gL'{}^2)$.

As already mentioned, it is known that there appear tachyonic states in 6D non-abelian gauge theories with magnetic flux background~\cite{KYS}.  For a 6D $SU(n)$ gauge theory, it was discussed that the WL phases have no contribution to the mass spectra, implying that they cannot stabilize the system.  By this reasoning, we set $\bm v_5=\bm v_6= 0$ in the background. In addition, we focus on a case with $\bm f_2=0$, leading to the following background
\begin{gather}\label{BG}
    \bm B_5(z)=-(1+\gamma)\bm fx^6/2, \qquad 
  \bm B_6(z)=(1-\gamma)\bm fx^5/2, \\\label{b78}
    \bm B_7(z)=\bm v_7, \qquad 
  \bm B_8(z)=\bm v_8, 
\end{gather}
where we have renamed $\bm f_1$ and $\gamma_1$ to $\bm{f}$ and $\gamma$. The background $\bm B_M(x^m)$ in eqs.~\eqref{bggensol1} and~\eqref{bggensol2} and the twist matrices in eq.~\eqref{AmBC} must be related by gauge transformations. In other words, the expressions of the twist matrices can vary depending on the choice of background. According to our choice above, the twist matrices can be taken as
\begin{align}\label{Twist}
  T_5=e^{ig(1-\gamma)\bm fx^6/2}, \qquad
  T_6=
  e^{-ig(1+\gamma) \bm fx^5/2}, \qquad T_7=T_8=I, 
\end{align}
where $I$ is the unit matrix.

We also introduce bulk matter fields. Let us take a field $\Phi_{\bm R}$ to be a complex scalar field of the representation $\bm R$ of $SU(n)$. Weyl fermions in 8D theories may give bulk gauge anomalies. To evade this, we introduce vector-like (Dirac) fermions. An 8D Dirac fermion of the representation $\bm R$, denoted by $\Psi_{\bm R}$, is a 16-component spinor having 16 real degrees of freedom (dof) on the mass shell. We suppose that they satisfy the following boundary conditions:
\begin{align}
  \label{matbc1}
  \Phi_{\bm R}({\cal T}_nx^m)=e^{2\pi i \eta_n(\Phi_{\bm R})}(T_n)_{\bm R}\Phi_{\bm R}(x^m),\qquad 
  \Psi_{\bm R}({\cal T}_nx^m)=e^{2\pi i \eta_n(\Psi_{\bm R})}(T_n)_{\bm R}\Psi_{\bm R}(x^m),
\end{align}
where $(T_n)_{\bm R}$ is a matrix of $T_n$ in a representation $\bm R$.  We have introduced real numbers $\eta_n(\Phi_{\bm R})$ and $\eta_n(\Psi_{\bm R})$, which are independently taken for each matter field.  Depending on a global symmetry of the full theory, allowed values of $\eta_n$ are constrained. Hereafter, we consider $\eta_5,\eta_6=0$ and $\eta_{m'}\in \{0,1/2\}$ $(m'=7,8)$. As discussed in the next section, for $\eta_{m'}(\phi)=1/2$, the discrete momentum labeled by $n_{m'}$ in masses of 4D modes appearing from a field $\phi$ is shifted from $n_{m'}$ to $n_{m'}+1/2$.

With the background configuration in eqs.~\eqref{BG} and~\eqref{b78}, the WL phases $\bm v_7$ and $\bm v_8$ can contribute to masses of 4D modes.  Thus, we expect that tachyonic states disappear in a low-energy theory at a vacuum with non-trivial values of flux and WL phases. Since the continuous WL phases have no potential at tree level, quantum corrections to their potential are crucial for examining the validity of vacua. The following sections discuss the vacuum structure in a concrete setup.

%
\section{An $SU(3)$ model}
\label{sec:su3}
%

\subsection{Background configuration}
\label{sec:backg}
As a concrete example, we will explore our setup for the gauge group $SU(3)$. Therefore, there are only two Cartan generators, and we choose $N^1 = 1$ and $N^2 = 2$ in eq.~\eqref{qcondhatb1}, corresponding to the flux background
\begin{align}\label{su3flux}
  \bm f=f^kH_k=\hat f
  \begin{pmatrix}
      1&0&0\\
      0&1&0\\
      0&0&-2
  \end{pmatrix}, \qquad {\rm where}\qquad \hat f={2\pi \over g}.
\end{align}
While keeping this flux background, the continuous WL phases can be diagonalized through a unitary transformation; hence, we write 
\begin{align}
  \bm v_{m'}=
  \begin{pmatrix}
      v_{m'}^1&0&0\\
      0&v_{m'}^2-v_{m'}^1&0\\
      0&0&-v_{m'}^2
  \end{pmatrix}, \qquad m'=7,8.
\end{align}
In the following discussions, we assume these choices of backgrounds.

We are interested in the theory at an energy scale sufficiently lower than the compactification scale $1/L$ and $1/L'$. In this case, we have a 4D effective theory where infinitely many 4D fields appear, coming from the mode expansions of 8D fields. Masses of 4D fields are determined by their charges concerning the Cartan generators and the helicity operator~\cite{Bachas}.  Let $\phi$ be an 8D field with definite charges associated with the generators $H_1$, $H_2$, and $H_3=H_1+2H_2$. We denote these charges of $\phi$ by $q_1(\phi)$, $q_2(\phi)$, and $q_3(\phi)$, respectively. Note that $q_3(\phi)=q_1(\phi)+2q_2(\phi)$ holds.  In addition, we denote the helicity of $\phi$ associated with the $x^5$--$x^6$ plane by $\Sigma_{56}(\phi)$. For example, linear combinations of $A_5$ and $A_6$ have $\Sigma_{56}=\pm 1$, whereas the other gauge fields, i.e., $A_\mu$ and $A_{m'}$, have $\Sigma_{56}=0$. In the following, we clarify field contents and their charges.  Then, we discuss the masses of 4D modes that appear in this setup.

\subsection{Gauge fields}
We first discuss gauge fields in this model.  The masses of 4D modes arising from these fields are determined by quadratic terms of the gauge-fixed Lagrangian
${\cal L}_{\rm YM}^{\rm gf}$ as
\begin{align}
  {\cal L}_{\rm YM}^{\rm gf}
  &\ni {\cal L}^{(2)}_{A_\mu}
  +{\cal L}^{(2)}_{A_m}
    +{\cal L}^{(2)}_{c}, \\
  {\cal L}^{(2)}_{A_\mu}
  &=\Tr[
    \bm A^\mu(\eta_{\mu\nu}\square
    +\eta_{\mu\nu}
    ({\cal D}^2)-(1-\xi^{-1})\der_\mu\der_\nu)
    \bm A^\nu],\\
  {\cal L}^{(2)}_{A_m}
  &=\Tr[
    \bm A^m(\delta_{mn}\square +\delta_{mn} ({\cal D}^2)
    -(1-\xi) {\cal D}_m{\cal D}_n
    -2ig(\delta_{m5}\delta_{n6}-\delta_{m6}\delta_{n5})\ad{\bm f}  )
    \bm A^n],
    \label{Lam2Tr1}
    \\
  {\cal L}^{(2)}_{c}
  &=-2\Tr[\bar {\bm c}(\square+\xi {\cal D}_m
    {\cal D}^m)\bm c], 
\end{align}
where $\delta_{mn}$ is the Kronecker delta function, $({\cal D}^2)={\cal D}_m{\cal D}^m$, and $\bm c\in su(3)$ is a ghost field. Note that the last term of eq.~\eqref{Lam2Tr1} is only nonzero for the $x^5$--$x^6$ directions where there is magnetic flux. After it is diagonalized, it is convenient to define
\begin{align}
&  \bm A_{z_1}={1\over 2}(\bm A_5-i\bm A_6), \qquad 
  {\bm A}_{{\bar z}_1}={1\over 2}(\bm A_5+i\bm A_6),\\
&  \bm A_{z_2}={1\over 2}(\bm A_7-i\bm A_8), \qquad 
  {\bm A}_{{\bar z}_2}={1\over 2}(\bm A_7+i\bm A_8).
\end{align}
In component form, they are written as
\begin{gather}
  \bm A_\mu=
  \begin{pmatrix}
      A_\mu^{(1)}&A_\mu^{(12)}&\bar A_\mu^{(31)}\\
      \bar A_\mu^{(12)}&-A_\mu^{(1)}+A_\mu^{(2)}&A_\mu^{(23)}\\
      A_\mu^{(31)}&\bar A_\mu^{(23)}&-A_\mu^{(2)}
  \end{pmatrix},
\end{gather}  
\begin{align}
  &\bm A_{z_p}=
  \begin{pmatrix}
      A_{z_p}^{(1)}&A_{z_p}^{(12)}&\bar A_{{\bar z}_p}^{(31)}\\
      \bar A_{{\bar z}_p}^{(12)}&-A_{z_p}^{(1)}+A_{z_p}^{(2)}&A_{z_p}^{(23)}\\
      A_{z_p}^{(31)}&\bar A_{{\bar z}_p}^{(23)}&-A_{z_p}^{(2)}
  \end{pmatrix},\quad 
  {\bm A}_{{\bar z}_p} =
  \begin{pmatrix}
      \bar A_{z_p}^{(1)}&A_{{\bar z}_p}^{(12)}&\bar A_{z_p}^{(31)}\\
      \bar A_{z_p}^{(12)}&-\bar A_{z_p}^{(1)}+\bar A_{z_p}^{(2)}&A_{{\bar z}_p}^{(23)}\\
      A_{{\bar z}_p}^{(31)}&\bar A_{z_p}^{(23)}&-\bar A_{z_p}^{(2)}
  \end{pmatrix},
\end{align}
where ${\bm A}_{{\bar z}_p}=(\bm A_{z_p})^\dag$ and $p=1,2$. 

We are interested in masses of 4D modes appearing from 8D fields in a low-energy theory.  These masses depend on quantum charges of 8D fields.  In table~\ref{tabgauge}, we have summarized dof and quantum numbers of independent 8D gauge fields.  In the table, we also show which fields couple to WL phases or flux backgrounds. As shown in table~\ref{tabgauge}, $A_{z_1}^{23}$, $A_{{\bar z}_1}^{23}$, $A_{z_1}^{31}$, and $A_{{\bar z}_1}^{31}$ receive tachyonic contributions in masses of their 4D modes as will be discussed later.  We also have ghost fields that cancel unphysical modes arising from gauge fields $\bm A_M$.
\begin{table}[]
    \centering
\begin{align*}
  \begin{array}{c|ccccc|ccc}\hline\hline
    \phi&{\rm dof}&\Sigma_{56}(\phi)&q_1(\phi)&q_2(\phi)&q_3(\phi)&{\rm WL}
                                                                       &{\rm flux}&{\rm tachyonic}
    \\\hline
    A_\mu^{(1)}&4&0&0&0&0&&&\\
    A_\mu^{(2)}&4&0&0&0&0&&&\\
    A_\mu^{(12)}&8&0&2&-1&0&\checkmark&&\\
    A_\mu^{(23)}&8&0&-1&2&3&\checkmark&\checkmark&\\
    A_\mu^{(31)}&8&0&-1&-1&-3&\checkmark&\checkmark&\\
    A_{z_1}^{(1)}&2&\pm 1&0&0&0&&&\\
    A_{z_1}^{(2)}&2&\pm 1&0&0&0&&&\\
    A_{z_1}^{(12)}&2&\pm 1&2&-1&0&\checkmark&&\\
    A_{z_1}^{(23)}&2&\pm 1&-1&2&3&\checkmark&\checkmark&\checkmark\\
    A_{z_1}^{(31)}&2&\pm 1&-1&-1&-3&\checkmark&\checkmark&\checkmark\\
    A_{{\bar z}_1}^{(12)}&2&\pm 1&2&-1&0&\checkmark&&\\
    A_{{\bar z}_1}^{(23)}&2&\pm 1&-1&2&3&\checkmark&\checkmark&\checkmark\\
    A_{{\bar z}_1}^{(31)}&2&\pm 1&-1&-1&-3&\checkmark&\checkmark&\checkmark\\
    A_{z_2}^{(1)}&2&0&0&0&0&&&\\
    A_{z_2}^{(2)}&2&0&0&0&0&&&\\
    A_{z_2}^{(12)}&2&0&2&-1&0&\checkmark&&\\
    A_{z_2}^{(23)}&2&0&-1&2&3&\checkmark&\checkmark&\\
    A_{z_2}^{(31)}&2&0&-1&-1&-3&\checkmark&\checkmark&\\
    A_{{\bar z}_2}^{(12)}&2&0&2&-1&0&\checkmark&&\\
    A_{{\bar z}_2}^{(23)}&2&0&-1&2&3&\checkmark&\checkmark&\\
    A_{{\bar z}_2}^{(31)}&2&0&-1&-1&-3&\checkmark&\checkmark&\\\hline\hline
  \end{array}
\end{align*}
\caption{Summary of independent 8D gauge fields, which are written in the first column. The following columns detail the real dof, helicity, and charges $q_1, q_2$ and $q_3$ of each field. The next two columns indicate which fields couple with WL phases or with the flux background. Finally, the last column shows which fields contain tachyonic contributions in their masses.}
\label{tabgauge}
\end{table}

\subsection{Matter fields}
\label{Matter fields}
In this section, we discuss matter fields. First, we consider matter fields in $\bm 3$, the fundamental representation of $SU(3)$. The scalar $\Phi_{\bm 3}$ has three components $\phi^{(\alpha)}_{\bm 3}$ $(\alpha=1,2,3)$. Their charges are given by
\begin{align}
\label{scalar1}
  q_1(\phi^{(1)}_{\bm 3})&=1, &&   q_2(\phi^{(1)}_{\bm 3})=0, &&   q_3(\phi^{(1)}_{\bm 3})=1, \\
  \label{scalar2}
  q_1(\phi^{(2)}_{\bm 3})&=-1, &&   q_2(\phi^{(2)}_{\bm 3})=1, &&   q_3(\phi^{(2)}_{\bm 3})=1, \\
  \label{scalar3}
  q_1(\phi^{(3)}_{\bm 3})&=0, &&   q_2(\phi^{(3)}_{\bm 3})=-1, &&   q_3(\phi^{(3)}_{\bm 3})=-2.
\end{align}
We also introduce fermion fields. The fermion $\Psi_{\bm 3}$ of the fundamental representation has three components $\psi^{(\alpha)}_{\bm 3}$, which have the same charges as $\phi^{(\alpha)}_{\bm 3}$. We note that fermion fields have non-trivial helicities $\Sigma_{56}=\pm 1/2$, while scalar fields have $\Sigma_{56}=0$. For the anti-fundamental representation $\bar {\bm 3}$, although charges change their signs, the mass spectrum of their 4D modes is the same as the one of $\bm 3$. 

One can introduce matter fields with representations other than the fundamental. The charges of any representation can be given by linear combinations of the charges of the fundamental representation.  For example, a scalar field belonging to the second rank symmetric tensor of $SU(3)$ has six components, which can be written as $\phi^{(\alpha,\beta)}_{\bm 6}$ $(1\leq \alpha \leq \beta\leq 3)$.  Their charges are written by
\begin{align}
  q_k(\phi^{(\alpha,\beta)}_{\bm 6})=q_k(\phi^{(\alpha)}_{\bm 3})+q_k(\phi^{(\beta)}_{\bm 3}).
\end{align}
For the adjoint representation $\bm 8$, there are eight components, two of which are neutral with respect to the gauge field background. The other components can be expressed by $\phi^{(\alpha,\beta)}_{\bm 8}$ $(1\leq \alpha,\beta\leq 3)$, whose charges are given by 
\begin{align}
  q_k(\phi^{(\alpha,\beta)}_{\bm 8})=q_k(\phi^{(\alpha)}_{\bm 3})-q_k(\phi^{(\beta)}_{\bm 3}).
\end{align}
As a final example, we consider the representation $\bm{10}$, the totally symmetric tensor product of three $\bm 3$. The components of $\Phi_{\bm{10}}$ are labeled by
\begin{align}\notag
  (\alpha,\beta,\delta)
  & \in (1,1,1),(2,2,2),(3,3,3),(1,1,2),(1,1,3),\\
  &\qquad (2,2,1),(2,2,3),(3,3,1),(3,3,2),(1,2,3), 
\end{align}
and their charges are given by
\begin{align}
  q_k(\phi^{(\alpha,\beta,\delta)}_{\bm{10}})=q_k(\phi^{(\alpha)}_{\bm 3})
  +q_k(\phi^{(\beta)}_{\bm 3})
  +q_k(\phi^{(\delta)}_{\bm 3}).
\end{align}

\subsection{Masses of 4D modes}
\label{sec:mass}
In this section, we discuss the mass spectra in a low-energy effective theory of the $SU(3)$ model. We perform the Kaluza-Klein (KK) expansions of the fields and calculate the eigenvalue of the mass operators in the Lagrangian given by eq.~\eqref{Lam2Tr1} acting on the corresponding mode function. In a 4D effective theory, infinite 4D modes appear from 8D fields discussed in the previous subsections. In the following discussions, the mass of a given 4D field $\phi$ will be expressed as $M^2(\phi)$, and we present the final expressions for the mass spectrum of the fields. 

\subsubsection{$q_3(\phi)=0$ case}
First, let us discuss the masses of 4D modes appearing from an 8D field $\phi$ having $q_3(\phi)=0$. They do not couple to the flux background. For example, $A_M^{(1)}$, $A_M^{(12)}$, and $\phi_{\bm 8}^{(1,2)}$ belong to this case.  To obtain masses of 4D modes from an 8D field $\phi$, the discussion in Ref.~\cite{KYS} is straightforwardly generalized. Their 4D modes are labeled by four integers $\hat n_m\in \mathbb Z$ $(m=5,6,7,8)$. We use $\hat {\bm n}=(\hat n_5,\hat n_6,\hat n_7,\hat n_8)$ and denote a 4D mode from $\phi$ by $\phi_{(\hat {\bm n})}$. It is convenient to introduce
\begin{align}
  \label{hatNdef}
    \hat N_{m'}(\phi)= \hat n_{m'}+\eta_{m'}(\phi)-q_1(\phi)a_{m'}^1-q_2(\phi)a_{m'}^2,
\end{align}
where we have used the parametrization of the WL phases as $a_{m'}^k=gL'v_{m'}^k/2\pi$. The parameter $\eta_{m'}(\phi)$ appears in the boundary condition of matter fields in eq.~\eqref{matbc1}. We imply $\eta_{m'}(\phi)=0$ if $\phi$ is a gauge field. In the following, we also use 
\begin{align}
  M_{56}^2(\phi)&=(2\pi )^2\left[\hat n_5^2+\hat n_6^2\right],
                  \qquad 
                  M_{78}^2(\phi)=(2\pi \hat M_w)^2\left[\hat N_{7}^2(\phi)+\hat N_{8}^2(\phi)\right].
                  \label{M78def}
\end{align}

For a matter field $\phi$, the tree-level mass spectrum of the 4D modes $\phi_{(\hat {\bm n})}$ is given by
\begin{align}\label{m001}
  M^2(\phi_{(\hat {\bm n})})
  &=M_{56}^2(\phi)+M_{78}^2(\phi),
\end{align}
for an arbitrary $\xi$. For a gauge field $\phi$ with $\xi =1$, masses of the 4D modes $\phi_{(\hat {\bm n})}$ are the same as in eq.~\eqref{m001}. For a gauge field $\phi$ with an arbitrary $\xi$, in addition to the above, there appear masses as 
\begin{align}\label{m002}
M_{\xi}^2(\phi_{(\hat {\bm n})})&=\xi\left(M_{56}^2(\phi)+M_{78}^2(\phi)\right).
\end{align}
Note that the 4D modes that have masses $M_{\xi}^2(\phi_{(\hat {\bm n})})$ are would-be Goldstone modes. They are eaten by massive 4D modes from $A_\mu$.

Furthermore, the masses of the 4D modes from the ghost fields also depend on $\xi$. It is observed that the masses of the 4D modes from the ghost fields $M_{\rm ghost}^2(\phi_{(\hat {\bm n})})$ are equal to the masses of $A_\mu$ as in eq.~\eqref{m001}, multiplied by $\xi$. Thus, we have
$M_{\rm ghost}^2(\phi_{(\hat {\bm n})})=M_{\xi}^2(\phi_{(\hat {\bm n})})$.

One sees that $\phi_{(0,0,0,0)}$ is a massless zero mode if $\phi$ has $\eta_{m'}(\phi)=q_k(\phi)=0$ $(k=1,2)$. For example, $A_\mu^{(1)}$ and $A_\mu^{(2)}$ have massless zero modes for any values of the WL phases. On the other hand, if $\phi$ has $q_k(\phi)\neq 0$, then $\phi$ couples to the WL phases, and their masses depend on the values of the WL phases. For example, $A_\mu^{(12)}$ have massless zero modes only if the combination $2a_{m'}^1-a_{m'}^2$ is an integer. Otherwise, $A_\mu^{(12)}$ has no massless mode.  As seen below, 4D gauge fields coupled to the flux background have no massless zero mode. Thus, in this case, the flux background in eq.~\eqref{su3flux} induces  the spontaneous breaking $SU(3)\to SU(2)\times U(1)$, and the WL phases can further break the gauge symmetry as $SU(2)\times U(1)\to U(1)^2$ depending on their values. 

\subsubsection{Matter fields in the $q_3(\phi)\neq 0$ case}
\label{sec:mattermass}
Next, we discuss the masses of 4D modes appearing from an 8D matter field $\phi$ having $q_3(\phi)\neq 0$. In this case, $\phi$ couples to the flux background. Then, their 4D modes receive mass contributions associated with Landau-level excitations, which we will call Landau-level contributions. Again, the discussion in Ref.~\cite{KYS} is straightforwardly generalized to obtain masses of 4D modes in the $q_3(\phi)\neq 0$ case.  Scalar and fermion fields have $\Sigma_{56}=0$ and $\Sigma_{56}=\pm 1/2$, respectively. Their 4D modes are labeled by four integers, $\hat \ell\geq 0$, $d\in \{1,\dots ,|q_3(\phi)|\}$, and $\hat n_7,\hat n_8\in \mathbb Z$. Hence, we denote the 4D mode as $\phi_{(\hat \ell,d,\hat n_7,\hat n_8)}$. Their masses are summarized as follows:
\begin{align}\label{mattermass}
  M^2(\phi_{(\hat \ell,d,\hat n_7,\hat n_8)})
  &=4\pi  |q_3(\phi)|\left[\hat \ell+1/2+  \Sigma_{56}(\phi)\right]
    +M_{78}^2(\phi), \qquad \hat \ell\geq 0, 
\end{align}
where we have used $\hat f=2\pi /g$ from eq.~\eqref{su3flux}. Note that these masses are consistent with the known mass formula~\cite{Bachas}. 

From eq.~\eqref{mattermass}, one sees that scalar fields have no massless modes at low energy. On the other hand, for fermions with $\Sigma_{56}=-1/2$, the $\hat \ell=0$ mode can be massless if $M_{78}^2(\phi)=0$ is satisfied. Such a massless mode has a degeneracy labeled by $d=1,\dots,|q_3(\phi)|$. 

\subsubsection{Gauge fields in the $q_3(\phi)\neq 0$ case}
\label{sec:q3neq0case}
Finally, we discuss the masses of 4D modes appearing from 8D gauge fields $\phi$ having $q_3(\phi)\neq 0$, where $\phi$ couples to the flux background. As in the matter case, the 4D modes are labeled by four integers, and we denote the 4D mode as $\phi_{(\hat \ell,d,\hat n_7,\hat n_8)}$.

The mass spectrum depends on the helicity $\Sigma_{56}(\phi)$ of the gauge fields. One sees that $\Sigma_{56}(A_\mu)=\Sigma_{56}(A_{m'})=0$, whereas $A_{5}$ and $A_{6}$ has the helicity $\pm 1$.  In addition, there appears to be a dependence on the gauge parameter $\xi$ in the mass spectrum of 4D modes in this case.  The mass spectrum of 4D modes from $A_\mu$ is determined independently to $\xi$ and is the same as in eq.~\eqref{mattermass}. On the other hand, the mass spectrum of 4D modes from $A_m$ depends on $\xi$.

We first discuss masses of 4D modes from $A_m$ in the $\xi=1$ case. From $A_5$ and $A_6$, we obtain 4D modes that have masses as 
\begin{align}\label{fluxmass56}
  M^2(\phi_{(\hat \ell,d,\hat n_7,\hat n_8)})&=4\pi |q_3(\phi)|(\hat \ell+1/2\pm 1)+
  M_{78}^2(\phi), \qquad \hat \ell\geq 0. 
\end{align}
On the other hand, from $A_{m'}$, we obtain 4D modes that have masses as 
\begin{align}\label{fluxmass78}
  M^2(\phi_{(\hat \ell,d,\hat n_7,\hat n_8)})&=4\pi |q_3(\phi)|(\hat \ell+1/2)+
  M_{78}^2(\phi), \qquad \hat \ell\geq 0. 
\end{align}
One sees that eqs.~\eqref{fluxmass56} and~\eqref{fluxmass78} are summarized as in eq.~\eqref{mattermass}. We note that 4D modes from ghost fields have the same mass as in eq.~\eqref{fluxmass78}.

For arbitrary $\xi$, 4D modes from $A_m$ mix. As shown in appendix~\ref{app:massesxi}, the mass spectrum of 4D modes from $A_m$ is given by eqs.~\eqref{fluxmass56} and~\eqref{fluxmass78}, and 
\begin{align}\label{q3neqmass}
  M^2_{\xi}(\phi_{(\hat \ell,d,\hat n_7,\hat n_8)})&=\xi\left[4\pi |q_3(\phi)|(\hat \ell+1/2)+
M_{78}^2(\phi)\right], \quad \hat \ell\geq 0. 
\end{align}
The 4D modes from ghost fields also have the same masses as in eq.~\eqref{q3neqmass}. 

As seen in eq.~\eqref{fluxmass56}, 4D modes $\phi_{(0,d,\hat n_7,\hat n_8)}$ receive a negative Landau-level contribution, which potentially makes some of the 4D modes tachyonic. The other 4D modes from the extra-dimensional gauge fields coupled with the flux are massive.  To eliminate tachyonic states in a low-energy theory, the values of WL phases included in $\hat N_7$ and $\hat N_8$ are constrained.

\subsection{Stabilizing potentially tachyonic states through Wilson line phases}
\label{sec:stab}
As seen in the previous subsection, some of the 4D modes from flux-coupled gauge fields can potentially be tachyonic due to negative Landau-level contributions. Since the existence of tachyonic states in a low-energy theory implies vacuum instability, we have to eliminate them from the 4D mass spectrum.  As noted, masses of 4D modes generally depend on values of WL phases.  A condition to eliminate tachyonic states can be regarded as a constraint on the values of WL phases.

In our setup, 4D modes from $A_{z_1}^{(23)}$, $A_{z_1}^{(31)}$, $A_{\bar z_1}^{(23)}$, and $A_{\bar z_1}^{(31)}$ include potentially tachyonic states. We examine their masses and derive constraints on WL phases in the $SU(3)$ model. For the lowest Landau-level excitations, their masses are given by
\begin{align}
  A_z^{(23)},A_{\bar z}^{(23)}:
  &\quad
M^2(\phi_{(0,d,\hat n_7,\hat n_8)}) =   -6\pi +
    (2\pi \hat M_w)^2 \left([\hat n_7-a_7^1+2a_7^2]^2
    +[\hat n_8-a_8^1+2a_8^2]^2\right),\\
  A_z^{(31)},A_{\bar z}^{(31)}:
  &\quad
M^2(\phi_{(0,d,\hat n_7,\hat n_8)}) =    -6\pi +
    (2\pi \hat M_w)^2 \left([\hat n_7-a_7^1-a_7^2]^2
    +[\hat n_8-a_8^1-a_8^2]^2\right). 
\end{align}
The WL phases contribute to the masses,  stabilizing the tachyonic states depending on their values. Sufficient conditions to make all masses non-negative are given by 
\begin{align}\label{constraint}
   [\hat n_7-a_7^1+2a_7^2]^2
  +[\hat n_8-a_8^1+2a_8^2]^2
  &\geq
  {3 \over 
  2\pi \hat M_w^2 },  \\ \label{constraint2}
 [\hat n_7-a_7^1-a_7^2]^2
  +[\hat n_8-a_8^1-a_8^2]^2
  &\geq
  {3 \over 
  2\pi \hat M_w^2 }, 
\end{align}
for any $\hat n_7$ and $\hat n_8$.

\begin{figure}
    \centering
    \includegraphics[width=0.5\textwidth]{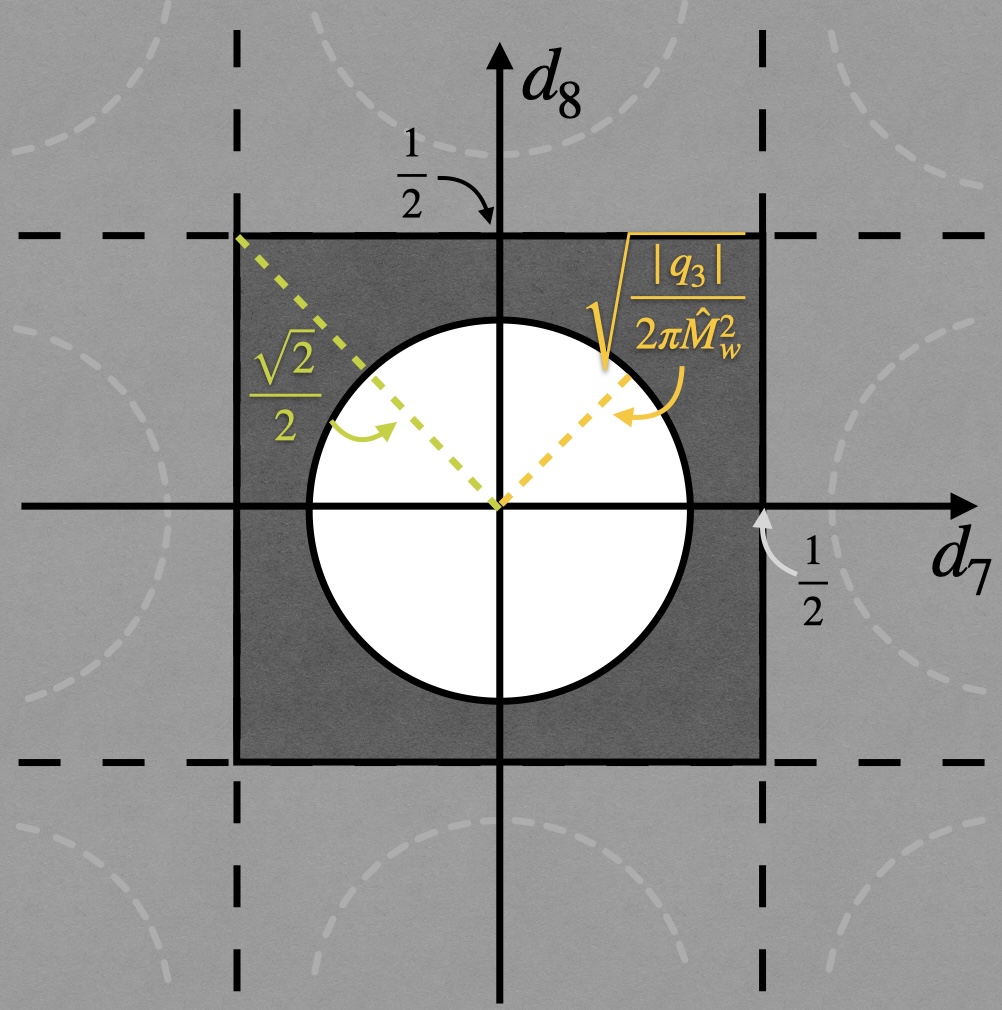}
    \caption{Tachyonic region. This illustration depicts the parameter space ($d_7$,$d_8$) where tachyonic states appear, represented by the white circle. The dark gray square delimits the region where no such modes are present.}\bigskip
    \label{fig1}
\end{figure}

To facilitate the discussion of the above constraints, we consider
\begin{align}
    & [\hat n_7+d_7]^2
  +[\hat n_8+d_8]^2
  \geq
  {3 \over 
      2\pi \hat M_w^2 }, \qquad {\rm for} \quad \hat n_7, \hat n_8\in \mathbb Z.
      \label{constraintd}
\end{align}
Since $d_7$ and $d_8$ have a shift symmetry modulo 1, we can choose the region $-1/2 \leq d_{7,8} \leq 1/2$ to simplify our analysis. For large values of $\hat n_7$ and $\hat n_8$, constraints on $d_7$ and $d_8$ from eq.~\eqref{constraintd} become weak. On the other hand, for $\hat n_7=\hat n_8=0$, constraints on $d_7$ and $d_8$ are stronger.  The tachyonic region is visually clarified in Figure~\ref{fig1}, where it is represented by the white circle. The dark gray zone represents the region where the previous constraints in eq.~\eqref{constraintd} and $-1/2 \leq d_{7,8} \leq 1/2$ are satisfied.  To obtain a parameter region where tachyonic states disappear, we obtain a constraint on the possible values of $\hat M_w^2$, given by
\begin{align}
 \hat M_w^2
  >
  {3 \over 
  \pi  }.
\end{align}

Solutions of the constraints on WL phases given by eqs.~\eqref{constraint} and~\eqref{constraint2} are not simple to be clarified analytically. We have checked that there are allowed parameter regions of the WL phases for ${\cal O}(1)$ values of $\hat M_w$. In the following, we constrain the values of the WL phases to satisfy the conditions in eqs.~\eqref{constraint} and~\eqref{constraint2}.

%
\section{One-loop effective potential in the $SU(3)$ model}
\label{sec:effpot}
%
As shown in previous sections, masses of 4D modes depend on values of the WL phases $a^k_{m'}$. An important consequence is that values of the WL phases are constrained as shown in eqs.~\eqref{constraint} and~\eqref{constraint2}. Although they are continuous moduli and have flat potential at tree level, quantum corrections generate effective potentials for the WL phases. Thus, a natural question is whether a vacuum that is consistent with the constraint exists. In this section, to discuss the vacuum structure, we present the one-loop effective potential for $a^k_{m'}$, showing contributions from each type of field in our setup.  For a detailed derivation of the potential contributions, please refer to appendix~\ref{app:surf}. For simplicity of our discussion, we hereafter fix the gauge fixing parameter as $\xi=1$.

\subsection{Contributions from flux-blind fields}
Let $\phi$ be a flux-blind field, that is, having $q_3(\phi)=0$. Their 4D modes masses are given by eq.~\eqref{m001}.  To simplify the notation, we define
\begin{align}
  d_{m'}(\phi)=-q_1(\phi)a_{m'}^1-q_2(\phi)a_{m'}^2
  +\eta_{m'}(\phi), \qquad m'=7,8.
  \label{d}
\end{align}
The effective potential contribution for the WL phases is generated by integrating out 4D modes from $\phi$ and depends on $q_1(\phi)$, $q_2(\phi)$, and $\eta_{m'}(\phi)$. We write this contribution as $V^{({\rm FB})}_{(\eta_7,\eta_8)}((-1)^{\hat F}N_{\rm deg},q_1,q_2)$, where $({\rm FB})$ refers to ``flux blind''.  Here, $N_{\rm deg}(\phi)$ is a positive integer that gives the real dof of $\phi$, and $\hat F$ is the fermion number of $\phi$.

Using the standard procedure, we obtain the effective potential contribution as 
\begin{align}
\notag  &
V^{({\rm FB})}_{(\eta_7,\eta_8)}((-1)^{\hat F}N_{\rm deg},q_1,q_2)\\
\label{VFB} & =
-(-1)^{\hat F}N_{\rm deg}{3\over \pi^4\hat M_w^2}
        \\
\notag  &\quad \times \sum_{\omega_5,\omega_6\in \mathbb Z}    \left(
    2\sum_{\omega\geq 1}  {\cos(2\pi \omega d_7)+\cos(2\pi \omega d_8)\over 
    \left[\omega_5^2+\omega_6^2+\omega^2/\hat M_w^2\right]^{4}}
    + 4\sum_{\omega_7,\omega_8\geq 1}
    {\cos(2\pi \omega_7 d_7)\cos(2\pi \omega_8 d_8)\over 
    \left[\omega_5^2+\omega_6^2+(\omega_7^2+\omega_8^2)/\hat M_w^2\right]^{4}}
    \right).
\end{align} 
Note that the potential contribution also depends on values of $\hat M_w$, the relative size of the extra dimensions. 

The potential in eq.~\eqref{VFB} is finite and has no ultraviolet (UV) and infrared (IR) divergences. The UV finiteness is expected since the WL phases are associated with non-contractible loops along extra dimensions and are intrinsically non-local dof. The integers $\omega$, $\omega_7$, and $\omega_8$ are often referred as winding numbers. One sees that local divergences are contained in the terms corresponding to vanishing winding numbers, which are independent of WL phases. We have discarded such constants in eq.~\eqref{VFB}. For more details, see appendix~\ref{app:surf}.

\subsection{Contributions from flux-coupled fields with  $\Sigma_{56}=0$ or $\pm 1/2$}
Now, let $\phi$ be a flux-coupled field with $\Sigma_{56}=0$ or $\pm 1/2$. Their 4D modes masses are given by eq.~\eqref{mattermass}. After integrating out the 4D modes, we obtain the contribution to the effective potential, denoted by $V^{(|\Sigma_{56}|)}_{(\eta_7,\eta_8)}((-1)^{\hat F}N_{\rm deg},q_1,q_2)$. Their contribution to the effective potential is given by 
\begin{align}\notag
  &
    V^{(0)}_{(\eta_7,\eta_8)}((-1)^{\hat F}N_{\rm deg},q_1,q_2)\\\notag
  &
    =
    -(-1)^{\hat F}{N_{\rm deg}|q_3| \over 128\pi^3\hat M_w^2}
    \bigg(  2\sum_{\omega\geq 1}[\cos(2\pi \omega d_7)+\cos(2\pi \omega d_8)]
\int_0^\infty dt\, t^{-4}
  { e^{-{\omega^2\over 4\pi \hat M_w^2t}}
\over 2\sinh(2\pi |q_3|t)}\\
&\qquad\qquad\qquad   + 4\sum_{\omega_7,\omega_8\geq 1}\cos(2\pi \omega_7 d_7)\cos(2\pi \omega_8 d_8)
  \int_0^\infty dt\,  t^{-4}
  { e^{-{\omega_7^2+\omega_8^2\over 4\pi \hat M_w^2t}}
\over 2\sinh(2\pi |q_3|t)
  }  \bigg),
  \label{effpSig01}
\end{align}
for $\Sigma_{56}=0$, and 
\begin{align}\notag
  &
    V^{(1/2)}_{(\eta_7,\eta_8)}((-1)^{\hat F}N_{\rm deg},q_1,q_2)\\\notag
  &=
    -(-1)^{\hat F}{N_{\rm deg}|q_3| \over 128\pi^3\hat M_w^2}  \bigg(  2\sum_{\omega\geq 1}[\cos(2\pi \omega d_7)+\cos(2\pi \omega d_8)]
\int_0^\infty dt\,  t^{-4}
  { e^{-{\omega^2\over 4\pi \hat M_w^2t}}
\over \tanh(2\pi |q_3|t)}\\
&\qquad\qquad\qquad   + 4\sum_{\omega_7,\omega_8\geq 1}\cos(2\pi \omega_7 d_7)\cos(2\pi \omega_8 d_8)
  \int_0^\infty dt\,  t^{-4}
  { e^{-{\omega_7^2+\omega_8^2\over 4\pi \hat M_w^2t}}
\over \tanh(2\pi |q_3|t)
                                    }  \bigg),
  \label{effpSighalf}
\end{align}
for $\Sigma_{56}=\pm 1/2$. In the above expression, $q_3=q_1+2q_2$ holds.  We note that the $\Sigma_{56}=\pm 1/2$ contribution is obtained from a pair of fields having $\Sigma_{56}=1/2$ and $-1/2$. 

As in the flux-blind case, the effective potential contributions in eqs.~\eqref{effpSig01} and~\eqref{effpSighalf} are free from UV and IR divergences. For fixed winding numbers, $\hat M_w$, and $q_3$, integrals with respect to $t$ in these contributions give numerical constants, which are suppressed for a large absolute value of winding numbers. 

\subsection{Contributions from flux-coupled fields with $\Sigma_{56}=\pm 1$}
\label{sec:pm1effsum}
If $\phi$ now corresponds to $A_{5,6}$, there appears a pair of fields having $\Sigma_{56}=\pm 1$. Let $V^{(1)}_{(\eta_7,\eta_8)}(N_{\rm deg},q_1,q_2)$ be a contribution from a pair of 4D modes having masses as in eq.~\eqref{fluxmass56}. The contribution is written as 
\begin{align}
&  V^{(1)}_{(\eta_7,\eta_8)}(N_{\rm
  deg},q_1,q_2)=
    -{N_{\rm deg}|q_3|\over 32\pi^2}\\\notag
&\qquad \times    \sum_{\hat n_7, \hat n_8 \in \mathbb{Z}} \int_{0}^{\infty} dt\, t^{-3} e^{-M_{78}^2 t}
    \left(   
        e^{-4\pi |q_3|(-1/2)t}+e^{-4\pi |q_3|(1/2)t}
    +2\sum_{\hat \ell\geq 1}              
    e^{-4\pi |q_3|(\hat \ell+1/2)t}\right), 
\end{align}
which corresponds to eq.~\eqref{dvphisigpm1}. This expression needs a more careful evaluation since it contains the contribution from the potentially tachyonic states as seen in section~\ref{sec:q3neq0case}.  Actually, tachyonic states are absent since we constrain the parameter region of the WL phases, as discussed in section~\ref{sec:stab}.  In appendix~\ref{sec:sig1potdel}, we derive an expression of the contribution, which is free from UV and IR divergences.

Here, we only show the result.  We introduce
\begin{align}
    \Delta V_{\rm tac}=
  -{N_{\rm deg}|q_3|\over 32\pi^2}
  \sum_{\hat n_7, \hat n_8 \in \mathbb{Z}} \int_{0}^{\infty} dt\, t^{-3} e^{-\left(M_{78}^2-2\pi |q_3|\right) t}, 
\end{align}
which corresponds to the contribution from potentially tachyonic states. We obtain the expression of $\Delta V_{\rm tac}$ as
\begin{align}\label{Vtacdef1}
  \Delta V_{\rm tac}=  -{N_{\rm deg}|q_3|\over 32\pi^2}I_{\rm T}, 
\end{align}
where
\begin{align}\notag
I_{\rm T}    &=
    2\sum_{\omega\geq 1} \left[\cos(2\pi \omega d_7)+\cos(2\pi \omega d_8)\right]
    \left({32\pi^2\hat M_w^4\over \omega^2}
    +{8\pi^2|q_3|\hat M_w^2\over \omega^4}
    +{2\pi^2|q_3|^2\over \omega^6}
    \right)\\\notag
  &+
    4\sum_{\omega_7,\omega_8\geq 1}
    \cos(2\pi \omega_7 d_7)\cos(2\pi \omega_8 d_8)
    \left({32\pi^2\hat M_w^4\over \omega_7^2+\omega_8^2}
    +{8\pi^2|q_3|\hat M_w^2\over (\omega_7^2+\omega_8^2)^2}
    +{2\pi^2|q_3|^2\over (\omega_7^2+\omega_8^2)^3}
    \right)\\ \label{tachyonic}
  &+\sum_{\hat n_7, \hat n_8 \in \mathbb{Z}} \sum_{k\geq 1} \frac{(2\pi |q_3|)^{2+k}}{(k+2)(k+1) k (M_{78}^2)^{k}}
    -\sum_{\left(\hat n_7, \hat n_8\right) \neq(0,0)} \frac{(2\pi |q_3|)^3}{6(2\pi \hat M_w)^2(\hat n_7^2+\hat n_8^2)}.
\end{align}
In the last term, the summations over $\hat n_7$ and $\hat n_8$ are taken for all integers except for $(\hat n_7, \hat n_8) =(0,0)$. Using the above, the potential is given by 
\begin{align}\notag
&  V^{(1)}_{(\eta_7,\eta_8)}(N_{\rm
  deg},q_1,q_2)=\Delta V_{\rm tac}
                     -{N_{\rm deg}|q_3|\over 128\pi^3\hat M_w^2}\\
  &\qquad \times 
\bigg(  2\sum_{\omega\in \geq 1}[\cos(2\pi \omega d_7)+\cos(2\pi \omega d_8)]
\int_0^\infty dt\,  t^{-4}
    { e^{-{\omega^2\over 4\pi \hat M_w^2t}}e^{- 2\pi |q_3|t}\over
     \tanh(2\pi |q_3|t)}
    \notag \\ 
&\qquad   + 4\sum_{\omega_7,\omega_8\geq 1}\cos(2\pi \omega_7 d_7)\cos(2\pi \omega_8 d_8)
  \int_0^\infty dt\,  t^{-4}
  { e^{-{\omega_7^2+\omega_8^2\over 4\pi \hat M_w^2t}}e^{- 2\pi |q_3|t}\over
     \tanh(2\pi |q_3|t)}
  \bigg),
  \label{effpSigone}
\end{align}
which is finite. We note that in the derivation of the potential, we have subtracted infinite constants that are independent of WL phases. 

\subsection{Total effective potential}
\label{sec:totalep}
From the above, we obtain the total effective potential. We first discuss the contribution from $A_M^{(12)}$.  We call these fields and related ghosts the (12)-sector. In this sector, there are ghost fields $c^{(12)}$ and $c^{(21)}$ that obey $(c^{(12)})^\dag \neq c^{(21)}$ since they are complex. Thus, the ghosts have, in total, four real dof. The contributions from each field in this sector are the same except for the overall sign, and the effective real bosonic dof of the contribution to the effective potential is given by $8\times 2-2\times2 =12$. Thus, the effective potential from the (12)-sector is
\begin{align}
  V^{[12]}(\hat M_w)&=
V^{({\rm FB})}_{(0,0)}(12,2,-1,\hat M_w).
\end{align}

Next, we discuss the contribution from $A_M^{(23)}$.  We call these fields and related ghosts as the (23)-sector.  In this sector, $A_\mu^{(23)}$, $A_{z_2}^{(23)}$, $A_{\bar{z}_2}^{(23)}$, and ghosts have $\Sigma_{56}=0$. Thus, the effective real bosonic dof of the contribution to the effective potential can be counted as $6\times 2-2\times 2=8$. On the other hand, $A_{z_1}^{(23)}$ and $A_{\bar {z}_1}^{(23)}$ have $\Sigma_{56}=\pm 1$. Thus, the contribution from the (23)-sector is given by
\begin{align}
  V^{[23]}(\hat M_w)
  &=
    V^{(0)}_{(0,0)}(8,-1,2,\hat M_w)
    +V^{(1)}(2,-1,2,\hat M_w).
\end{align}

Finally, we discuss the contribution from $A_M^{(31)}$.  We call these fields and related ghosts as the (31)-sector. From a similar discussion as done above, we have
\begin{align}
  V^{[31]}(\hat M_w)
  &=     V^{(0)}_{(0,0)}(8,-1,-1,\hat M_w)
    +V^{(1)}(2,-1,-1,\hat M_w).
\end{align}

There are no fields that couple to the WL phases in the gauge sector other than the above. Thus, from the gauge fields and ghosts, we obtain the effective potential $V^{[\rm pYM]}(a_{m'}^k; \hat M_w)$ for the WL phases $a_{m'}^k$ as follows  
\begin{align}
  V^{[\rm pYM]}(a_{m'}^k; \hat M_w)
  &=V^{[12]}(\hat M_w)+V^{[23]}(\hat M_w)+V^{[31]}(\hat M_w)\\\notag
  &=
V^{({\rm FB})}_{(0,0)}(12,2,-1,\hat M_w)
  +V^{(0)}_{(0,0)}(8,-1,2,\hat M_w)+V^{(0)}_{(0,0)}(8,-1,-1,\hat M_w)\\
  & +V^{(1)}(2,-1,2,\hat M_w) +V^{(1)}(2,-1,-1,\hat M_w).
    \label{pYMpot}
\end{align}

Next, we discuss the effective potentials generated by bulk matter fields. Let $V^{[\phi]}_{(\eta_7,\eta_8)}(a_{m'}^k; \hat M_w)$ be a contribution from a matter field $\phi$, where $\eta_7,\eta_8\in \{0,1/2\}$ indicates the periodicity of $\phi$. For scalar fields, we obtain
\begin{align}
  V_{(\eta_7,\eta_8)}^{[\Phi_{\bm 3}]}(a_{m'}^k; \hat M_w)
  &=
    V^{(0)}_{(\eta_7,\eta_8)}(2,1,0,\hat M_w)
    +V^{(0)}_{(\eta_7,\eta_8)}(2,-1,1,\hat M_w)
    +V^{(0)}_{(\eta_7,\eta_8)}(2,0,-1,\hat M_w), \\\notag
  V^{[\Phi_{\bm 6}]}_{(\eta_7,\eta_8)}(a_{m'}^k; \hat M_w)
  &=
    V^{(0)}_{(\eta_7,\eta_8)}(2,2,0,\hat M_w)
    +V^{(0)}_{(\eta_7,\eta_8)}(2,-2,2,\hat M_w)
    +V^{(0)}_{(\eta_7,\eta_8)}(2,0,-2,\hat M_w)\\ 
  &
    + V^{(0)}_{(\eta_7,\eta_8)}(2,0,1,\hat M_w)
    + V^{(0)}_{(\eta_7,\eta_8)}(2,-1,0,\hat M_w)
    + V^{(0)}_{(\eta_7,\eta_8)}(2,1,-1,\hat M_w)  ,\\
  V^{[\Phi_{\bm 8}]}_{(\eta_7,\eta_8)}(a_{m'}^k; \hat M_w)
  & =
    V^{({\rm FB})}_{(\eta_7,\eta_8)}(4,2,-1,\hat M_w)
    +V^{(0)}_{(\eta_7,\eta_8)}(4,-1,2,\hat M_w)+
    V^{(0)}_{(\eta_7,\eta_8)}(4,-1,-1,\hat M_w), \\\notag
  V^{[\Phi_{\bm{10}}]}_{(\eta_7,\eta_8)}(a_{m'}^k; \hat M_w)
  &=
    V^{({\rm FB})}_{(\eta_7,\eta_8)}(4,2,-1,\hat M_w)
    +V^{(0)}_{(\eta_7,\eta_8)}(4,1,1,\hat M_w)
    +V^{(0)}_{(\eta_7,\eta_8)}(4,-1,2,\hat M_w)\\
  & +V^{(0)}_{(\eta_7,\eta_8)}(2,3,0,\hat M_w)
    +V^{(0)}_{(\eta_7,\eta_8)}(2,-3,3,\hat M_w)
    +V^{(0)}_{(\eta_7,\eta_8)}(2,0,-3,\hat M_w).
\end{align}
For 8D Dirac fermions, we obtain
\begin{align}
  V_{(\eta_7,\eta_8)}^{[\Psi_{\bm 3}]}&(a_{m'}^k; \hat M_w)
  =
    V^{(1/2)}_{(\eta_7,\eta_8)}(-8,1,0,\hat M_w)
    +V^{(1/2)}_{(\eta_7,\eta_8)}(-8,-1,1,\hat M_w)
    +V^{(1/2)}_{(\eta_7,\eta_8)}(-8,0,-1,\hat M_w), \\\notag
  V^{[\Psi_{\bm 6}]}_{(\eta_7,\eta_8)}&(a_{m'}^k; \hat M_w)
  =
    V^{(1/2)}_{(\eta_7,\eta_8)}(-8,2,0,\hat M_w)
    +V^{(1/2)}_{(\eta_7,\eta_8)}(-8,-2,2,\hat M_w)
    +V^{(1/2)}_{(\eta_7,\eta_8)}(-8,0,-2,\hat M_w)\\
  &\qquad
    + V^{(1/2)}_{(\eta_7,\eta_8)}(-8,0,1,\hat M_w)
    + V^{(1/2)}_{(\eta_7,\eta_8)}(-8,-1,0,\hat M_w)
    + V^{(1/2)}_{(\eta_7,\eta_8)}(-8,1,-1,\hat M_w)  ,\\ \notag
  V^{[\Psi_{\bm 8}]}_{(\eta_7,\eta_8)}&(a_{m'}^k; \hat M_w)
  =
    V^{({\rm FB})}_{(\eta_7,\eta_8)}(-32,2,-1,\hat M_w)
    +V^{(1/2)}_{(\eta_7,\eta_8)}(-16,-1,2,\hat M_w)\\\label{fermion8}
  &\qquad
    + V^{(1/2)}_{(\eta_7,\eta_8)}(-16,-1,-1,\hat M_w), \\\notag
  V^{[\Psi_{\bm{10}}]}_{(\eta_7,\eta_8)}&(a_{m'}^k; \hat M_w)
  =V^{({\rm FB})}_{(\eta_7,\eta_8)}(-32,2,-1,\hat M_w)
      +V^{(1/2)}_{(\eta_7,\eta_8)}(-16,1,1,\hat M_w)
  +V^{(1/2)}_{(\eta_7,\eta_8)}(-16,-1,2,\hat M_w)\\
&\qquad
    +V^{(1/2)}_{(\eta_7,\eta_8)}(-8,3,0,\hat M_w)
    +V^{(1/2)}_{(\eta_7,\eta_8)}(-8,-3,3,\hat M_w)
    +V^{(1/2)}_{(\eta_7,\eta_8)}(-8,0,-3,\hat M_w).
\end{align}
In the above, we have used the fact that the mass spectrum of 4D modes from $\phi$ is unchanged under $(q_1(\phi),q_2(\phi))\to (-q_1(\phi),-q_2(\phi))$. Namely, $V^{(s)}_{(\eta_7,\eta_8)}((-1)^{\hat F}N_{\rm deg},q_1,q_2,\hat M_w) =V^{(s)}_{(\eta_7,\eta_8)}((-1)^{\hat F}N_{\rm deg},-q_1,-q_2,\hat M_w)$ holds for $s=0,1/2,1$.

%
\section{Vacuum structure in the $SU(3)$ model}
\label{sec:vacuum}
%
In this section, we will explore the vacuum structure. We aim to find a minimum point in the effective potential, which will suggest the existence of a stable vacuum configuration. We start by doing an analytical discussion to understand qualitative features of the effective potential. Some critical points are naturally characterized by simple fractional numbers and have periodic properties. Through this discussion, it was possible to identify candidates for minimum points. To facilitate the analysis, we proceed to numerical calculations. Although there seems to be no stable vacuum in the pure Yang-Mills case, by adding matter fields, we can find minimum points.  

\subsection{Analytical discussion of the potentials}
\label{sec:anpot}
To see the qualitative features of the effective potentials, we first examine them analytically. We see that critical points of the potentials are expected to exist in the field space of the WL phases satisfying the constraint to eliminate tachyonic states in eqs.~\eqref{constraint} and~\eqref{constraint2}. It is convenient to notice that the expressions contributing to the effective potential from a field $\phi$, that is
eqs.~\eqref{VFB},~\eqref{effpSig01},~\eqref{effpSighalf}, and~\eqref{effpSigone}, have the following structure:
\begin{align} \label{potstructure} \notag
  V(\phi)&=
 A(\phi)\sum_{\omega\in \mathbb Z_+}( \cos(2\pi \omega d_7)+\cos(2\pi \omega d_8))
+ B(\phi)\sum_{\omega_7,\omega_8\in \mathbb Z_+}\cos(2\pi \omega_7 d_7)\cos(2\pi \omega_8 d_8) \\
&\qquad + \sum_{\hat{n}_{7}, \hat{n}_{8} \in \mathbb{Z}} \sum_{k=1}^{\infty}C(\phi) \left[\frac{1}{\left(\hat{n}_{7}+d_{7}\right)^{2}+\left(\hat{n}_{8}+d_{8}\right)^{2}}\right]^{k}
\end{align}
where $A(\phi)$, $B(\phi)$ and $C(\phi)$ are constants that depend on the field $\phi$. 

Let us begin with the pure Yang-Mills setup. The critical points are found when the first derivatives of eq.~\eqref{pYMpot} with respect to the WL phases vanish. As can be seen from eq.~\eqref{potstructure}, the derivative of the first line always generates sine functions; the derivative can be factorized by the following functions:
\begin{align} \label{pYMderivative}
\sin \left(2 \pi \omega\left(-2 {a}_{m'}^{1}+{a}_{m'}^2\right)\right),\quad \sin \left(2 \pi \omega\left({a}_{m'}^{1}-2 {a}_{m'}^2\right)\right), \quad \sin \left(2 \pi \omega\left({a}_{m'}^{1}+{a}_{m'}^2\right)\right), 
\end{align}
where $\omega$ is an integer. From the second line in eq.~\eqref{potstructure}, we obtain $2d_7=2d_8=0$ mod~1 as the condition for an extremum.

\begin{figure}
    \centering
    \includegraphics[width=0.6\textwidth]{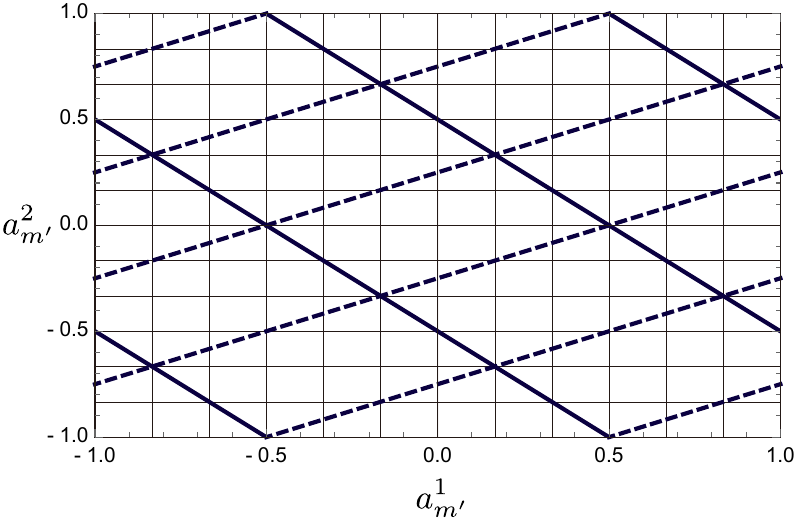} 
    \caption{Solutions of eq.~\eqref{WLcond1}. The horizontal and vertical axes show
      the values of $a_{m'}^1$ and $a_{m'}^2$, respectively. Intersection points of
      the solid and dashed lines on this figure correspond to solutions of eq.~\eqref{WLcond1}.}
    \label{figWLcon1}
\end{figure}

One possible solution of an extremum of the potential is to analyze the case where the sine functions in eq.~\eqref{pYMderivative} and the derivatives of the last line in eq.~\eqref{potstructure} vanish simultaneously. Starting with the latter condition, we find that the only possible critical points outside the tachyonic region satisfy $d_{m'} = 1/2$ mod 1, which implies
\begin{align}\label{WLcond1}
  a_{m'}^1-2a_{m'}^2=1/2~~{\rm mod}~1, \quad {\rm and} \quad
  a_{m'}^1+a_{m'}^2=1/2~~{\rm mod}~1.
\end{align}
The solution of this condition can be found in fig.~\ref{figWLcon1}. The WL phases have mod 1 property. If we restrict their values as $0\leq a_{m'}^k< 1$, the solutions are given by
\begin{align}\label{WLsol1}
  (a_{m'}^1,a_{m'}^2)=(1/2,0), ~
  (1/6,1/3),~
  (5/6,2/3).
\end{align}
More generally, we can write all solutions as
\begin{align}\label{WLsol2}
  (a_{m'}^1,a_{m'}^2)=\left((3-2n_{m'})/6,2n_{m'}/6+n'_{m'}\right), \qquad n_{m'},n'_{m'}\in \mathbb Z.
\end{align}

One sees that the solutions in eq.~\eqref{WLsol2} also satisfy the condition that the sine functions in eq.~\eqref{pYMderivative} vanish. Therefore, the values of the WL phases in eq.~\eqref{WLsol2} are critical points, candidates for minima. A notable point is that any solution in eq.~\eqref{WLsol2} gives the same physical consequences in the pure Yang-Mills case. To see this, it is convenient to examine the WL phase factors $W_{m'}=\exp(igL'\vev{A_{m'}})$. Note that we can always make VEVs of WL phases $\vev{A_{m'}}$ vanish by a gauge transformation without changing physical consequences. After eliminating $\vev{A_{m'}}$, the boundary conditions in eqs.~\eqref{AmBC} and~\eqref{matbc1} change and contain the WL phase factors.  As a result, the low-energy mass spectrum remains unchanged~\cite{KYS,Haba:2002py,Haba:2003ux,Kawamura:2022ecd}.

If the WL phases take the values in eq.~\eqref{WLsol2}, we find that
\begin{align}
  W_{m'}&={\rm diag}(e^{2\pi i (3-2n_{m'})/6},e^{2\pi i (4n_{m'}-3)/6},e^{2\pi i (-2n_{m'}/6)})\\
  &=e^{-2\pi i n_{m'}/3}{\rm diag}(-1,-1,1).\label{wmcm}
\end{align}
Let us introduce 
\begin{align}
  C_{m'}=e^{-2\pi i n_{m'}/3}{\rm diag}(1,1,1)\in \mathbb Z_3,
  \qquad \hat W_{m'}={\rm diag}(-1,-1,1)=e^{2\pi i H_3/2} ,
\end{align}
where $\mathbb Z_3$ is the center subgroup of $SU(3)$. Then, we obtain $W_{m'}=C_{m'}\hat W_{m'}$.  The center element $C_{m'}$ depends on $n_{m'}$, but $\hat W_{m'}$ does not.  Since the adjoint representation of $SU(3)$ is neutral under the subgroup $\mathbb Z_3$, the solutions in eq.~\eqref{WLsol2} are not discriminated for any values of $n_{m'}$ in the pure Yang-Mills case.

One also sees from eq.~\eqref{wmcm} that the Wilson line phase factors under the solution in eq.~\eqref{WLsol2} are along the same direction to the flux background, $\bm f\propto H_3$, up to a center element $C_{m'}$. Thus, the contribution to the effective potential from the potentially tachyonic states tends to align the WL phases with the flux background at an extremum. At a vacuum, the remaining gauge symmetry is spanned by the generators   $t_a$ that satisfy $[\bm f,t_a]=[W_{m'},t_a]=0$. Thus, around extrema in eq.~\eqref{WLsol2}, the WL phases do not induce further gauge symmetry breaking.

We can now investigate the effects of including matter fields. Their contribution to the effective potential was summarized in Section~\ref{sec:totalep}. It can be shown that the solutions in eq.~\eqref{WLsol2} also give extrema of matter contributions. For example, when adding fermions in the $\bm 8$ representation, as given by eq.~\eqref{fermion8}, and following the same procedure as done for the pure Yang-Mills case, we obtain the same result as the one in eq.~\eqref{WLsol2}. Hence, adding matter fields of $\bm 8$ with the periodic boundary condition gives no change in candidates for minimum points obtained from the pure Yang-Mills discussion above. However, if we consider general matter fields, candidates for minimum points might change.

\subsection{Potential structure with ansatzes}
To see the potential structure more closely, let us examine the potential numerically. There are four independent WL phases in this setup. Here, we examine the potential structure with some ansatzes. These ansatzes make it easier to see the potential structure since the independent values of WL phases are reduced.

We examine the cases where WL phase factors are aligned along $H_1$, $H_2$, and $H_3$. We call them ansatz 1, 2, and 3, respectively. Ansatz 3 is a particular case since the WL phase factors and the flux background are aligned as was discussed in the previous subsection.  In addition, we also examine the case with $a_7^k= a_8^k$, called ansatz 4. This is motivated by a symmetry of the potential.  Since the potential is unchanged under the exchange between $a_7^k$ and $a_8^k$, a 2D hypersurface defined by $a_7^k= a_8^k$ in the 4D field space of the WL phases seems to tend to have extrema. These ansatzes are summarized as follows:
\begin{align}
  &\textrm{Ansatz 1}:\qquad (a_7^1,a_7^2,a_8^1,a_8^2)=(b_1,0,b_2,0),\\
  &\textrm{Ansatz 2}:\qquad (a_7^1,a_7^2,a_8^1,a_8^2)=(0,b_1,0,b_2),\\
  &\textrm{Ansatz 3}:\qquad (a_7^1,a_7^2,a_8^1,a_8^2)=(b_1,2b_1,b_2,2b_2),\\
  &\textrm{Ansatz 4}:\qquad (a_7^1,a_7^2,a_8^1,a_8^2)=(b_1,b_2,b_1,b_2), 
\end{align}
where we have introduced $b_1,b_2\in\mathbb R$. With the above ansatzes, the WL phase factors in the fundamental representation are given by
\begin{align}
  &\textrm{Ansatz 1}:\quad
    W_7={\rm diag}(e^{2\pi i b_1},e^{-2\pi i b_1},1), \qquad
    W_8={\rm diag}(e^{2\pi i b_2},e^{-2\pi i b_2},1), 
  \\
  &\textrm{Ansatz 2}:\quad
    W_7={\rm diag}(1,e^{2\pi i b_1},e^{-2\pi i b_1}), \qquad
    W_8={\rm diag}(1,e^{2\pi i b_2},e^{-2\pi i b_2}), 
  \\
  &\textrm{Ansatz 3}:\quad 
    W_7={\rm diag}(e^{2\pi i b_1},e^{2\pi i b_1},e^{-4\pi i b_1}), \qquad
    W_8={\rm diag}(e^{2\pi i b_2},e^{2\pi i b_2},e^{-4\pi i b_2}), 
  \\
  &\textrm{Ansatz 4}:\quad 
    W_7=W_8={\rm diag}(e^{2\pi i b_1},e^{2\pi i (b_2-b_1)},e^{-2\pi i b_2}).
\end{align}

\begin{figure}[]
 \centering
     \includegraphics[width=0.24\linewidth,clip]{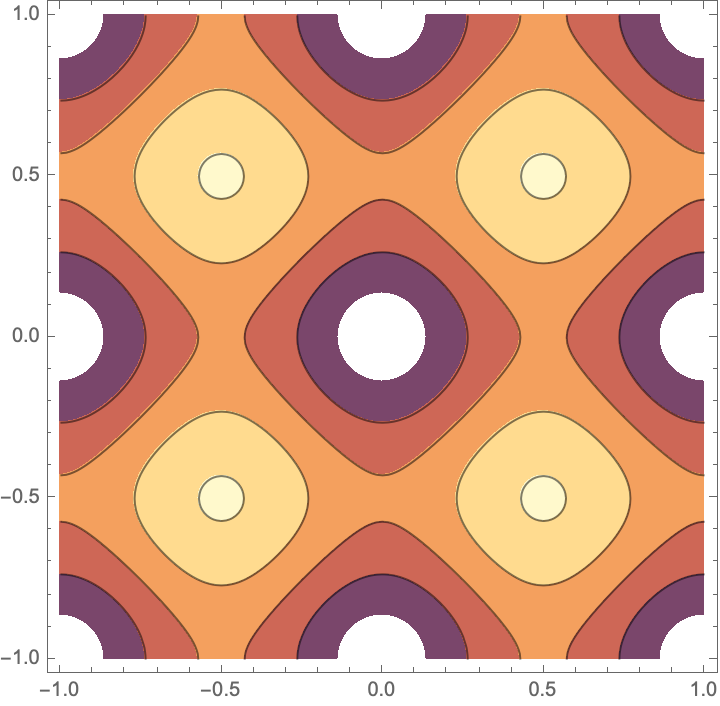}\hspace{0.1cm}
     \includegraphics[width=0.24\linewidth,clip]{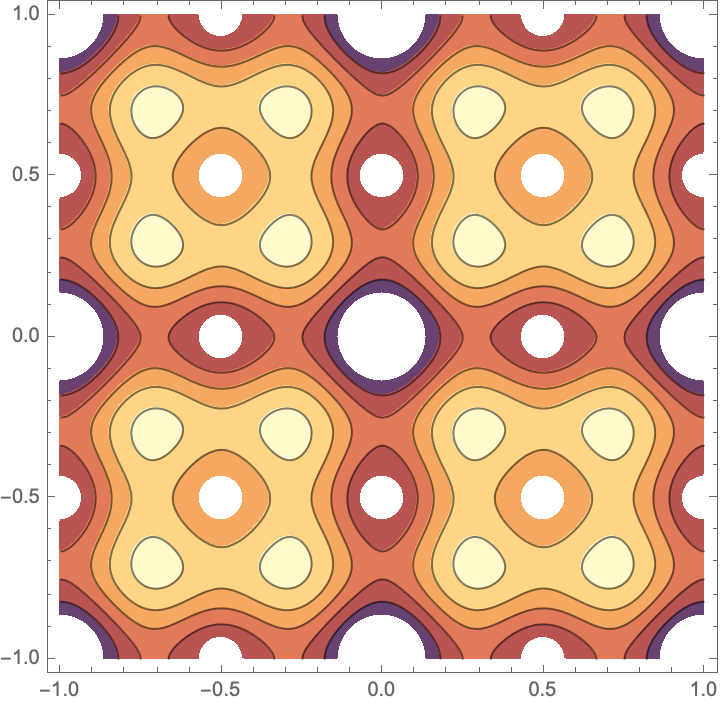}\hspace{0.1cm} 
     \includegraphics[width=0.24\linewidth,clip]{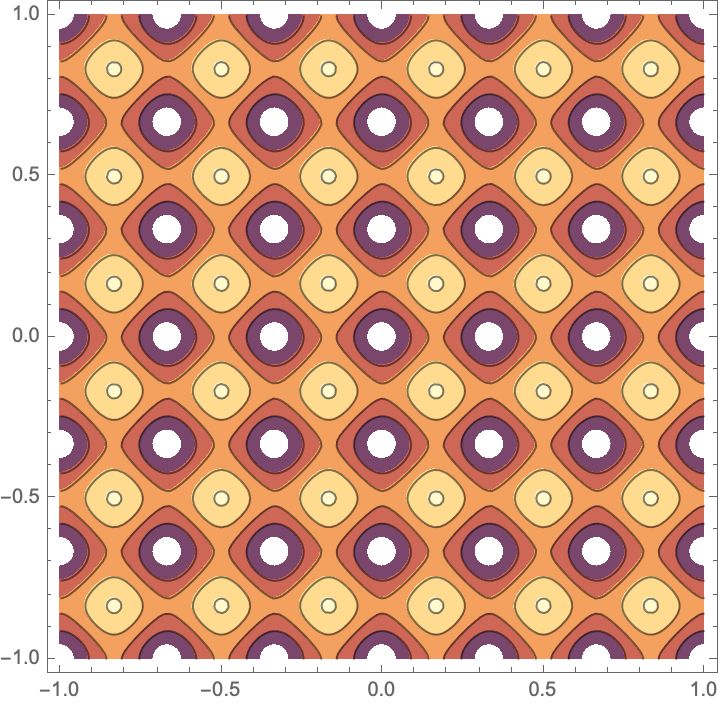}\hspace{0.1cm}
    \includegraphics[width=0.24\linewidth,clip]{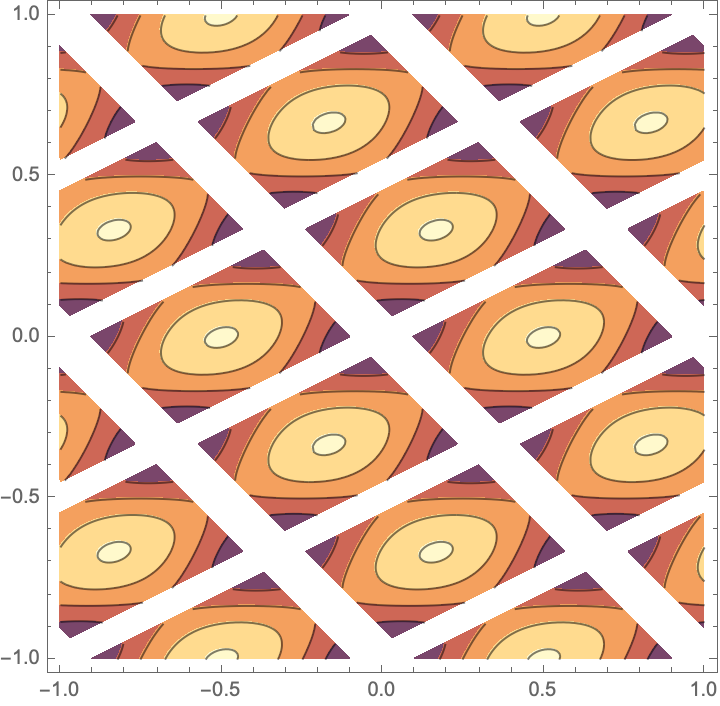} 
    \caption{Contour plots of the effective potential in the pure
      Yang-Mills case. From left to right, we assume Ansatz 1 to 4.
      In these contour plots, the horizontal (vertical) axis shows the
      value of $b_1$ ($b_2$). From light to dark colors in the plots,
      the potential decreases.  The white regions are excluded by
      the constraint on the WL phases in eqs.~\eqref{constraint}
      and~\eqref{constraint2}.  }\bigskip
 \label{fig_pYM}
\end{figure}

We numerically examine the effective potentials with these ansatzes. From now on, we take $ \hat M_w= 5.0$ as an example.  We begin by plotting the effective potential for the pure Yang-Mills case, as shown in Figure~\ref{fig_pYM}. In these contour plots, the horizontal (vertical) axis shows the value of $b_1$ ($b_2$).  From light to dark colors in the plots, the potential decreases. We also introduce the constraint on the WL phases in eqs.~\eqref{constraint} and~\eqref{constraint2}. In the contour plots, the excluded region from the constraint is shown by the white area.  In the numerical calculations, we introduce some cutoffs for the infinite summations of winding numbers in the potentials. In addition, the infinite summations of KK numbers $\hat n_{m'}$ and $k$ in $\Delta V_{\rm tac}$ in eq.~\eqref{Vtacdef1} are also truncated at some finite terms. The results are less sensitive to these cutoffs.

\begin{figure}[]
    \centering
\includegraphics[width=0.24\linewidth,clip]{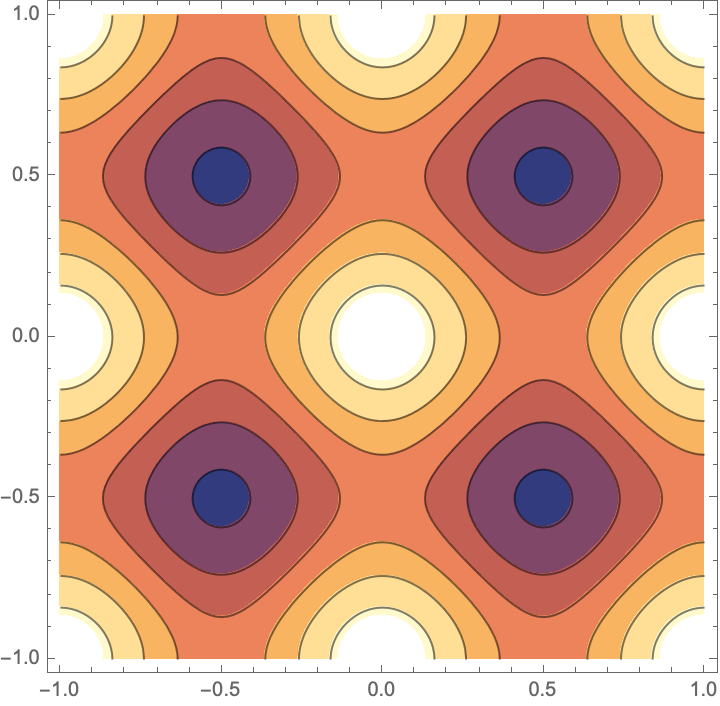}\hspace{0.1cm}
\includegraphics[width=0.24\linewidth,clip]{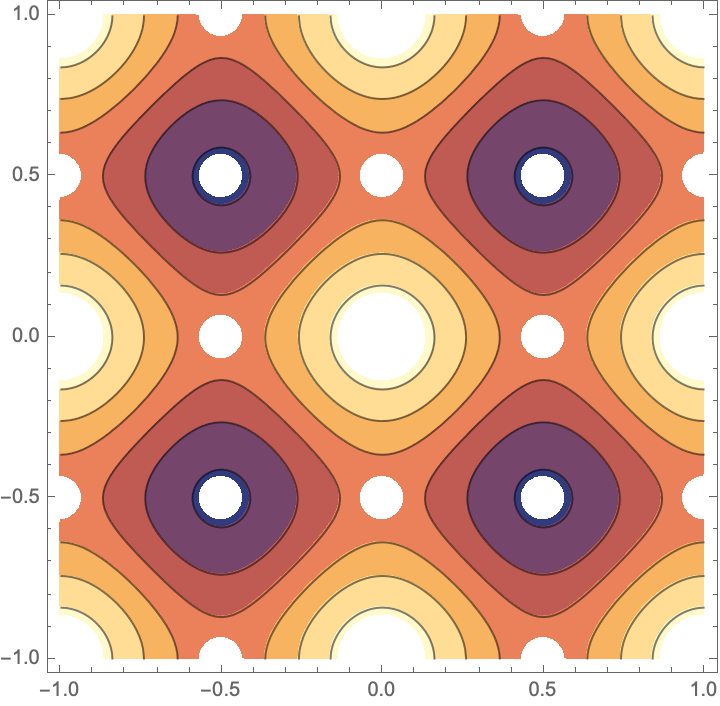}\hspace{0.1cm}
\includegraphics[width=0.24\linewidth,clip]{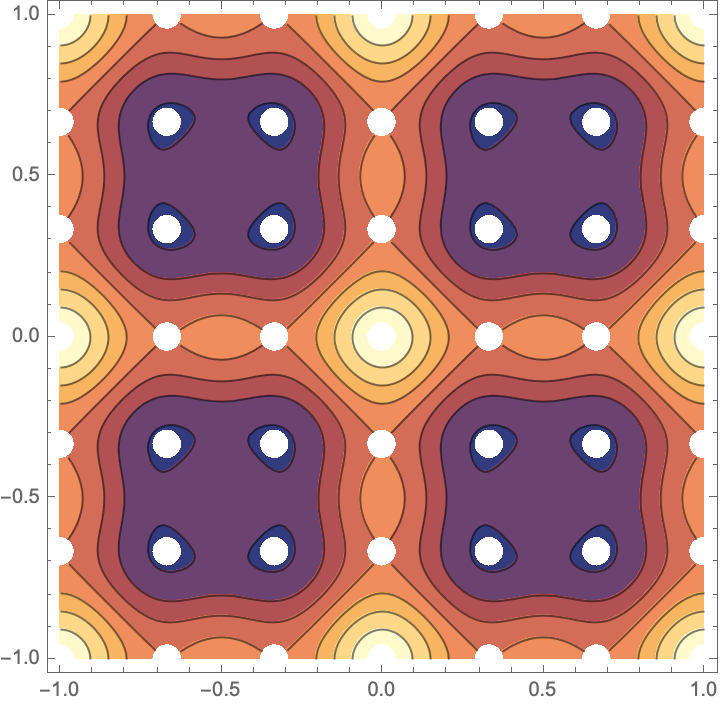}\hspace{0.1cm}
    \includegraphics[width=0.24\linewidth,clip]{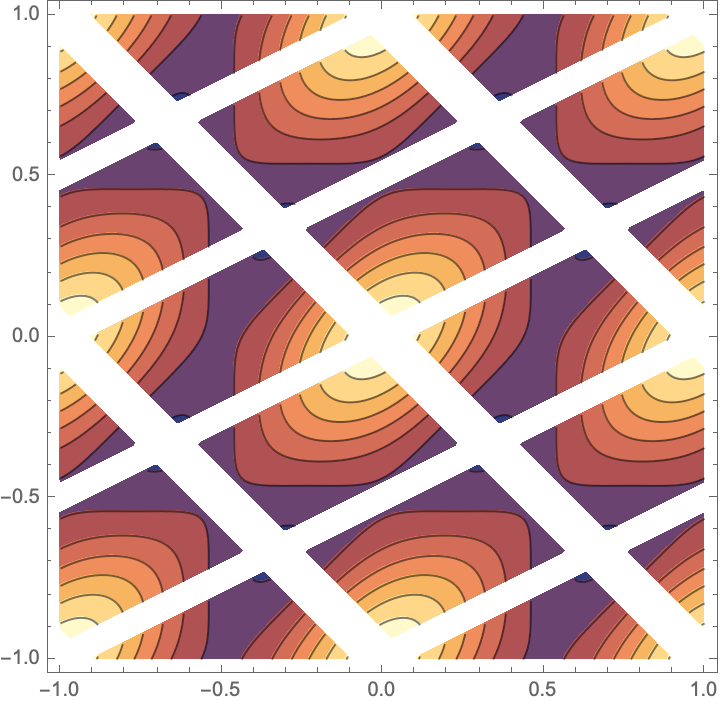}\\[0.3cm]
\includegraphics[width=0.24\linewidth,clip]{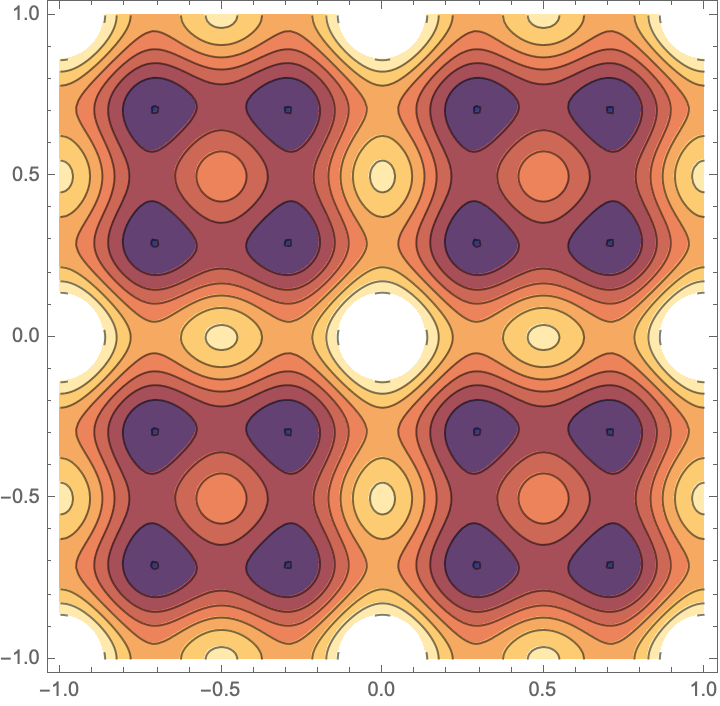}\hspace{0.1cm}
\includegraphics[width=0.24\linewidth,clip]{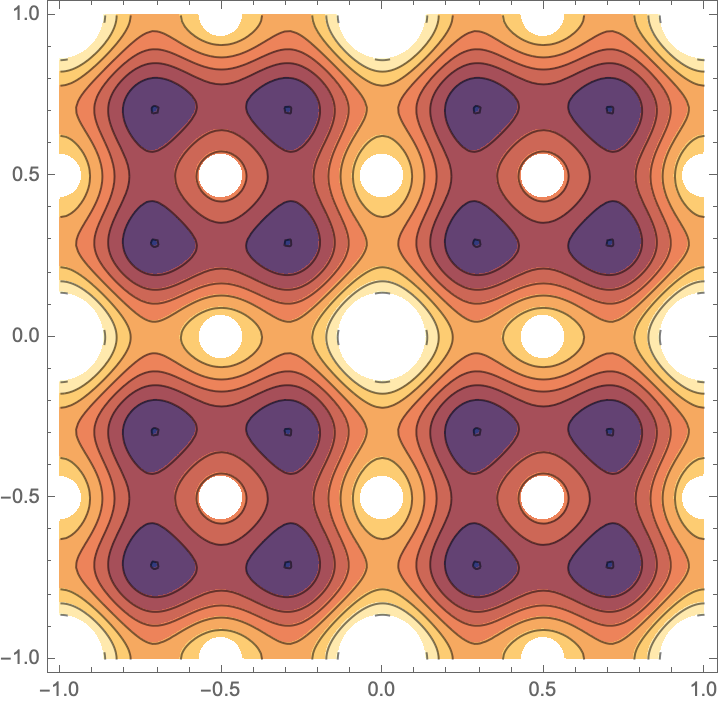}\hspace{0.1cm}
\includegraphics[width=0.24\linewidth,clip]{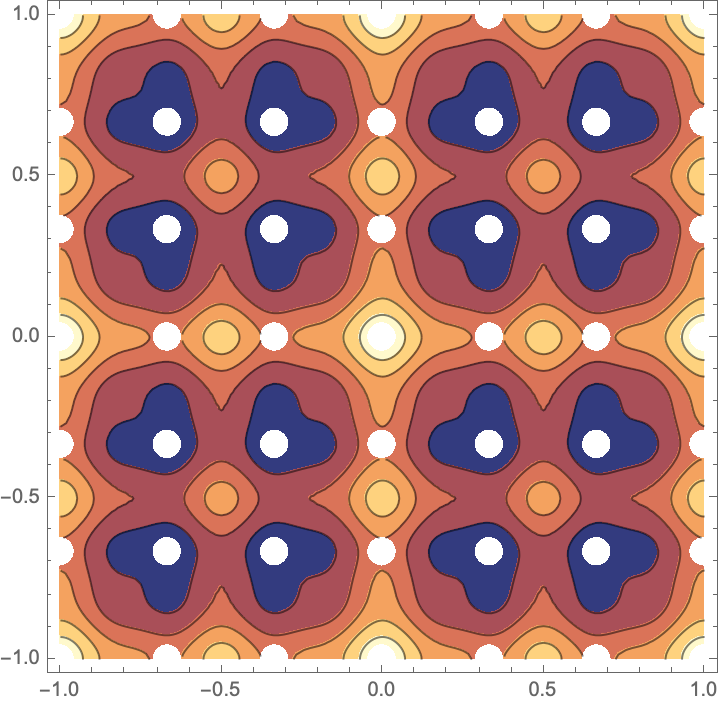}\hspace{0.1cm}
    \includegraphics[width=0.24\linewidth,clip]{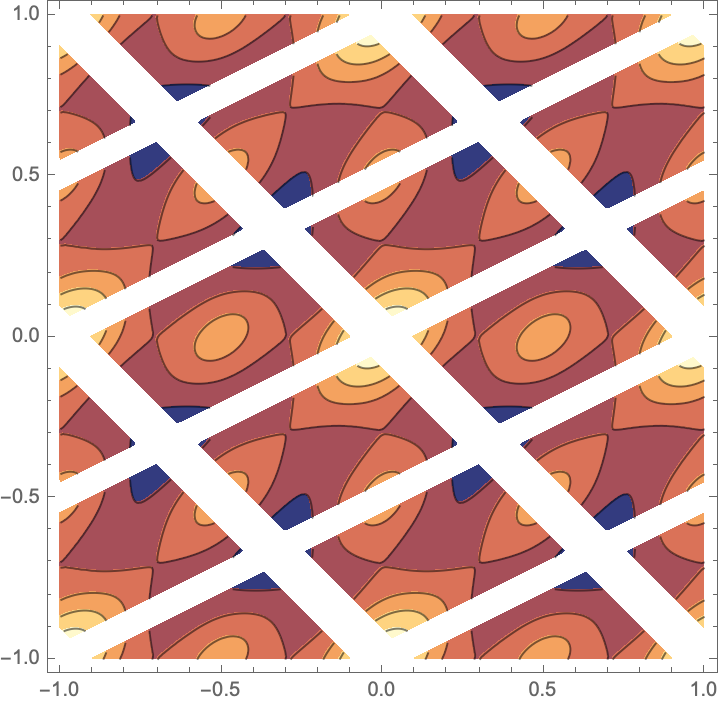}\\[0.3cm]
\includegraphics[width=0.24\linewidth,clip]{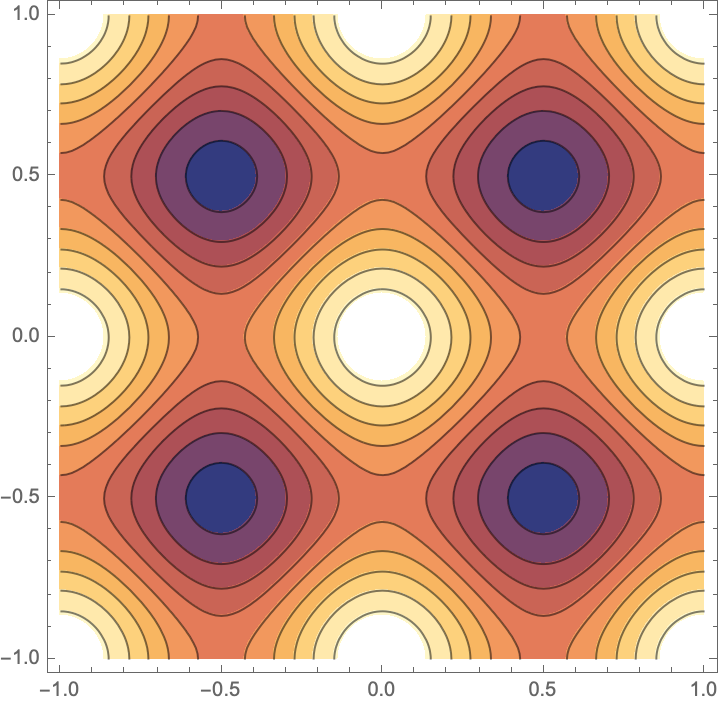}\hspace{0.1cm}
\includegraphics[width=0.24\linewidth,clip]{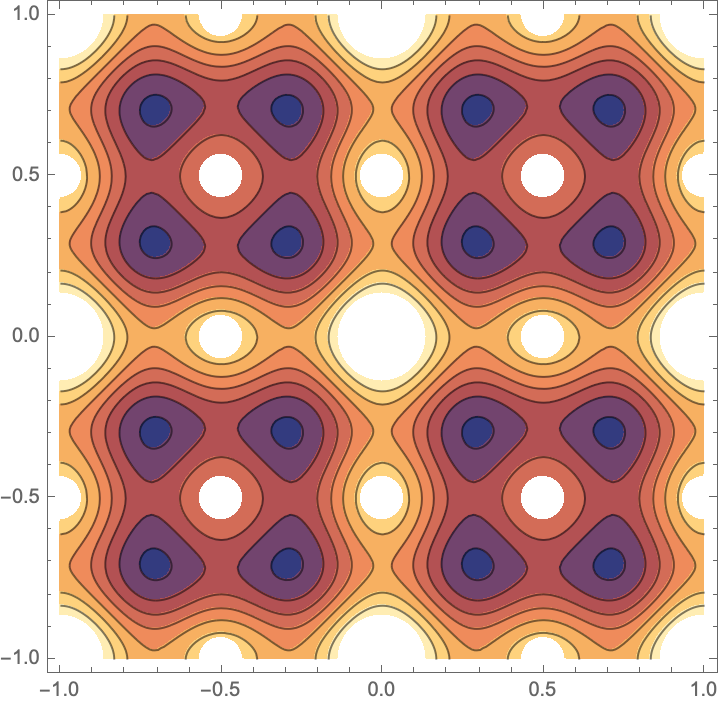}\hspace{0.1cm}
\includegraphics[width=0.24\linewidth,clip]{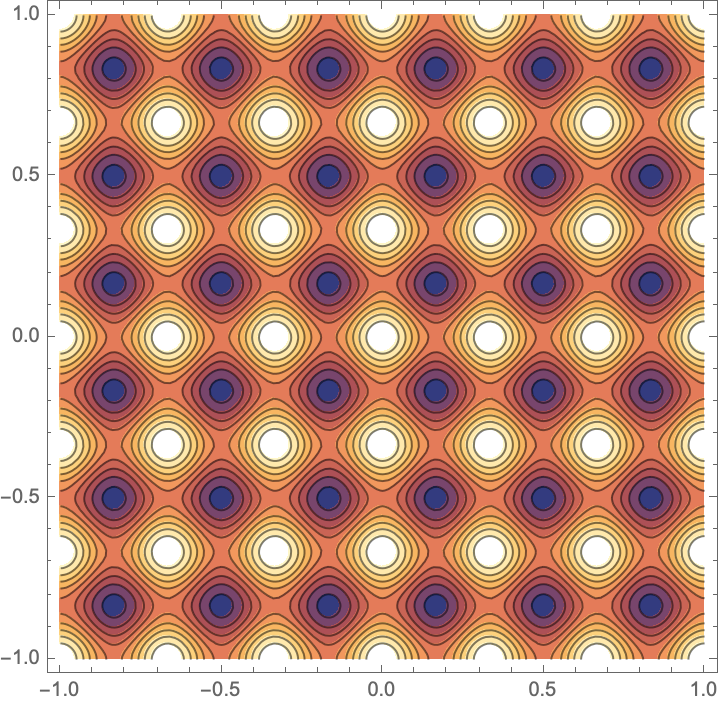}\hspace{0.1cm}
    \includegraphics[width=0.24\linewidth,clip]{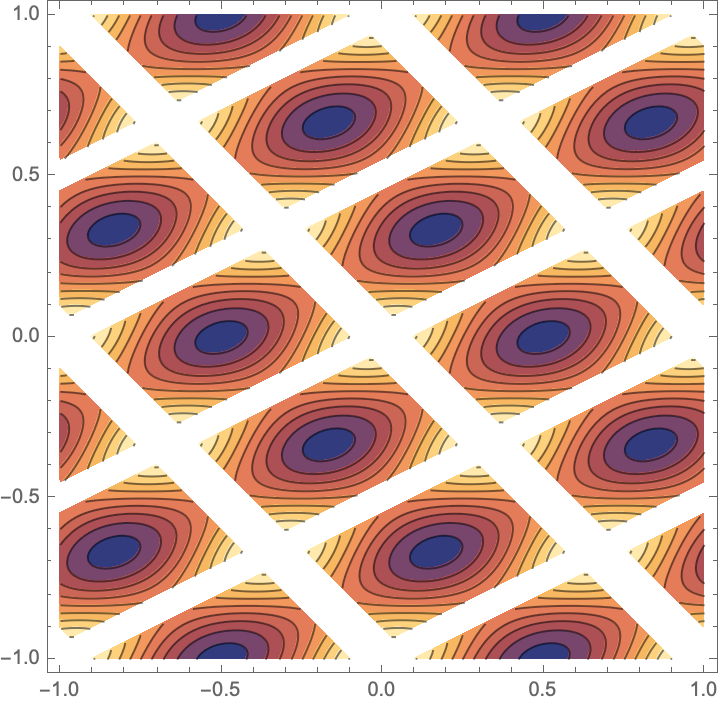}\\[0.3cm]
\includegraphics[width=0.24\linewidth,clip]{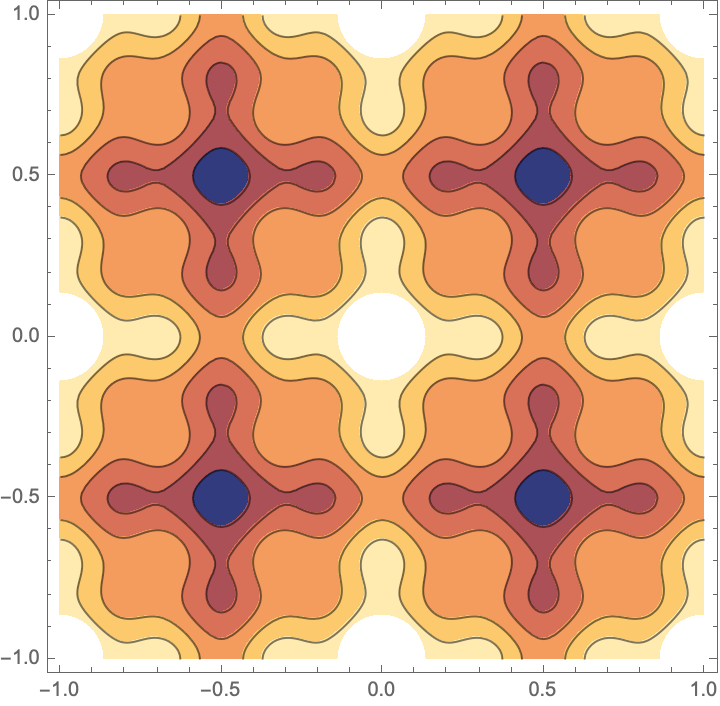}\hspace{0.1cm}
\includegraphics[width=0.24\linewidth,clip]{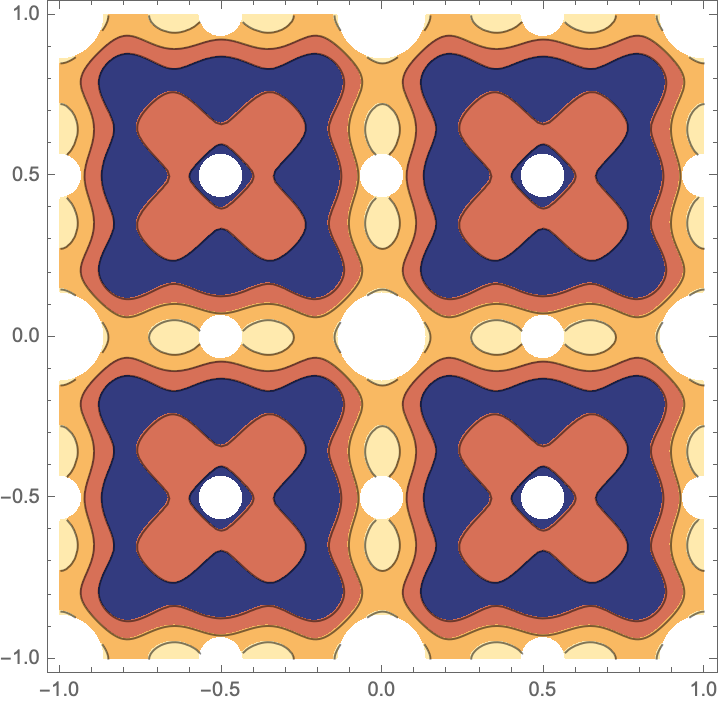}\hspace{0.1cm}
\includegraphics[width=0.24\linewidth,clip]{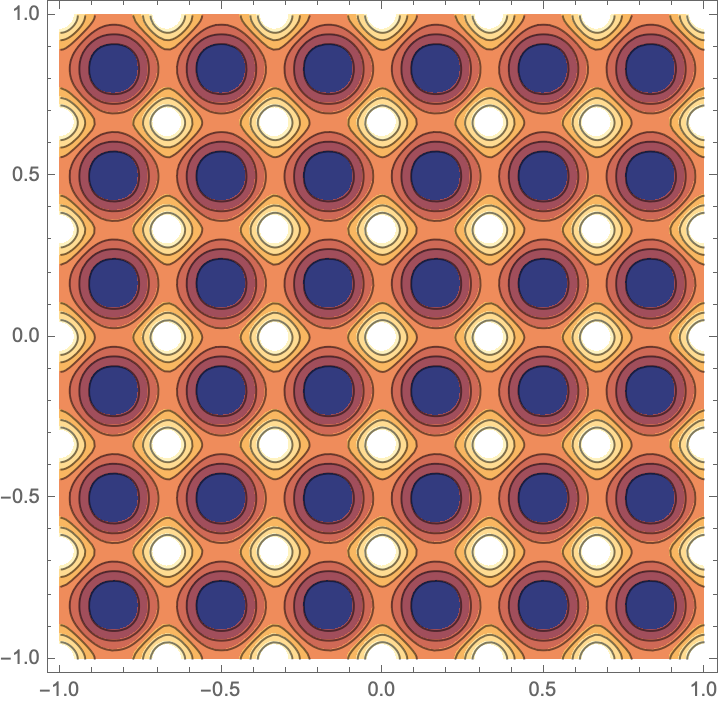}\hspace{0.1cm}
    \includegraphics[width=0.24\linewidth,clip]{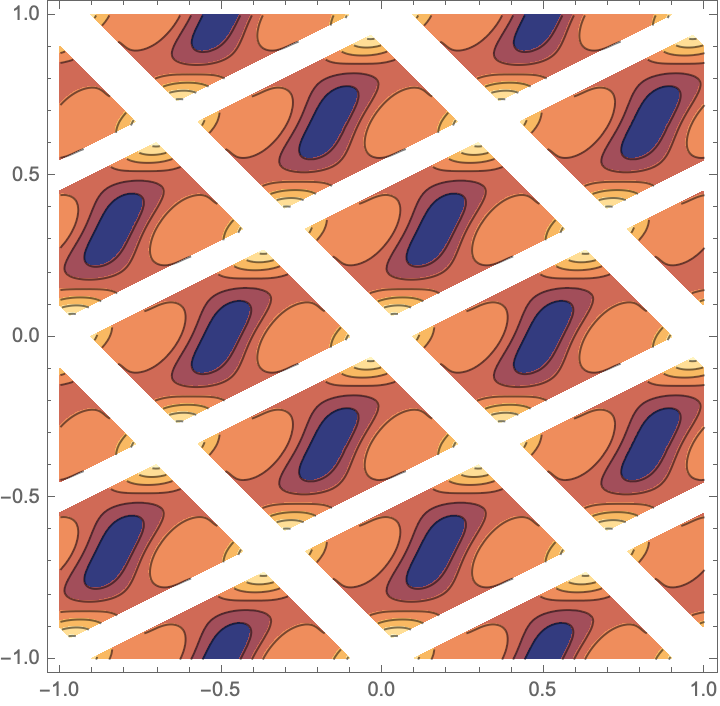}
\caption{Contour plots of the effective potential contributions from fermion fields. 
  From left to right, we assume Ansatz 1 to 4. From top to bottom, we plot contributions from $\Psi_{\bm 3}$, $\Psi_{\bm 6}$, $\Psi_{\bm 8}$, and $\Psi_{\bm{10}}$ with $(\eta_7,\eta_8)=(0,0)$. In these contour plots, the horizontal (vertical) axis shows the value of $b_1$ ($b_2$). From light to dark colors in the plots, the potential decreases.  The white regions are excluded by the constraint on the WL phases in eqs.~\eqref{constraint} and~\eqref{constraint2}.  }\bigskip
 \label{fig_fer}
\end{figure}
 
We see that local minima of the potentials in the pure Yang-Mills case are located in the excluded parameter region where tachyonic states appear in the low-energy mass spectrum. We also analyze the behavior of the potential contributions from matter fields. If we introduce fermion fields as mentioned in section~\ref{Matter fields}, we obtain the plots shown in Figure~\ref{fig_fer}. Now, we can see that there are some possible minimum points in regions where tachyonic states are absent.

\subsection{An example of local minima}
As seen above, in the pure Yang-Mills case, there seems to be no local minimum under the condition discussed in section~\ref{sec:stab}. On the other hand, including matter fields, we can find local minima where tachyonic states disappear in the low-energy mass spectrum.

As an example, we discuss local minima appearing with an adjoint fermion with the periodic boundary condition. The potential for the WL phases is given by
\begin{align}\label{vmodel1}
V(a_{m'}^k)=  V^{[\rm pYM]}(a_{m'}^k; \hat M_w)
  +V^{[\Psi_{\bm 8}]}_{(0,0)}&(a_{m'}^k; \hat M_w). 
\end{align}
In this potential, we have numerically checked that there are degenerate local minima where the WL phases take 
\begin{align}\label{WLmin1}
  (a_7^1,a_7^2,a_8^1,a_8^2)=  ((3-2n_{7})/6,2n_{7}/6+n'_{7},
  (3-2n_{8})/6,2n_{8}/6+n'_{8}),
  \qquad n_{m'},n'_{m'}\in \mathbb Z.
\end{align}
There are degeneracies that are expected from the discussion in section~\ref{sec:anpot} since the adjoint matter field is also neutral under the center $\mathbb Z_3$.  If we introduce matter fields that are charged under $\mathbb Z_3$, the degeneracy of the local minima is generally disturbed. For example, if we add the contribution from a fermion in the fundamental representation with the periodic boundary condition to the potential in eq.~\eqref{vmodel1}, we have confirmed that only a subset of the solution shown in eq.~\eqref{WLmin1}, where $n_{m'},n'_{m'}\in 3\mathbb Z$ is satisfied, corresponds to the local minima.  

Let us discuss the case with the potential in eq.~\eqref{vmodel1}. As a representative of local minima, we take
\begin{align}
  (a_7^1,a_7^2,a_8^1,a_8^2)=(1/2,0,1/2,0). 
\end{align}
At the minimum, the WL phase factors take
\begin{align}\label{WL78def}
  W_7=W_8={\rm diag}(-1,-1,1)=e^{i\pi H_3}, 
\end{align}
where $H_3$ is defined in section~\ref{sec:backg}. In our setup, the flux background satisfies $\bm f\propto H_3$.  Thus, at the minimum, both the WL phases and the flux are along the $H_3$ direction.  The WL phases contribute to masses of all flux-coupled 4D modes since their direction is the same as the one of the flux background.  At the minimum, no tachyonic states exist in the low-energy mass spectrum. In Ref.~\cite{Franken:2022pug}, a 6D $SU(2)$ model with an adjoint scalar field is studied, focusing on the mechanism to eliminate tachyonic states with a flux background. In the 6D model, it was shown that tachyonic states do not appear if an appropriate VEV of the scalar field is developed.  In our setup, although there are no elementary scalar fields, the WL phases play a similar role to the adjoint scalar in the 6D model in the sense that they give contributions to masses of 4D modes from flux-coupled fields.

Around the minimum, let us derive the mass spectrum of the fluctuations of WL phases at low energy. The WL phases are zero modes from $A_{m'}^{k}$. The normalization of the zero modes is chosen to be
\begin{align}\label{zeromodenorm1}
  A_{m'}^{k}= {1\over LL'}A_{m'}^{(0)k}+(\textrm{non-zero modes}), 
\end{align}
where $A_{m'}^{(0)k}$ are implied to be zero modes, which are independent of the extra-dimensional coordinates. Then, in the 4D Lagrangian ${\cal L}_{\rm 4D}$, the kinetic terms for the zero modes take
\begin{align}
  {\cal L}_{\rm 4D}\ni
  \delta^{m'n'}
  \begin{pmatrix}
      A_{m'}^{(0)1}&A_{m'}^{(0)2}
  \end{pmatrix}
                     \begin{pmatrix}
                         2&-1\\-1&2
                     \end{pmatrix}
                                   (\der_\mu)^2
                                   \begin{pmatrix}
      A_{n'}^{(0)1}\\A_{n'}^{(0)2}
  \end{pmatrix}. 
\end{align}
Let $\phi^{(\alpha)}$ $(\alpha=1,\dots,4)$ be  canonically normalized real scalar fields, which are defined by 
\begin{align}
  \phi^{(1)}&=2 A_{7}^{(0)1}-A_{7}^{(0)2},   \qquad 
  \phi^{(2)}=\sqrt{3}A_{7}^{(0)2},\\
  \phi^{(3)}&=2 A_{8}^{(0)1}-A_{8}^{(0)2},   \qquad 
  \phi^{(4)}=\sqrt{3}A_{8}^{(0)2}. 
\end{align}
These fields are massless at tree level. Their non-vanishing masses arise from the one-loop corrections. Let ${\cal M}^2_{\alpha\beta}$ be the mass matrix for
$\phi^{(\alpha)}$. We see that 
\begin{align}\label{massmat1}
  {\cal M}^2_{\alpha\beta}={\der^2V(a_{m'}^k)\over \der \phi^{(\alpha)}\der\phi^{(\beta)}}\bigg|_{(a_7^1,a_7^2,a_8^1,a_8^2)=(1/2,0,1/2,0)}.
\end{align}

To evaluate the mass matrix in eq.~\eqref{massmat1}, we remind that the VEVs of the canonically normalized fields $\phi^{(\alpha)}$ are rewritten by the WL phases $a_{m'}^{k}$ as
\begin{align}
  a_{m'}^k&={gL'\over 2\pi}\vev{A_{m'}^k}={g\over 2\pi L}\vev{A_{m'}^{(0)k}}.
\end{align}
Thus, we obtain 
\begin{align}
  {\der \over \der \phi^{(1)}}&={g_4L\over 2\pi \hat M_w}
  {1\over 2}{\der \over \der a_7^1},&&
  {\der \over \der \phi^{(2)}}={g_4L\over 2\pi \hat M_w}\left(
  {1\over 2\sqrt{3}}{\der \over \der a_7^1}
  +{1\over \sqrt{3}}{\der \over \der a_7^2}\right)  ,\\
  {\der \over \der \phi^{(3)}}&={g_4L\over 2\pi \hat M_w}
  {1\over 2}{\der \over \der a_8^1},&&
  {\der \over \der \phi^{(4)}}={g_4L\over 2\pi \hat M_w}\left(
  {1\over 2\sqrt{3}}{\der \over \der a_8^1}
  +{1\over \sqrt{3}}{\der \over \der a_8^2}\right), 
\end{align}
where we have defined the 4D gauge coupling $g_4$ as $g_4=g/(LL')$. Let us define the dimensionless potential $\tilde V(a_{m'}^k)$ as $\tilde V(a_{m'}^k)=L^4V(a_{m'}^k) $. The mass scale of ${\cal M}^2_{\alpha\beta}$ is roughly estimated as 
\begin{align}
  {\cal O}( {\cal M}^2_{\alpha\beta})\sim
  {1\over
      L^2}\left({g_4\over 2\pi \hat M_w}\right)^2{\der^2 \tilde V(a_{m'}^k)\over
      \der (a_{m'}^k)^2} ={1\over L'{}^2}{1\over \hat M_w^2}\left({g_4\over
          2\pi \hat M_w}\right)^2{\der^2 \tilde V(a_{m'}^k)\over \der (a_{m'}^k)^2}. 
\end{align}
From the above, one sees that the mass scale depends on $g_4$ and $\hat M_w$.

For $\hat M_w=5$, we numerically evaluate the mass matrix in eq.~\eqref{massmat1}. Around the minimum, we find 
\begin{align}\label{massmatmin1}
  {\cal M}^2_{\alpha\beta}\simeq {g_4^2\over L'{}^2}
  {\rm diag}
  (1.307,3.945,1.307,3.945)\times 10^3. 
\end{align}
For comparison, we also show the mass matrix in the pure Yang-Mills case, where $(a_7^1,a_7^2,a_8^1,a_8^2)=(1/2,0,1/2,0)$ is a local maximum as
\begin{align}
  {\cal M}^2_{\alpha\beta}\bigg|_{\textrm{pure Yang-Mills}}\simeq -{g_4^2\over L'{}^2}
  {\rm diag}
  (0.760,2.345,0.760,2.345)\times 10^3. 
\end{align}
Although the eigenvalues of the mass matrix are negative in the pure Yang-Mills case, they become positive if we include the potential contribution from the matter field.  For a small value of the 4D gauge coupling, mass scales of eigenvalues of the matrix in eq.~\eqref{massmatmin1} are smaller than ${\cal O}(1/L')$. On the other hand, for a moderate size of the coupling, the mass scale exceeds ${\cal O}(1/L')$. In the 4D effective theory, there are infinite massive modes. The one-loop effective potential is a sum of the contributions from the infinite modes and tends to be large because of the infinite summation in the presence of the flux background.

Let us discuss the masses of the 4D modes from the other fields around the minimum.  For flux-blind fields, masses of 4D modes are determined by $M_{56}^2$ and $M_{78}^2$ in eq.~\eqref{M78def}. As implied from the WL phase factors in eq.~\eqref{WL78def}, there are no contributions from the VEV of the WL phases to the tree-level masses of 4D modes from flux-blind fields at the minimum. The massless 4D gauge fields appearing from zero modes of $A_\mu^k$ correspond to the remaining gauge symmetry at low energy. In addition, there are 4D scalars that are massless at tree level, originating from zero modes of the extra-dimensional components of the flux-blind gauge fields. As the WL phases discussed above, they could obtain non-vanishing masses from the quantum corrections. However, in this setup, zero modes of $A_{5}^k$ and $A_{6}^k$ contain the NG bosons, which are intrinsically massless scalars related to the breaking of the translational symmetry by the flux background. Since light scalars would induce cosmological problems, an explicit breaking of the translational symmetry may be required to complete a phenomenologically viable setup, for example, introducing an orbifold in place of the torus as extra dimensions.
In an extended setup, these light scalars may play the role of Higgs scalars in GHU and GUT models.  

For flux-coupled fields, masses of the 4D modes consist of the KK mass contribution $M_{78}^2$ and the Landau-level contribution, as discussed in section~\ref{sec:mass}. At the minimum, $M_{78}^2$ contains the non-zero contribution of the WL phases. The Landau-level contribution can vanish only for the fermion case.  An interesting feature is that 4D modes from fermion fields with the anti-periodic boundary condition $(\eta_7,\eta_8)=(1/2,1/2)$ are massless at the minimum. These massless states have degeneracy, characterized by $q_3$. For example, in the adjoint fermions, there are flux-coupled components, which have $|q_3|=3$. At the minimum, they lead to three massless states. Massless fermionic states with degeneracy may be useful for understanding the generation structure in the standard model, as often discussed in models with $U(1)$ flux. 

%
\section{Conclusions}
\label{sec:conclusions}
%
In this work, we have explored the vacuum configurations of an 8D non-abelian gauge theory. The extra dimensions consist of a 4D torus, having a flux background in two of them. The WL phases along the remaining two compactified dimensions are treated dynamically. Their values contribute to the masses of low-energy 4D modes. Thus, to obtain phenomenological implications of this setup, it is crucial to clarify the vacuum structure of the potential of the WL phases.

As a concrete example, in an $SU(3)$ model, we have performed the analysis of the vacuum structure evaluating the quantum corrections of the potential. We have introduced matter fields and began by deriving the masses of the 4D modes emerging in the low-energy effective theory. As expected, some 4D modes can be tachyonic, coming from flux-coupled fields. However, the masses of these 4D modes also contain contributions depending on the WL phases, which can stabilize the system by taking appropriate values. We have shown the constraints on the parameter region of the WL phases where tachyonic states disappear at low energy.

To discuss the vacuum structure, we have derived the one-loop effective potential for the WL phases, which have no tree-level potential. In the search for minimum points of the one-loop potential, we have shown that critical points naturally appear where the WL phases take a simple fractional form. The WL phase factors at these extrema of the potential were shown to be aligned with the flux background in the $SU(3)$ space. We also have plotted the effective potential as functions of the WL phases with some ansatzes. For the pure Yang-Mills case, the local minima were found to be located only on the parameter region excluded by the condition to eliminate tachyonic states from the 4D mass spectra. On the other hand, in the fermionic contributions to the potential, some local minima were found in the allowed parameter region of the WL phases.

In models including matter fields, we have numerically found that local minima exist without any ansatz, and at the minima, tachyonic states disappear from the low-energy mass spectrum. As discussed, the WL phase factors at these points are aligned with the flux background. As an illustrative example, we have examined the low-energy mass spectrum around a minimum point $(a_7^1,a_7^2,a_8^1,a_8^2)=(1/2,0,1/2,0)$. The fluctuations of the WL phases around the minimum obtain positive mass squared, which are generated by the one-loop effective potential and are proportional to $g_4^2/L'{}^2$. Massless 4D gauge fields appear corresponding to the remaining gauge symmetry $SU(2)\times U(1)$. In addition, massless 4D scalars regarded as the NG bosons appear from the flux-blind fields. We also have discussed that chiral fermions can be obtained from flux-coupled fields at the minimum point if we introduce fermion fields with the anti-periodic boundary condition. 

The above results imply that several higher-dimensional gauge theories with flux backgrounds related to a simply-connected gauge group can have phenomenologically viable meta-stable vacua. Thus, we expect new possibilities of diverse models beyond the SM, such as GHU and GUT, in this framework. The discussions on vacuum stability concerning tunneling processes, realistic model constructions, and their predictions are intriguing research topics left for future studies.

\bigskip \bigskip

\begin{center}
    {\bf Acknowledgements}
\end{center}
The authors would like to thank K.~Takenaga for the fruitful discussion and comments. The work of Y.O. is supported in part by the Kyushu University Leading Human Resources Development Fellowship Program. The work of C.T. is supported by the Ministry of Education, Culture, Sports, Science and Technology (MEXT) of Japan.

\bigskip \bigskip
\appendix
%

%
\section{Mass spectrum of 4D modes from extra-dimensional gauge fields with arbitrary $\xi$}
\label{app:massesxi}
%
In section~\ref{sec:q3neq0case}, we show the masses of 4D modes from the extra-dimensional gauge fields having $q_3(\phi)\neq 0$. Here, we explain the derivation of their masses in an arbitrary $\xi$ case. To obtain the mass spectrum of the 4D mode from the gauge fields that couple to the flux, we have to diagonalize the Lagrangian corresponding to eq.~(5.34) in Ref.~\cite{KYS}.  The gauge parameter dependence appears with $A_m$, and the relevant part of the Lagrangian is given by
\begin{gather}\label{amlad1}
  {\cal L}_{A_m}=2\bar A_m
  [\delta^{mn}\square +\delta^{mn} (D_{l})^2
    -(1-\xi) D^{m}D^{n}
                  -2ig\hat f q_3
                  (\delta_{m5}\delta_{n6}-\delta_{m6}\delta_{n5})
                  ] A_n, \\
  D_5=\der_5+ig\hat f q_3(1+\gamma)x^6/2,
  \qquad 
       D_6=\der_6-ig\hat f q_3(1-\gamma)x^5/2,\\
  D_7=\der_7-ig (q_1v_7^1+q_2v_7^2), \qquad 
  D_8=\der_8-ig (q_1v_8^1+q_2v_8^2), 
\end{gather}
where $q_k$ is the charge of $\phi$ with respect to $H_k$.  One sees that the terms including $A_{5,6}$ and $A_{7,8}$ are completely separated in eq.~\eqref{amlad1} for $\xi =1$. For an arbitrary $\xi$, there are mixing terms.

To diagonalize ${\cal L}_{A_m}$,  we change the basis from $(A_5,A_6)$ to $(A_-,A_+)$ as
\begin{align}
  \begin{pmatrix}
      A_-\\A_+
  \end{pmatrix}
&=
{1\over \sqrt{2}}  \begin{pmatrix}
    A_5-iA_6\\
    A_5+iA_6
                  \end{pmatrix}. 
\end{align}
Note that $[D_5,D_6]=-ig\hat fq_3$ holds. We hereafter take $\hat f,q_3>0$ for simplicity. Then, we can introduce the annihilation and creation operators as
\begin{align}
D_5-iD_6=i\sqrt{2g\hat fq_3}  \hat a^\dag, \qquad 
D_5+iD_6=i\sqrt{2g\hat fq_3}  \hat a, \qquad [\hat a,\hat a^\dag]=1, 
\end{align}
and 
\begin{align}
  (D_5)^2+(D_6)^2=-2g\hat fq_3(\hat a^\dag \hat a+1/2). 
\end{align}

We can rearrange the Lagrangian in eq.~\eqref{amlad1} as
\begin{align}\label{4dmasslag1}
{\cal L}_{A_m}=-2
  \begin{pmatrix}
      \bar A_-~\bar A_+~\bar A_7~\bar A_8
  \end{pmatrix}
(-\square {\cal I}+  \Gamma_{\rm mass})
  \begin{pmatrix}
    A_-\\A_+\\A_7\\A_8
  \end{pmatrix} ,
\end{align}
where
\begin{align}
  \Gamma_{\rm mass}=
 [2g\hat fq_3(\hat a^\dag \hat a+1/2)
  -D_7^2-D_8^2]{\cal I}+
  (1-\xi)
                   {(DD)}
-2g\hat f q_3
     \begin{pmatrix}
     1&0&0&0\\    
     0&-1&0&0\\    
     0&0&0&0\\    
     0&0&0&0
     \end{pmatrix}, 
\end{align}
and
\begin{align}
  (DD)
  &=\begin{pmatrix}
    {D_5^2+ D_6^2 + g\hat f q_3 \over 2}
    & {(D_5-iD_6)^2 \over 2}& 
  {(D_5 - i D_6)D_7\over \sqrt{2}}& {(D_{5} - i D_{6})D_8\over \sqrt{2}}\\
 {(D_{5}+i  D_{6})^2 \over 2}& 
 {D_{5}^2+ D_{6}^2 - g\hat f q_3  \over 2}& {(D_{5} + i D_{6})D_7\over \sqrt{2}} &
 {  (D_{5} + i D_{6})D_8\over \sqrt{2}}\\
 {(D_{5} + i D_{6})D_7\over \sqrt{2}}& {(D_{5} - i D_{6})D_7\over \sqrt{2}}& D_7^2& D_7D_{8}\\
 {(D_{5} + i D_{6})D_8\over \sqrt{2}}& {(D_{5} - i D_{6})D_8\over \sqrt{2}}& D_8D_7& D_8^2
\end{pmatrix}.
\label{DDop1}  
\end{align}
Here, ${\cal I}$ is the $4\times 4$ unit matrix.  

To evaluate the eigenvalues of the operator $\Gamma_{\rm mass}$, let us introduce the mode expansion as
\begin{align}
  \phi(x^\mu,z_1,x^7,x^8)&=\sum_{l\geq 0} \sum_{d=1}^{q_3}
\sum_{\hat n_7,\hat n_8\in\mathbb Z}
                         \phi_{(l,d)}^{(\hat n_7,\hat n_8)}(x^\mu)
                         \zeta_{l,d}(z_1)f_{\hat n_7,\hat n_8}(x^7,x^8), 
\end{align}
where $z_1=x^5+ix^6$, and $f_{\hat n_7,\hat n_8}(x^7,x^8)$ is defined as
\begin{align}
  f_{\hat n_7,\hat n_8}(x^7,x^8)=e^{2\pi i \hat n_7 \hat M_w x^7}e^{2\pi  i \hat n_8 \hat M_w x^8}.
\end{align}
On the other hand, the mode functions
$\zeta_{l,d}(z_1)$
satisfy~\cite{KYS}
\begin{align}
  \hat a \zeta_{0,d}(z_1)=0, \qquad 
   \zeta_{l,d}(z_1)={1\over \sqrt{l!}}(\hat a^\dag)^l\zeta_{0,d}(z_1). 
\end{align}
We obtain
\begin{align}
\hat a^\dag \zeta_{l,d}=\sqrt{l+1}  \zeta_{l+1,d},\qquad
  \hat a \zeta_{l,d}=\sqrt{l}  \zeta_{l-1,d},
  \qquad
  \hat a^\dag \hat a  \zeta_{l,d}=l  \zeta_{l,d}.
\end{align}

{\allowdisplaybreaks
Using this mode expansion, we can deduce the low-energy mass spectrum from the Lagrangian in eq.~\eqref{4dmasslag1} integrating over the extra dimensions.  Let $(\Gamma_{\rm mass})_{ij}$ be an $(i,j)$ component of the matrix $\Gamma_{\rm mass}$.  From the diagonal entries of $\Gamma$, we obtain 
\begin{align}
\int_{\rm ED}  \bar A_- (\Gamma_{\rm mass})_{11}A_-
  &=\sum{}'
    (\bar A_-)_{(l,d)}^{(\hat n_7,\hat n_8)}
\left[
2g\hat fq_3\left( {1+\xi\over 2}l-{1\over 2}\right)
  +  (2\pi \hat M_w)^2(\hat N_7^2+\hat N_8^2)\right]
    ( A_-)_{(l,d)}^{(\hat n_7,\hat n_8)},\\
\int_{\rm ED}  \bar A_+ (\Gamma_{\rm mass})_{22}A_+
  &=\sum{}'
    (\bar A_+)_{(l,d)}^{(\hat n_7,\hat n_8)}
\left[
2g\hat fq_3\left( {1+\xi\over 2}l+{2+\xi\over 2}\right)
  +  (2\pi \hat M_w)^2(\hat N_7^2+\hat N_8^2)\right]
    ( A_+)_{(l,d)}^{(\hat n_7,\hat n_8)},\\
\int_{\rm ED}  \bar A_7 (\Gamma_{\rm mass})_{33}A_7
  &=\sum{}'
    (\bar A_7)_{(l,d)}^{(\hat n_7,\hat n_8)}
\left[
2g\hat fq_3\left( l+1/2\right)
  +  (2\pi \hat M_w)^2(\xi \hat N_7^2+\hat N_8^2)\right]
    ( A_7)_{(l,d)}^{(\hat n_7,\hat n_8)},\\
\int_{\rm ED}  \bar A_8 (\Gamma_{\rm mass})_{44}A_8
  &=\sum{}'
    (\bar A_8)_{(l,d)}^{(\hat n_7,\hat n_8)}
\left[
2g\hat fq_3\left( l+1/2\right)
  +  (2\pi \hat M_w)^2(\hat N_7^2+\xi \hat N_8^2)\right]
    ( A_8)_{(l,d)}^{(\hat n_7,\hat n_8)},
\end{align}
where we have used $\sum{}'=\sum_{l\geq 0} \sum_{d=1}^{q_3} \sum_{\hat n_7,\hat n_8\in\mathbb Z}$ and $\int_{\rm ED}=\int d^4y$. The WL phases are contained in $\hat N_7$ and $\hat N_8$, defined in eq.~\eqref{hatNdef}.  From the off-diagonal entries of $\Gamma$, we find 
\begin{align}
\int_{\rm ED}  \bar A_- (\Gamma_{\rm mass})_{12}A_+
  &=\sum{}'(\bar A_-)_{(l+2,d)}^{(\hat n_7,\hat n_8)}
    \left[
    (\xi-1)g\hat f q_3\sqrt{(l+1)(l+2)}
    \right] (A_+)_{(l,d)}^{(\hat n_7,\hat n_8)},\\
\int_{\rm ED}  \bar A_- (\Gamma_{\rm mass})_{13}A_7
  &=\sum{}'(\bar A_-)_{(l+1,d)}^{(\hat n_7,\hat n_8)}  
    \left[
(\xi-1)\sqrt{g\hat fq_3(l+1)} (2\pi\hat M_w)\hat N_7
    \right] (A_7)_{(l,d)}^{(\hat n_7,\hat n_8)},\\
\int_{\rm ED}  \bar A_- (\Gamma_{\rm mass})_{14}A_8
  &=\sum{}'(\bar A_-)_{(l+1,d)}^{(\hat n_7,\hat n_8)}  
    \left[
(\xi-1)\sqrt{g\hat fq_3(l+1)} (2\pi\hat M_w)\hat N_8
    \right] (A_8)_{(l,d)}^{(\hat n_7,\hat n_8)},\\
\int_{\rm ED}  \bar A_+ (\Gamma_{\rm mass})_{21}A_-
  &=\sum{}'(\bar A_+)_{(l,d)}^{(\hat n_7,\hat n_8)}
    \left[
    (\xi-1)g\hat f q_3\sqrt{(l+1)(l+2)}
    \right] (A_-)_{(l+2,d)}^{(\hat n_7,\hat n_8)}  ,\\
\int_{\rm ED}  \bar A_+ (\Gamma_{\rm mass})_{23}A_7
  &=\sum{}'(\bar A_+)_{(l,d)}^{(\hat n_7,\hat n_8)}  
    \left[
(\xi-1)\sqrt{g\hat fq_3(l+1)} (2\pi\hat M_w)\hat N_7
    \right] (A_7)_{(l+1,d)}^{(\hat n_7,\hat n_8)}  ,\\
\int_{\rm ED}  \bar A_+ (\Gamma_{\rm mass})_{24}A_8
  &=\sum{}'(\bar A_+)_{(l,d)}^{(\hat n_7,\hat n_8)}  
    \left[
(\xi-1)\sqrt{g\hat fq_3(l+1)} (2\pi\hat M_w)\hat N_8
    \right] (A_8)_{(l+1,d)}^{(\hat n_7,\hat n_8)}  ,\\
\int_{\rm ED}  \bar A_7 (\Gamma_{\rm mass})_{31}A_-
  &=\sum{}'(\bar A_7)_{(l,d)}^{(\hat n_7,\hat n_8)}  
    \left[
(\xi-1)\sqrt{g\hat fq_3(l+1)} (2\pi\hat M_w)\hat N_7
    \right] (A_-)_{(l+1,d)}^{(\hat n_7,\hat n_8)},\\  
\int_{\rm ED}  \bar A_7 (\Gamma_{\rm mass})_{32}A_+
  &=\sum{}'(\bar A_7)_{(l+1,d)}^{(\hat n_7,\hat n_8)}  
    \left[
(\xi-1)\sqrt{g\hat fq_3(l+1)} (2\pi\hat M_w)\hat N_7
    \right] (A_+)_{(l,d)}^{(\hat n_7,\hat n_8)},\\
\int_{\rm ED}  \bar A_7 (\Gamma_{\rm mass})_{34}A_8
  &=\sum{}'(\bar A_7)_{(l,d)}^{(\hat n_7,\hat n_8)}  
    \left[
(\xi-1)(2\pi \hat M_w)^2\hat N_7\hat N_8
    \right] (A_8)_{(l,d)}^{(\hat n_7,\hat n_8)},\\
\int_{\rm ED}  \bar A_8 (\Gamma_{\rm mass})_{41}A_-
  &=\sum{}'(\bar A_8)_{(l,d)}^{(\hat n_7,\hat n_8)}  
    \left[
(\xi-1)\sqrt{g\hat fq_3(l+1)} (2\pi\hat M_w)\hat N_8
    \right] (A_-)_{(l+1,d)}^{(\hat n_7,\hat n_8)},\\  
\int_{\rm ED}  \bar A_8 (\Gamma_{\rm mass})_{42}A_+
  &=\sum{}'(\bar A_8)_{(l+1,d)}^{(\hat n_7,\hat n_8)}  
    \left[
(\xi-1)\sqrt{g\hat fq_3(l+1)} (2\pi\hat M_w)\hat N_8
    \right] (A_+)_{(l,d)}^{(\hat n_7,\hat n_8)},\\
\int_{\rm ED}  \bar A_8 (\Gamma_{\rm mass})_{43}A_7
  &=\sum{}'(\bar A_8)_{(l,d)}^{(\hat n_7,\hat n_8)}  
    \left[
(\xi-1)(2\pi \hat M_w)^2\hat N_7\hat N_8
    \right] (A_7)_{(l,d)}^{(\hat n_7,\hat n_8)}.
\end{align}
}

We hereafter suppress the indices $n_{m'}$ and $d$ of 4D modes since there are no mixing terms with respect to them. It is convenient to introduce 
\begin{align}
 K_{m'}=2\pi \hat M_w\hat N_{m'},
  \qquad
  L_k=g\hat f q_3 (l+k).
\end{align}
Note that $g\hat f=2\pi$ holds under the assumption in eq.~\eqref{su3flux}. Let us define 
\begin{align}
&  \int d^4y
  \begin{pmatrix}
      \bar A_-~\bar A_+~\bar A_7~\bar A_8
  \end{pmatrix}
  \Gamma
  \begin{pmatrix}
    A_-\\A_+\\A_7\\A_8
  \end{pmatrix}
\equiv  \sum_{d=1}^{q_3}\sum_{\hat n_7,\hat n_8\in\mathbb Z} {\cal L}_\Gamma, 
\end{align}
where
{\allowdisplaybreaks
\begin{align} \notag
{\cal L}_\Gamma  &=(\bar A_-)_{(0)}
\left[
-2 \pi q_3
  +  K_7^2+K_8^2\right]
                   ( A_-)_{(0)}\\
  &\quad +    
(\bar A_-)_{(1)}
\left[
2 \pi q_3
  +  K_7^2+K_8^2+(\xi-1)2 \pi q_3\right]
    ( A_-)_{(1)}+ \sum_{l\geq 0}({\cal L}_{0(l)}+(\xi-1){\cal L}_{1(l)}), \\
\notag
  {\cal L}_{0(l)}&=
    (\bar A_-)_{(l+2)}
\left[
L_1+L_2
  +  K_7^2+K_8^2\right]
    ( A_-)_{(l+2)}
+
    (\bar A_+)_{(l)}
\left[
L_1+L_2
  +  K_7^2+K_8^2\right]
    ( A_+)_{(l)}\\
&\quad+
      (\bar A_7)_{(l)}
\left[
L_0+L_1
  + K_7^2+K_8^2\right]
    ( A_7)_{(l)}
+
      (\bar A_8)_{(l)}
\left[
L_0+L_1
  + K_7^2+K_8^2\right]
  ( A_8)_{(l)},\\\notag
  {\cal L}_{1(l)}&=
  (\bar A_-)_{(l+2)}  L_2( A_-)_{(l+2)}
                   +  (\bar A_+)_{(l)}  L_1( A_+)_{(l)}\\\notag
  &\quad+(\bar A_-)_{(l+2)}    \sqrt{L_1L_2}  (A_+)_{(l)}
    +(\bar A_+)_{(l)}    \sqrt{L_1L_2}    (A_-)_{(l+2)}\\\notag
 &\quad+(\bar A_7)_{(l)}  K_7^2( A_7)_{(l)}
   +(\bar A_8)_{(l)}    K_8^2( A_8)_{(l)}
   +(\bar A_7)_{(l)}  K_7K_8    (A_8)_{(l)}
    +(\bar A_8)_{(l)}  K_7K_8 (A_7)_{(l)}\\\notag
  &\quad+(\bar A_7)_{(l)}  \sqrt{L_1} K_7    (A_-)_{(l+1)}
    +(\bar A_-)_{(l+1)}  \sqrt{L_1} K_7  (A_7)_{(l)}\\\notag
&\quad    +(\bar A_8)_{(l)}  \sqrt{L_1} K_8    (A_-)_{(l+1)}
  +(\bar A_-)_{(l+1)}  \sqrt{L_1} K_8    (A_8)_{(l)}\\\notag
&\quad    +(\bar A_7)_{(l+1)}  \sqrt{L_1} K_7    (A_+)_{(l)}
  +(\bar A_+)_{(l)}  \sqrt{L_1} K_7    (A_7)_{(l+1)}\\
  &\quad+(\bar A_8)_{(l+1)}  \sqrt{L_1} K_8    (A_+)_{(l)}
    +(\bar A_+)_{(l)}  \sqrt{L_1} K_8    (A_8)_{(l+1)}. 
\end{align}
}
In this expression, mixing terms only appear in ${\cal L}_{1(l)}$.
{\allowdisplaybreaks
After a straightforward calculation, we find the mass eigenstates as 
\begin{align}
  (B_0)_{(l+2)}
  &={1\over \sqrt{L_1+L_2}}\left(\sqrt{L_2}(A_+)_{(l)}-\sqrt{L_1}(A_-)_{(l+2)}\right),\\
  (C_0)_{(l)}
  &={1\over \sqrt{M_{78}^2}}\left(K_8(A_7)_{(l)}-K_7(A_8)_{(l)}\right),\\
  (D_0)
  &={1\over \sqrt{2 \pi q_3+M_{78}^2}}
    \left[
    \sqrt{2 \pi q_3\over M_{78}^2}\left(K_7(A_7)_{(0)}+K_8(A_8)_{(0)}\right)
    -\sqrt{M_{78}^2}(A_-)_{(1)}\right],\\
  (D_\xi)
  &={1\over \sqrt{2 \pi q_3+M_{78}^2}}
    \left(
    K_7(A_7)_{(0)}+K_8(A_8)_{(0)}
    +\sqrt{2 \pi q_3}(A_-)_{(1)}\right),\\\notag
  (E_0)_{(l)}
  &={1\over \sqrt{L_1+L_2+M_{78}^2}}
    \bigg[\sqrt{M_{78}^2\over L_1+L_2}
    \left(\sqrt{L_1}(A_+)_{(l)}+\sqrt{L_2}(A_-)_{(l+2)}\right)
  \\
  &\qquad\qquad\qquad\qquad\qquad -\sqrt{L_1+L_2\over M_{78}^2}
    \left(K_7(A_7)_{(l+1)}+K_8(A_8)_{(l+1)}\right)
    \bigg],\\
  (E_\xi)_{(l)}
  &={1\over \sqrt{L_1+L_2+M_{78}^2}}
    \left(\sqrt{L_1}(A_+)_{(l)}+\sqrt{L_2}(A_-)_{(l+2)}
    +K_7(A_7)_{(l+1)}+K_8(A_8)_{(l+1)}\right),
\end{align}
where $M_{78}^2$ is defined in eq.~\eqref{M78def}. A diagonalized form of the Lagrangian ${\cal L}_\Gamma$ is given by 
\begin{align}\notag
{\cal L}_\Gamma
  &=
    (\bar A_-)_{(0)}\left[4\pi q_3(-1/2)  +  M_{78}^2\right]( A_-)_{(0)}
    + \sum_{l\geq 0}
        (\bar B_0)_{(l+2)}\left[4\pi q_3(l+3/2)+  M_{78}^2\right] ( B_0)_{(l+2)}
  \\\notag
  & + \sum_{l\geq 0} (\bar C_0)_{(l)}\left[4\pi q_3(l+1/2) + M_{78}^2\right]  ( C_0)_{(l)}\\\notag
  &+(\bar D_0)\left[4\pi q_3(1/2)  +  M_{78}^2\right](D_0)
  +\sum_{l\geq 0}(\bar E_0)_{(l)}\left[4\pi q_3(l+3/2)+  M_{78}^2\right] ( E_0)_{(l)}\\
  &+(\bar D_\xi)\xi \left[4\pi q_3(1/2)  +  M_{78}^2\right](D_\xi)
    +\sum_{l\geq 0}(\bar E_\xi)_{(l)}\xi \left[4\pi q_3(l+3/2)+  M_{78}^2\right] ( E_\xi)_{(l)}.
    \label{diaglagxi}
\end{align}
}

This expression shows that the masses of the 4D modes are given as discussed in section~\ref{sec:q3neq0case}. It also shows that the mass eigenvalues consist of different contributions; one is from the Landau-level excitations, and the other depends on the WL phases contained in $M_{78}^2$. The former contributions are specified by half integers appearing in coefficients of $4\pi q_3$. These contributions for each 4D mode are schematically expressed as follows.
\begin{align}
  \begin{array}{l|cccc}\hline
    \{( A_-)_{(0)},( B_0)_{(2)},( B_0)_{(3)},\dots \} & -1/2 & & 3/2 & \dots\\
    \{( C_0)_{(0)},( C_0)_{(1)},\dots\}&  &1/2 & 3/2 & \dots\\
    \{(D_0),( E_0)_{(0)},( E_0)_{(1)},\dots\}&  &1/2 & 3/2 & \dots\\
    \{(D_\xi),( E_\xi)_{(0)},( E_\xi)_{(1)},\dots\}&  &\xi 1/2 &\xi 3/2 & \dots\\\hline
  \end{array}
\end{align}
The ghost sector has the same masses as $\{(D_\xi),( E_\xi)_{(0)},( E_\xi)_{(0)},\dots\}$. For $\xi =1$, contributions from ${\cal L}_{1(l)}$ vanish. We obtain simplified mass eigenstates as follows.
\begin{align}
  \begin{array}{l|cccc}\hline
    \{(A_{-})_{(0)},(A_{-})_{(1)},\dots\} & -1/2 & & 3/2 & \dots\\
    \{(A_{+})_{(0)},(A_{+})_{(1)},\dots\}&  &1/2 & 3/2 & \dots\\
    \{(A_{7})_{(0)},(A_{7})_{(1)},\dots\}&  &1/2 & 3/2 & \dots\\
    \{(A_{8})_{(0)},(A_{8})_{(1)},\dots\}&  &1/2 & 3/2 & \dots\\\hline
  \end{array}
\end{align}
The physical mass spectrum corresponds to the $\xi$-independent ones. 
\begin{align}
  \begin{array}{l|cccc}\hline
    \{( A_-)_{(0)},( B_0)_{(2)},( B_0)_{(3)},\dots \} & -1/2 & & 3/2 & \dots\\
    \{( C_0)_{(0)},( C_0)_{(1)},\dots\}&  &1/2 & 3/2 & \dots\\
    \{(D_0),( E_0)_{(0)},( E_0)_{(1)},\dots\}&  &1/2 & 3/2 & \dots\\
    \hline
  \end{array}
\end{align}

%
\section{Derivation of the effective potential}
\label{app:surf}
%
In this section, we derive contributions to the one-loop effective potential for the WL phases $a_{m'}^k$ $(m'=7,8$ and $k=1,2)$ in the $SU(3)$ model in section~\ref{sec:effpot}.

\subsection{Flux-blind case}
Let $\phi$ be a flux-blind field. As discussed in section~\ref{sec:mass}, their 4D modes $\phi_{(\hat {\bm n})}$ have masses $M^2(\phi_{(\hat {\bm n})})$ as in eq.~\eqref{m001}. The effective potential contribution for the WL phases generated by 4D modes from $\phi$ is given by
\begin{align}
  \Delta V(\phi)
  &=(-1)^{\hat F}{N_{\rm deg}\over 2}
    \sum_{\hat {\bm n}}\int{d^4p_{\rm E}\over (2\pi)^4}\ln(p_{\rm E}^2+M^2(\phi_{(\hat {\bm n})}))\\
  \label{delphiva1}
  &=-(-1)^{\hat F}{N_{\rm deg}\over 32\pi^2}
\sum_{\hat {\bm n}}\int^\infty_0dt\,  t^{-3}e^{-M^2(\phi_{(\hat {\bm n})})t}, 
\end{align}
where $N_{\rm deg}$ is a positive integer that gives the number of real dof of $\phi$, and $\hat F$ is the fermion number of $\phi$. The summation for $\hat {\bm n}=(\hat n_5,\hat n_6,\hat n_7,\hat n_8)$ is taken over all integers for each $\hat n_m$. The expression above is divergent for small values of the integration variable $t$. Since $t$ has dimension of $M^{-2}$, it is a UV divergence. It is useful to rewrite this expression using the Poisson resummation formula, which is given by
\begin{align}\label{poissongen}
  \sum_{n_i\in \mathbb Z}e^{-\pi(n_i+d_i)(A^{-1})^{ij}(n_j+d_j)}=\sqrt{\det A}\sum_{\omega^i\in \mathbb Z} e^{-\pi \omega^iA_{ij}\omega^j}e^{2\pi i \omega^kd_k}, 
\end{align}
for a $d$-dimensional invertible matrix $A$ $(i,j=1,\dots,d)$~\cite{Kojima:2023mew}. In our case, we have
\begin{gather}
  A={1\over 4\pi t}{\rm diag  }(1,1,1/\hat M_w^2,1/\hat M_w^2), \quad
          \sqrt{\det A}=1/|4\pi t \hat M_w|^2, \\
  d_5=d_6=0, \qquad d_7=-q_1a_7^1-q_2a_7^2+\eta_7, \qquad
  d_8=-q_1a_8^1-q_2a_8^2+\eta_8.
  \label{d78}
\end{gather}
Thus, we rewrite eq.~\eqref{delphiva1} as
\begin{align}
  \Delta V(\phi)
  &=-(-1)^{\hat F}{N_{\rm deg}\over 32\pi^2}
    \sum_{\bm \omega}\int^\infty_0dt {1\over (4\pi \hat M_w)^2}t^{-5}
    e^{-{1\over 4t}(\omega_5^2+\omega_6^2+(\omega_7^2+\omega_8^2)/\hat M_w^2)}e^{2\pi i (\omega_7 d_7+\omega_8d_8)}\\
  &=-(-1)^{\hat F}{3 N_{\rm deg}\over \pi^4\hat M_w^2}
    \sum_{{\bm \omega}}
{e^{2\pi i (\omega_7 d_7+\omega_8d_8)}\over 
    \left[\omega_5^2+\omega_6^2+(\omega_7^2+\omega_8^2)/\hat M_w^2\right]^{4}}, 
    \label{delphiva2}
\end{align}
where we have used
\begin{align}
  \int_0^\infty dt\,  t^{-5}e^{-X/t}={6\over X^4} \qquad {\rm for} \quad X>0, 
\end{align}
and the summation is taken over $\omega_5,\omega_6,\omega_7,\omega_8\in \mathbb Z$ in $\sum_{{\bm \omega}}$.

In eq.~\eqref{delphiva2}, the UV divergence became more evident, now being expressed by the term $(\omega_5,\omega_6,\omega_7,\omega_8)=(0,0,0,0)$. The contributions from $(\omega_7,\omega_8)=(0,0)$ have no dependence on the WL phases. Therefore, they can be disregarded since we are only interested in the potential for the WL phases. Then, the remaining part is finite. Hereafter, we replace the summations in eq.~\eqref{delphiva2} with the new definition
\begin{align}\label{sumdef}
  \sum_{{\bm \omega}}{}'=\sum_{{\bm \omega}}-\left(
\textrm{contributions of}~
  (\omega_7,\omega_8)=(0,0)\right). 
\end{align}
This summation is written more explicitly as 
\begin{align}
  \sum_{{\bm \omega}}{}'=\sum_{\omega_5,\omega_6\in \mathbb Z}
  \bigg(&
  \sum_{\omega_7\geq 1}\bigg|_{\omega_8=0}
  +\sum_{\omega_7\leq -1}\bigg|_{\omega_8=0}
  +\sum_{\omega_8\geq 1}\bigg|_{\omega_7=0}
  +\sum_{\omega_8\leq -1}\bigg|_{\omega_7=0}\\ \notag
&\quad  +\sum_{\omega_7\geq 1}\sum_{\omega_8\geq 1}
  +\sum_{\omega_7\geq 1}\sum_{\omega_8\leq -1}
  +\sum_{\omega_7\leq -1}\sum_{\omega_8\geq 1}
  +\sum_{\omega_7\leq -1}\sum_{\omega_8\leq -1}
  \bigg).
\end{align}
We find
\begin{align}
&  \left(\sum_{\omega_7\geq 1}
  +\sum_{\omega_7\leq -1}\right)\bigg|_{\omega_8=0}
  {e^{2\pi i (\omega_7 d_7+\omega_8d_8)}\over 
  \left[\omega_5^2+\omega_6^2+(\omega_7^2+\omega_8^2)/\hat M_w^2\right]^{4}}
  =\sum_{\omega_7\geq 1}{2\cos(2\pi \omega_7d_7)\over 
                \left[\omega_5^2+\omega_6^2+\omega_7^2/\hat M_w^2\right]^{4}},\\ 
&  \left(\sum_{\omega_8\geq 1}
  +\sum_{\omega_8\leq -1}\right)\bigg|_{\omega_7=0}
  {e^{2\pi i (\omega_7 d_7+\omega_8d_8)}\over 
  \left[\omega_5^2+\omega_6^2+(\omega_7^2+\omega_8^2)/\hat M_w^2\right]^{4}}
  =\sum_{\omega_8\geq 1} {2\cos(2\pi \omega_8d_8)\over 
                \left[\omega_5^2+\omega_6^2+\omega_8^2/\hat M_w^2\right]^{4}}. 
\end{align}
Thus, we obtain
\begin{align}\notag
  \bigg(  \sum_{\omega_7\geq 1}\bigg|_{\omega_8=0}
  +\sum_{\omega_7\leq -1}\bigg|_{\omega_8=0}&
  +\sum_{\omega_8\geq 1}\bigg|_{\omega_7=0}
  +\sum_{\omega_8\leq -1}\bigg|_{\omega_7=0}\bigg)
  {e^{2\pi i (\omega_7 d_7+\omega_8d_8)}\over 
  \left[\omega_5^2+\omega_6^2+(\omega_7^2+\omega_8^2)/\hat M_w^2\right]^{4}} \\ 
  &=2\sum_{\omega\geq 1}  {\cos(2\pi \omega d_7)+\cos(2\pi \omega d_8)\over 
    \left[\omega_5^2+\omega_6^2+\omega^2/\hat M_w^2\right]^{4}},
\end{align}
and 
\begin{align}
  \bigg(
  \sum_{\omega_7\geq 1}\sum_{\omega_8\geq 1}
  +\sum_{\omega_7\geq 1}\sum_{\omega_8\leq -1}&
  +\sum_{\omega_7\leq -1}\sum_{\omega_8\geq 1}
  +\sum_{\omega_7\leq -1}\sum_{\omega_8\leq -1}
 \bigg)  {e^{2\pi i (\omega_7 d_7+\omega_8d_8)}\over 
                \left[\omega_5^2+\omega_6^2+(\omega_7^2+\omega_8^2)/\hat M_w^2\right]^{4}}\\ \notag
  &=4\sum_{\omega_7,\omega_8\geq 1}
  {\cos(2\pi \omega_7 d_7)\cos(2\pi \omega_8 d_8)\over 
                \left[\omega_5^2+\omega_6^2+(\omega_7^2+\omega_8^2)/\hat M_w^2\right]^{4}}. 
\end{align}
Finally, the contribution to the effective potential obtained from flux-blind fields is summarized as
\begin{align}
&  \Delta V(\phi)
  =-(-1)^{\hat F}{6N_{\rm deg} \over \pi^4\hat M_w^2}
    \sum_{\omega_5,\omega_6\in \mathbb Z}    \\ \notag
  &\qquad \times     \bigg(
    \sum_{\omega\geq 1}  {\cos(2\pi \omega d_7)+\cos(2\pi \omega d_8)\over 
    \left[\omega_5^2+\omega_6^2+\omega^2/\hat M_w^2\right]^{4}}
    + 2\sum_{\omega_7,\omega_8\geq 1}
    {\cos(2\pi \omega_7 d_7)\cos(2\pi \omega_8 d_8)\over 
    \left[\omega_5^2+\omega_6^2+(\omega_7^2+\omega_8^2)/\hat M_w^2\right]^{4}}
    \bigg).
\end{align}

\subsection{Flux-coupled case with $\Sigma_{56}=0$ or $\Sigma_{56}=\pm 1/2$ }
\label{app:efsig0}
Now, let $\phi$ be a flux-coupled field. As shown in section~\ref{sec:mattermass}, if $\phi$ has $\Sigma_{56}=0$ or $\pm 1/2$, the masses of 4D modes from $\phi$, denoted by $M^2(\phi_{(\hat \ell,d,\hat n_7,\hat n_8)})$, are given by eq.~\eqref{mattermass}. From a discussion similar to the one in the previous subsection, we find that 4D modes from $\phi$ generate the effective potential contribution $\Delta V(\phi)$, which is given by 
\begin{align}
&  \Delta V(\phi)
    =
  -(-1)^{\hat F}{N_{\rm deg}|q_3|\over 32\pi^2}
  \sum_{\hat \ell\geq 0}
  \sum_{\hat n_7,\hat n_8\in \mathbb Z}\int^\infty_0dt\,  t^{-3}e^{-M^2(\phi_{(\hat \ell,d,\hat n_7,\hat n_8)})t}\\ 
  &=-(-1)^{\hat F}{N_{\rm deg}|q_3|\over 32\pi^2}
   \int^\infty_0dt\,  t^{-3}
   \sum_{\hat n_7,\hat n_8\in \mathbb Z}  e^{-(2\pi \hat M_w)^2 \left(\hat N_7^2  +\hat N_8^2  \right)t}
    \sum_{\hat \ell\geq 0}e^{-4\pi |q_3|(\hat \ell+1/2+\Sigma_{56})t},
    \label{epsig0h1}
\end{align}
where the overall factor $|q_3|$ is a result of the degeneracy labeled by $d$ in $\phi_{(\hat \ell,d,\hat n_7,\hat n_8)}$ and the order of the integration and the summations were exchanged in the second line. 

The infinite sum over $\hat n_{7}$ and $\hat n_{8}$ in eq.~\eqref{epsig0h1} can be rearranged by using the Poisson resummation formula given by eq.~\eqref{poissongen} with
\begin{align}
  A^{-1}={4\pi \hat M_w^2 t}
  \begin{pmatrix}
    1&0\\0&1
  \end{pmatrix}, \qquad 
  A={1\over {4\pi \hat M_w^2 t}}
  \begin{pmatrix}
    1&0\\0&1
  \end{pmatrix}, \qquad
  \sqrt{\det A}={1\over {|4\pi \hat M_w^2 t|}}, 
\end{align}
and $d_7$ and $d_8$ in eq.~\eqref{d78}. We obtain
\begin{align}\label{Poiswflcase}
  \sum_{\hat n_7,\hat n_8\in \mathbb Z}  e^{-(2\pi \hat M_w)^2 \left(\hat N_7^2  +\hat N_8^2  \right)t}
  =\sum_{\omega_7,\omega_8\in \mathbb Z}{1\over {|4\pi \hat M_w^2 t|}}
  e^{-{1\over {4\pi \hat M_w^2 t}}(\omega_7^2+\omega_8^2)}
  e^{2\pi i (\omega_7d_7+\omega_8d_8)}.
\end{align}

We focus on the infinite sum over $\hat \ell$ in eq.~\eqref{epsig0h1}. In the $\Sigma_{56}(\phi)=0$ case, the summation is rewritten using
\begin{align}
\sum_{\hat \ell\geq 0}e^{-S(2\hat \ell +1)t}=e^{-St}+e^{-3St}+e^{-5St}+\dots=  {1\over 2\sinh St}, 
\end{align}
where $S=2\pi |q_3|$ is implied. In $\Sigma_{56}(\phi)=\pm 1/2$ cases, we combine contributions from 4D modes from $\Sigma_{56}(\phi)=1/2$ and $\Sigma_{56}(\phi)=-1/2$ fields. Then, the summation is rewritten using
\begin{align}
  \left(\sum_{\hat \ell\geq 1}
  +\sum_{\hat \ell\geq 0}
  \right)e^{-2S\hat \ell t}=\left(  e^{-2St}+e^{-4St}+\dots\right)
  +
  \left(1+ e^{-2St}+e^{-4St}+\dots\right)=
{1\over \tanh St}.
\end{align}

From the discussions above, the effective potential contribution for the $\Sigma_{56}(\phi)=0$ case is given by
\begin{align}
\Delta V(\phi)\big|_{\Sigma_{56}=0}
  =-(-1)^{\hat F}{N_{\rm deg}|q_3|\over 128\pi^3\hat M_w^2}
  \sum_{\omega_7,\omega_8\in\mathbb Z}
e^{2\pi i(\omega_7d_7+\omega_8d_8)}
  \int_0^\infty dt\,  t^{-4}
  { e^{-{\omega_7^2+\omega_8^2\over 4\pi \hat M_w^2t}}
\over 2\sinh(2\pi |q_3|t)
  }. 
\end{align}
As was done in the previous subsection, we subtract the divergent $(\omega_7,\omega_8)=(0,0)$ contribution. Finally, the contribution to the effective potential from flux-coupled fields with $\Sigma_{56}=0$ is given by
\begin{align}
\Delta V(\phi)\big|_{\Sigma_{56}=0}  
  &=-(-1)^{\hat F}{N_{\rm deg}|q_3|\over 128\pi^3\hat M_w^2}
\bigg(  2\sum_{\omega\geq 1}[\cos(2\pi \omega d_7)+\cos(2\pi \omega d_8)]
\int_0^\infty dt\,  t^{-4}
  { e^{-{\omega^2\over 4\pi \hat M_w^2t}}
\over 2\sinh(2\pi |q_3|t)} \notag \\ 
&\qquad   + 4\sum_{\omega_7,\omega_8\geq 1}\cos(2\pi \omega_7 d_7)\cos(2\pi \omega_8 d_8)
  \int_0^\infty dt\,  t^{-4}
  { e^{-{\omega_7^2+\omega_8^2\over 4\pi \hat M_w^2t}}
\over 2\sinh(2\pi |q_3|t).
                                    }  \bigg)
\end{align}
On the other hand, the effective potential contribution for the $\Sigma_{56}(\phi)=\pm 1/2$ case is given by
\begin{align}\notag
\Delta V(\phi)&\big|_{\Sigma_{56}=\pm 1/2}  
  =-(-1)^{\hat F}{N_{\rm deg}|q_3|\over 128\pi^3\hat M_w^2}
  \sum_{\omega_7,\omega_8\in\mathbb Z}
e^{2\pi i(\omega_7d_7+\omega_8d_8)}
  \int_0^\infty dt\,  t^{-4}
  { e^{-{\omega_7^2+\omega_8^2\over 4\pi \hat M_w^2t}}
\over \tanh(2\pi |q_3|t)
  }\\
  &=-(-1)^{\hat F}{N_{\rm deg}|q_3|\over 128\pi^3\hat M_w^2}
\bigg(  2\sum_{\omega\geq 1}[\cos(2\pi \omega d_7)+\cos(2\pi \omega d_8)]
\int_0^\infty dt\,  t^{-4}
  { e^{-{\omega^2\over 4\pi \hat M_w^2t}}
\over \tanh(2\pi |q_3|t)} \notag \\ 
&\qquad   + 4\sum_{\omega_7,\omega_8\geq 1}\cos(2\pi \omega_7 d_7)\cos(2\pi \omega_8 d_8)
  \int_0^\infty dt\,  t^{-4}
  { e^{-{\omega_7^2+\omega_8^2\over 4\pi \hat M_w^2t}}
\over \tanh(2\pi |q_3|t)
                                    }  \bigg).
\end{align}
We note that the expression above comes from a pair of $\Sigma_{56}(\phi)=\pm 1/2$ fields. Thus, $N_{\rm deg}$ corresponds to half of the real dof of the pair.

\subsection{Flux-coupled case with $\Sigma_{56}=\pm 1$}
\label{sec:sig1potdel}
Here, we consider the $\xi=1$ case. If $\phi$ now corresponds to $A_{5,6}$, there appears a pair of fields having $\Sigma_{56}=\pm 1$. Their 4D modes have masses as in eq.~\eqref{fluxmass56}. Hence, the effective potential contribution from a pair of $|\Sigma_{56}|=1$ fields is given by 
\begin{align}\label{dvphisigpm1}
  \Delta V(\phi)
  &=
    -{N_{\rm deg}|q_3|\over 32\pi^2}
    \sum_{\hat n_7, \hat n_8 \in \mathbb{Z}} \int_{0}^{\infty} dt\, t^{-3} e^{-M_{78}^2 t}
    \left(   
        e^{-2S(-1/2)t}+e^{-2S(1/2)t}
    +2\sum_{\hat \ell\geq 0}              
    e^{-2S(\hat \ell+1/2)t}\right), 
\end{align}
where we have used $M_{78}^2$ as in eq.~\eqref{M78def} and 
\begin{align}
  S=
  2 \pi\left|q_{3}\right|>0, 
\end{align}
for simplicity of the expressions. 

The contribution in eq.~\eqref{dvphisigpm1} contains UV divergences corresponding to the singularity of the integrand in the $t\to 0$ limit. On the other hand, there are no IR divergences since we only consider the WL phases that satisfy the conditions in eqs.~\eqref{constraint} and~\eqref{constraint2}. We can use the Poisson resummation as done in previous subsections to isolate the UV divergent contribution, which is independent of the WL phases. However, in this case, the Poisson resummation may cause a worse IR behavior. 

To see this, we first show an evaluation of the contributions that have a worse IR behavior. Using the formula
\begin{align}
  e^{St}+e^{-St}+2(e^{-3St}+e^{-5St}+\dots)={\cosh(2St)\over \sinh(St)}, 
\end{align}
we can formally rewrite eq.~\eqref{dvphisigpm1} as 
\begin{align}\label{dvphisigpm12}
  \Delta V(\phi)
  &=
    -{N_{\rm deg}|q_3|\over 32\pi^2}
    \sum_{\hat n_7, \hat n_8 \in \mathbb{Z}} \int_{0}^{\infty} dt\, t^{-3} e^{-M_{78}^2 t}
    {\cosh(4\pi |q_3|t)\over \sinh(2\pi |q_3|t)}.
\end{align}
As in previous subsections, using the Poisson resummation formula, we obtain 
\begin{align}\label{dvphisigpm13}
\Delta V(\phi)  
  &=-{N_{\rm deg}|q_3|\over 128\pi^3\hat M_w^2}
\bigg(  2\sum_{\omega\geq 1}[\cos(2\pi \omega d_7)+\cos(2\pi \omega d_8)]
\int_0^\infty dt\,  t^{-4}
  { e^{-{\omega^2\over 4\pi \hat M_w^2t}}}
{\cosh(4\pi |q_3|t)\over \sinh(2\pi |q_3|t)}
    \notag \\ 
&\qquad   + 4\sum_{\omega_7,\omega_8\geq 1}\cos(2\pi \omega_7 d_7)\cos(2\pi \omega_8 d_8)
  \int_0^\infty dt\,  t^{-4}
  { e^{-{\omega_7^2+\omega_8^2\over 4\pi \hat M_w^2t}}}
{\cosh(4\pi |q_3|t)\over \sinh(2\pi |q_3|t)}
                 \bigg).
\end{align}
In this expression, one sees that the integrands badly diverge for $t\to \infty$, namely the IR limit. This behavior cannot be evaded as long as we use the Poisson resummation formula to separate unwounded local divergences in the contributions from potentially tachyonic states, the first term in the parenthesis in eq.~\eqref{dvphisigpm1}. Alternatively, we can regularize the local divergences, which are independent of the WL phases, by subtracting an infinite constant, leading to the final expression for the regularized contribution to the effective potential.

To give a more appropriate evaluation of the potential contribution $\Delta V(\phi)$, we give a careful treatment of the contribution from potentially tachyonic states in the 4D modes. Let us define
\begin{align}\label{vtac1}
    \Delta V_{\rm tac}=
  -{N_{\rm deg}|q_3|\over 32\pi^2}
  \sum_{\hat n_7, \hat n_8 \in \mathbb{Z}} \int_{0}^{\infty} dt\, t^{-3} e^{-\left(M_{78}^2-S\right) t}.   
\end{align}
The total contribution is rewritten as
\begin{align}
  \Delta V(\phi)
  &=\Delta V_{\rm tac}
    -{N_{\rm deg}|q_3|\over 32\pi^2}
    \sum_{\hat n_7, \hat n_8 \in \mathbb{Z}} \int_{0}^{\infty} dt\, t^{-3} e^{-M_{78}^2 t}
    \left(    \sum_{\hat \ell\geq 0}
    +\sum_{\hat \ell\geq 0}              
    \right)e^{-2S(\hat \ell+1/2)t}.
\end{align}
Let us use
\begin{align}
\left(    \sum_{\hat \ell\geq 0}
    +\sum_{\hat \ell\geq 0}              
  \right)e^{-S(2\hat \ell+1)t}
  &=  \left(e^{-St}+e^{-3St}+\dots\right)+
    \left(e^{-3St}+e^{-5St}+\dots\right)
    ={e^{-St}\over \tanh(St)}.
\end{align}
Except for $\Delta V_{\rm tac}$, we calculate the potential contributions as in the previous sections. The result is given by 
\begin{align}
\notag
  \Delta V(\phi)
  &=
\Delta V_{\rm tac}
    -{N_{\rm deg}|q_3|\over 128\pi^3\hat M_w^2}\\
  &\qquad \times 
\bigg(  2\sum_{\omega\geq 1}[\cos(2\pi \omega d_7)+\cos(2\pi \omega d_8)]
\int_0^\infty dt\,  t^{-4}
    { e^{-{\omega^2\over 4\pi \hat M_w^2t}}e^{- 2\pi |q_3|t}\over
     \tanh(2\pi |q_3|t)}
    \notag \\ 
&\qquad   + 4\sum_{\omega_7,\omega_8\geq 1}\cos(2\pi \omega_7 d_7)\cos(2\pi \omega_8 d_8)
  \int_0^\infty dt\,  t^{-4}
  { e^{-{\omega_7^2+\omega_8^2\over 4\pi \hat M_w^2t}}e^{- 2\pi |q_3|t}\over
     \tanh(2\pi |q_3|t)}
                 \bigg).     
\end{align}

Let us discuss the evaluation of $\Delta V_{\rm tac}$. As discussed in section~\ref{sec:stab}, we only consider values of WL phases that eliminate tachyonic states. In this case, the relation $M_{78}^2-S\geq 0$ is ensured, and the integrand in eq.~\eqref{vtac1} converges for $t\to \infty$.
We evaluate $\Delta V_{\rm tac}$ under the condition $M_{78}^2-S\geq 0$.  Let us first define
\begin{align}
  I_{\rm T}&=\sum_{\hat n_7, \hat n_8 \in \mathbb{Z}} \int_{0}^{\infty} dt\, t^{-3} e^{-\left(M_{78}^2-S\right) t},
       \qquad
       \Delta V_{\rm tac}=
  -{N_{\rm deg}|q_3|\over 32\pi^2}I_{\rm T}.
\end{align}
We consider the parameter region of our interest $M_{78}^2-S \geq 0$, which is rewritten as 
\begin{align}
0 \leq S / M_{78}^2 \leq 1,
\end{align}
and expand the factor $e^{S t}$ in $I_{\rm T}$ as
\begin{align}
  I_{\rm T}&=\sum_{k\geq 0} {S^k\over k!}I_{\mathrm{T}}^{(k)},
             \qquad 
I_{\mathrm{T}}^{(k)}=\sum_{\hat n_7, \hat n_8 \in \mathbb{Z}} \int_{0}^{\infty} dt\,  t^{-3+k} e^{-M_{78}^2 t}. 
\end{align}

One sees that $I_{\mathrm{T}}^{(k)}$ for $k=0,1,2$  contains UV divergences. Let us use the Poisson resummation formula as in eq.~\eqref{poissongen} to separate the UV divergent parts,
\begin{align}
\sum_{\hat n_7, \hat n_8 \in \mathbb{Z}} e^{-M_{78}^2 t}=\sum_{\omega_{7}, \omega_{8} \in \mathbb{Z}} \frac{1}{\left|4 \pi \hat M_w^{2} t\right|} e^{-\frac{1}{4 \hat M_w^{2} t}\left(\omega_{7}^{2}+\omega_{8}^{2}\right)} e^{2 \pi i\left(\omega_{7} d_{7}+\omega_{8} d_{8}\right)}
\end{align}
and evaluate $I_{\mathrm{T}}^{(k\leq 2)}$ as
\begin{align}
I_{\mathrm{T}}^{(k \leq 2)} =\sum_{{\omega}_{7}, {\omega}_{8} \in \mathbb{Z}} \frac{e^{2 \pi i\left(\omega_{7} d_{7}+\omega_{8} d_{8}\right)}}{\left|4 \pi \hat M_w^{2}\right|} \int_{0}^{\infty} dt \  \frac{e^{-\frac{1}{4 \pi \hat M_w^{2} t}\left(\omega_{7}^{2}+\omega_{8}^{2}\right)}}{t^{4-k}}  
 =\frac{\Gamma(3-k)}{(4 \pi \hat M_w^{2})^{k-2}} \sum_{{\omega}_{7}, {\omega}_{8} \in \mathbb{Z}} \frac{e^{2 \pi i\left(\omega_{7} d_{7}+\omega_{8} d_{8}\right)}}{\left(\omega_{7}^{2}+\omega_{8}^{2}\right)^{3-k}} .
\end{align}
Thus, we have that the expressions for each value of $k \leq 2$ are given by
\begin{align}\notag
  I_{\mathrm{T}}^{(0)}
&=32\pi^2\hat M_w^4
             \left(
    2\sum_{\omega\geq 1}  {\cos(2\pi \omega d_7)+\cos(2\pi \omega d_8)\over 
    \omega^2}
    + 4\sum_{\omega_7,\omega_8\geq 1}
    {\cos(2\pi \omega_7 d_7)\cos(2\pi \omega_8 d_8)\over 
    \omega_7^2+\omega_8^2}
    \right)\\
&\qquad  +({\rm constant}), \\\notag
  I_{\mathrm{T}}^{(1)}
  &=4\pi\hat M_w^2
             \left(
    2\sum_{\omega\geq 1}  {\cos(2\pi \omega d_7)+\cos(2\pi \omega d_8)\over 
    (\omega^2)^2}
    + 4\sum_{\omega_7,\omega_8\geq 1}
    {\cos(2\pi \omega_7 d_7)\cos(2\pi \omega_8 d_8)\over 
    (\omega_7^2+\omega_8^2)^2}
    \right)\\
&\qquad+({\rm constant}), \\
I_{\mathrm{T}}^{(2)}
  &=
    2\sum_{\omega\geq 1}  {\cos(2\pi \omega d_7)+\cos(2\pi \omega d_8)\over 
    (\omega^2)^3}
    + 4\sum_{\omega_7,\omega_8\geq 1}
    {\cos(2\pi \omega_7 d_7)\cos(2\pi \omega_8 d_8)\over 
    (\omega_7^2+\omega_8^2)^3}+({\rm constant}) .
\end{align}
In these expressions, the UV divergent part originating from zero winding terms is separated as ``(constant)''.  Since they are independent of the WL phases, we hereafter discard these constants.

For $k \geq 3$, we obtain a simple expression of $I_{\mathrm{T}}^{(k)}$ as 
\begin{align}
I_{\mathrm{T}}^{(k \geq 3)}=\sum_{\hat n_7, \hat n_8 \in \mathbb{Z}} \frac{\Gamma(k-2)}{\left(M_{78}^2\right)^{k-2}}.
\end{align}
Then, the $k \geq 3$ contributions are expressed as 
\begin{align}
  \sum_{k\geq 3} {S^k\over k!}I_{\mathrm{T}}^{(k)}
  &=\sum_{\hat n_7, \hat n_8 \in \mathbb{Z}} \sum_{k\geq 1} \frac{S^2\left(S / M_{78}^2\right)^{k}}{(k+2)(k+1) k}.
    \label{k3sum}
\end{align}
In the above expression, there is a divergent contribution contained in the $k=1$ term on the right-hand side, which originates from $I_{\rm T}^{(3)}$. It is possible to regularize it by a procedure similar to the Pauli-Villars regularization.  We subtract the infinite constant that is independent of the WL phases from $I_{\rm T}$ as
\begin{align}
I_{\rm T}\to I_{\rm T}-\sum_{\left(\hat n_7, \hat n_8\right) \neq(0,0)} \frac{S^3}{6(2\pi \hat M_w)^2(\hat n_7^2+\hat n_8^2)},
\end{align}
which exactly cancels the divergence in eq.~\eqref{k3sum}. We note that the summation is taken over integers $\hat n_7$ and $\hat n_8$ except for $(\hat n_7, \hat n_8)=(0,0)$ in the regulator.

Consequently, we obtain the expression of $I_{\rm T}$ as
\begin{align}\notag
    I_{\rm T}
  &=  
I_{\rm T}^{(0)}+SI_{\rm T}^{(1)}+{S^2\over 2}I_{\rm T}^{(2)}\\
&\qquad    +\sum_{\hat n_7, \hat n_8 \in \mathbb{Z}} \sum_{k\geq 1} \frac{S^2\left(S / M_{78}^2\right)^{k}}{(k+2)(k+1) k}
    -\sum_{\left(\hat n_7, \hat n_8\right) \neq(0,0)} \frac{S^3}{6(2\pi \hat M_w)^2(\hat n_7^2+\hat n_8^2)}\\\notag
  &=
    2\sum_{\omega\geq 1} \left[\cos(2\pi \omega d_7)+\cos(2\pi \omega d_8)\right]
    \left({32\pi^2\hat M_w^4\over \omega^2}
    +{8\pi^2|q_3|\hat M_w^2\over \omega^4}
    +{2\pi^2|q_3|^2\over \omega^6}
    \right)\\\label{tacexp1}
  &+
    4\sum_{\omega_7,\omega_8\geq 1}
    \cos(2\pi \omega_7 d_7)\cos(2\pi \omega_8 d_8)
    \left({32\pi^2\hat M_w^4\over \omega_7^2+\omega_8^2}
    +{8\pi^2|q_3|\hat M_w^2\over (\omega_7^2+\omega_8^2)^2}
    +{2\pi^2|q_3|^2\over (\omega_7^2+\omega_8^2)^3}
    \right)\\\notag
  &+\sum_{\hat n_7, \hat n_8 \in \mathbb{Z}} \sum_{k\geq 1} \frac{(2\pi |q_3|)^{2+k}}{(k+2)(k+1) k (M_{78}^2)^{k}}
    -\sum_{\left(\hat n_7, \hat n_8\right) \neq(0,0)} \frac{(2\pi |q_3|)^3}{6(2\pi \hat M_w)^2(\hat n_7^2+\hat n_8^2)}+({\rm constant}). 
\end{align}
Discarding the irrelevant constant contribution in the last equation, this expression is finite.


\bigskip\bigskip


\begin{thebibliography}{99}

\bibitem{manton}
    N.~S.~Manton,
    ``A New Six-Dimensional Approach To The Weinberg-Salam Model,''
    Nucl.\ Phys.\  B {\bf 158}, (1979), 141.
\bibitem{fair}
    D.~B.~Fairlie,
    ``Higgs' Fields And The Determination Of The Weinberg Angle,''
    Phys.\ Lett.\  B {\bf 82}, (1979), 97.
\bibitem{hosotani1}
    Y.~Hosotani,
    ``Dynamical Mass Generation by Compact Extra Dimensions,''
    Phys.\ Lett.\ B {\bf 126}, (1983), 309.
\bibitem{Hosotani:1988bm}
    Y.~Hosotani,
    ``Dynamics of Nonintegrable Phases and Gauge Symmetry Breaking,''
    Annals Phys. \textbf{190} (1989), 233.
\bibitem{Hatanaka:1998yp}
    H.~Hatanaka, T.~Inami and C.~S.~Lim,
    ``The Gauge hierarchy problem and higher dimensional gauge theories,''
    Mod. Phys. Lett. A \textbf{13} (1998), 2601-2612
    [arXiv:hep-th/9805067].
\bibitem{Maru:2006wa}
    N.~Maru and T.~Yamashita,
    ``Two-loop Calculation of Higgs Mass in Gauge-Higgs Unification: 5D Massless QED Compactified on S**1,''
    Nucl. Phys. B \textbf{754} (2006), 127-145
    [arXiv:hep-ph/0603237].
\bibitem{Hosotani:2007kn}
    Y.~Hosotani, N.~Maru, K.~Takenaga and T.~Yamashita,
    ``Two Loop finiteness of Higgs mass and potential in the gauge-Higgs unification,''
    Prog. Theor. Phys. \textbf{118} (2007), 1053-1068
    [arXiv:0709.2844].
\bibitem{Hall:2001zb}
    L.~J.~Hall, Y.~Nomura and D.~Tucker-Smith,
    ``Gauge Higgs unification in higher dimensions,''
    Nucl. Phys. B \textbf{639} (2002), 307-330
    [arXiv:hep-ph/0107331].
\bibitem{Antoniadis:2001cv}
    I.~Antoniadis, K.~Benakli and M.~Quiros,
    ``Finite Higgs mass without supersymmetry,''
    New J. Phys. \textbf{3} (2001), 20
    [arXiv:hep-th/0108005].
\bibitem{Kubo:2001zc}
    M.~Kubo, C.~S.~Lim and H.~Yamashita,
    ``The Hosotani mechanism in bulk gauge theories with an orbifold extra space S**1 / Z(2),''
    Mod. Phys. Lett. A \textbf{17} (2002), 2249-2264
    [arXiv:hep-ph/0111327].
\bibitem{Csaki:2002ur}
    C.~Csaki, C.~Grojean and H.~Murayama,
    ``Standard model Higgs from higher dimensional gauge fields,''
    Phys. Rev. D \textbf{67} (2003), 085012
     [arXiv:hep-ph/0210133].
\bibitem{Burdman:2002se}
    G.~Burdman and Y.~Nomura,
    ``Unification of Higgs and Gauge Fields in Five Dimensions,''
    Nucl. Phys. B \textbf{656} (2003), 3-22
    [arXiv:hep-ph/0210257].
\bibitem{Gogoladze:2003bb}
    I.~Gogoladze, Y.~Mimura and S.~Nandi,
     ``Gauge Higgs unification on the left right model,''
    Phys. Lett. B \textbf{560} (2003), 204-213
     [arXiv:hep-ph/0301014].
\bibitem{Scrucca:2003ra}
    C.~A.~Scrucca, M.~Serone and L.~Silvestrini,
    ``Electroweak symmetry breaking and fermion masses from extra dimensions,''
    Nucl. Phys. B \textbf{669} (2003), 128-158
     [arXiv:hep-ph/0304220].
\bibitem{Scrucca:2003ut}
    C.~A.~Scrucca, M.~Serone, L.~Silvestrini and A.~Wulzer,
    ``Gauge Higgs unification in orbifold models,''
    JHEP \textbf{02} (2004), 049
    [arXiv:hep-th/0312267].
\bibitem{Haba:2004qf}
    N.~Haba, Y.~Hosotani, Y.~Kawamura and T.~Yamashita,
    ``Dynamical symmetry breaking in gauge Higgs unification on orbifold,''
    Phys. Rev. D \textbf{70} (2004), 015010
    [arXiv:hep-ph/0401183].
\bibitem{Haba:2004jd}
    N.~Haba and T.~Yamashita,
     ``Dynamical symmetry breaking in gauge Higgs unification of 5-D N=1 SUSY theory,''
    JHEP \textbf{04} (2004), 016
    [arXiv:hep-ph/0402157].
\bibitem{GG}
    H.~Georgi and S.~L.~Glashow,
    ``Unity of All Elementary Particle Forces,''
    Phys.\ Rev.\ Lett.\  {\bf 32}, (1974), 438.
\bibitem{Lim:2007jv}
    C.~S.~Lim and N.~Maru,
    ``Towards a realistic grand gauge-Higgs unification,''
    Phys. Lett. B \textbf{653} (2007), 320-324
    [arXiv:0706.1397].
\bibitem{Kojima:2011ad}
    K.~Kojima, K.~Takenaga and T.~Yamashita,
    ``Grand Gauge-Higgs Unification,''
    Phys. Rev. D \textbf{84} (2011), 051701
    [arXiv:1103.1234].
\bibitem{Yamashita:2011an}
    T.~Yamashita,
    ``Doublet-Triplet Splitting in an SU(5) Grand Unification,''
    Phys. Rev. D \textbf{84} (2011), 115016
    [arXiv:1106.3229].
\bibitem{Hosotani:2015hoa}
    Y.~Hosotani and N.~Yamatsu,
    ``Gauge\textendash{}Higgs grand unification,''
    PTEP \textbf{2015} (2015), 111B01
    [arXiv:1504.03817].
\bibitem{Yamatsu:2015oit}
    N.~Yamatsu,
    ``Gauge coupling unification in gauge\textendash{}Higgs grand unification,''
    PTEP \textbf{2016} (2016) no.4, 043B02
    [arXiv:1512.05559].
\bibitem{Furui:2016owe}
    A.~Furui, Y.~Hosotani and N.~Yamatsu,
    ``Toward Realistic Gauge-Higgs Grand Unification,''
    PTEP \textbf{2016} (2016) no.9, 093B01
    [arXiv:1606.07222].
\bibitem{Kojima:2016fvv}
    K.~Kojima, K.~Takenaga and T.~Yamashita,
    ``Gauge symmetry breaking patterns in an SU(5) grand gauge-Higgs unification model,''
    Phys. Rev. D \textbf{95} (2017) no.1, 015021
    [arXiv:1608.05496].
\bibitem{Kojima:2017qbt}
    K.~Kojima, K.~Takenaga and T.~Yamashita,
    ``The Standard Model Gauge Symmetry from Higher-Rank Unified Groups in Grand Gauge-Higgs Unification Models,''
    JHEP \textbf{06} (2017), 018
    [arXiv:1704.04840].
\bibitem{Hosotani:2017edv}
    Y.~Hosotani and N.~Yamatsu,
    ``Electroweak symmetry breaking and mass spectra in six-dimensional gauge\textendash{}Higgs grand unification,''
    PTEP \textbf{2018} (2018) no.2, 023B05
    [arXiv:1710.04811].
\bibitem{Maru:2019lit}
    N.~Maru and Y.~Yatagai,
    ``Fermion Mass Hierarchy in Grand Gauge-Higgs Unification,''
    PTEP \textbf{2019} (2019) no.8, 083B03
    [arXiv:1903.08359].
\bibitem{Englert:2019xhz}
    C.~Englert, D.~J.~Miller and D.~D.~Smaranda,
    ``Phenomenology of GUT-inspired gauge-Higgs unification,''
    Phys.\ Lett.\  B {\bf 802}, (2020), 135261
    [arXiv:1911.05527 [hep-ph]].
\bibitem{Angelescu:2021nbp}
    A.~Angelescu, A.~Bally, S.~Blasi and F.~Goertz,
    ``Minimal SU(6) gauge-Higgs grand unification,''
    Phys. Rev. D \textbf{105} (2022) no.3, 035026
    [arXiv:2104.07366].
\bibitem{Nakano:2022lyt}
    H.~Nakano, M.~Sato, O.~Seto and T.~Yamashita,
    ``Dirac gaugino from grand gauge-Higgs unification,''
    PTEP \textbf{2022} (2022) no.3, 033B06
    [arXiv:2201.04428].
\bibitem{Angelescu:2022obm}
    A.~Angelescu, A.~Bally, F.~Goertz and S.~Weber,
    ``SU(6) gauge-Higgs grand unification: minimal viable models and flavor,''
    JHEP \textbf{04} (2023), 012
    [arXiv:2208.13782 [hep-ph]].
\bibitem{Kojima:2023mew}
    K.~Kojima, K.~Takenaga and T.~Yamashita,
    ``Grand gauge-Higgs unification on T2/Z3 via the diagonal embedding method,''
    Phys. Rev. D \textbf{108} (2023) no.3, 035031
    [arXiv:2304.05701 [hep-ph]].
\bibitem{Maru:2024ghd}
    N.~Maru and R.~Nago,
    ``Attempt Constructing a Model of Grand Gauge-Higgs Unification with Family Unification,''
    arXiv:2403.02731 [hep-ph].
\bibitem{Generations} 
    E.~Witten,
    ``Some Properties of O(32) Superstrings,''
    Phys. Lett. B \textbf{149} (1984), 351-356.
\bibitem{Cremades} 
    D.~Cremades, L.~E.~Ibanez and F.~Marchesano,
    ``Computing Yukawa couplings from magnetized extra dimensions,''
    JHEP \textbf{05} (2004), 079
    [arXiv:hep-th/0404229].
\bibitem{Abe:2008sx}
    H.~Abe, K.~S.~Choi, T.~Kobayashi and H.~Ohki,
    ``Three generation magnetized orbifold models,''
    Nucl. Phys. B \textbf{814}, (2009), 265-292 
    [arXiv:0812.3534].
    \bibitem{Kobayashi:2010an}
    T.~Kobayashi, R.~Maruyama, M.~Murata, H.~Ohki and M.~Sakai,
    ``Three-generation Models from $E_8$ Magnetized Extra Dimensional Theory,''
    JHEP \textbf{05} (2010), 050
    [arXiv:1002.2828].
\bibitem{Abe:2015yva}
    T.~h.~Abe, Y.~Fujimoto, T.~Kobayashi, T.~Miura, K.~Nishiwaki, M.~Sakamoto and Y.~Tatsuta,
    ``Classification of three-generation models on magnetized orbifolds,''
    Nucl. Phys. B \textbf{894}, (2015), 374-406
    [arXiv:1501.02787].
\bibitem{Abe:2015mua}
    H.~Abe, T.~Kobayashi, H.~Otsuka and Y.~Takano,
    ``Realistic three-generation models from SO(32) heterotic string theory,''
    JHEP \textbf{09} (2015), 056
    [arXiv:1503.06770].
\bibitem{Sakamoto:2020pev}
    M.~Sakamoto, M.~Takeuchi and Y.~Tatsuta,
    ``Zero-mode counting formula and zeros in orbifold compactifications,''
    Phys. Rev. D \textbf{102}, (2020) no.2, 025008 
    [arXiv:2004.05570].
\bibitem{Sakamoto:2020vdy}
    M.~Sakamoto, M.~Takeuchi and Y.~Tatsuta,
    ``Index theorem on $T^2/\mathbb{Z}_N$ orbifolds,''
    Phys. Rev. D \textbf{103}, (2021) no.2, 025009 
    [arXiv:2010.14214].
\bibitem{Kobayashi:2022tti}
    T.~Kobayashi, H.~Otsuka, M.~Sakamoto, M.~Takeuchi, Y.~Tatsuta and H.~Uchida,
    ``Index theorem on magnetized blow-up manifold of T2/ZN,''
    Phys. Rev. D \textbf{107}, (2023) no.7, 075032 
    [arXiv:2211.04595].
\bibitem{Imai:2022bke}
    H.~Imai, M.~Sakamoto, M.~Takeuchi and Y.~Tatsuta,
    ``Index and winding numbers on T2/ZN orbifolds with magnetic flux,''
    Nucl. Phys. B \textbf{990}, (2023) 116189 
    [arXiv:2211.15541].
\bibitem{Imai:2023yuf}
    H.~Imai and N.~Maru,
    ``Toward Realistic Models in $T^2/\mathbb{Z}_2$ Flux Compactification,''
    arXiv:2311.10324 [hep-ph].
\bibitem{Abe:2014vza}
    H.~Abe, T.~Kobayashi, K.~Sumita and Y.~Tatsuta,
    ``Gaussian Froggatt-Nielsen mechanism on magnetized orbifolds,''
    Phys. Rev. D \textbf{90}, (2014) no.10, 105006 
    [arXiv:1405.5012].
\bibitem{Fujimoto:2016zjs}
    Y.~Fujimoto, T.~Kobayashi, K.~Nishiwaki, M.~Sakamoto and Y.~Tatsuta,
    ``Comprehensive analysis of Yukawa hierarchies on $T^2/Z_N$ with magnetic fluxes,''
    Phys. Rev. D \textbf{94}, (2016) no.3, 035031 
    [arXiv:1605.00140].
\bibitem{Abe:2016eyh}
    H.~Abe, T.~Kobayashi, H.~Otsuka, Y.~Takano and T.~H.~Tatsuishi,
    ``Flavor structure in $SO(32)$ heterotic string theory,''
    Phys. Rev. D \textbf{94} (2016) no.12, 126020
    [arXiv:1605.00898].
\bibitem{Kobayashi:2016qag}
    T.~Kobayashi, K.~Nishiwaki and Y.~Tatsuta,
    ``CP-violating phase on magnetized toroidal orbifolds,''
    JHEP \textbf{04} (2017), 080 
    [arXiv:1609.08608].
\bibitem{Buchmuller:2017vut}
    W.~Buchmuller and K.~M.~Patel,
    ``Flavor physics without flavor symmetries,''
    Phys. Rev. D \textbf{97}, (2018) no.7, 075019 
    [arXiv:1712.06862].
\bibitem{Buchmuller:2017vho}
    W.~Buchmuller and J.~Schweizer,
    ``Flavor mixings in flux compactifications,''
    Phys. Rev. D \textbf{95}, (2017) no.7, 075024 
    [arXiv:1701.06935].
\bibitem{Neutrino} 
    M.~Ishida, K.~Nishiwaki and Y.~Tatsuta,
    ``Seesaw mechanism in magnetic compactifications,''
    JHEP \textbf{07} (2018), 125
    [arXiv:1802.06646].
\bibitem{Bachas} 
    C.~Bachas,
    ``A Way to break supersymmetry,''
    arXiv:hep-th/9503030.
\bibitem{Buchmuller} 
    W.~Buchmuller, M.~Dierigl, E.~Dudas and J.~Schweizer,
    ``Effective field theory for magnetic compactifications,''
    JHEP \textbf{04} (2017), 052
    [arXiv:1611.03798].
\bibitem{Ghilencea:2017jmh}
    D.~M.~Ghilencea and H.~M.~Lee,
    ``Wilson lines and UV sensitivity in magnetic compactifications,''
    JHEP \textbf{06} (2017), 039 
    [arXiv:1703.10418].
\bibitem{B1}
    W.~Buchmuller, M.~Dierigl and E.~Dudas,
    ``Flux compactifications and naturalness,''
    JHEP \textbf{08} (2018), 151
    [arXiv:1804.07497].
\bibitem{Honda:2019ema}
    M.~Honda and T.~Shibasaki,
    ``Wilson-line Scalar as a Nambu-Goldstone Boson in Flux Compactifications and Higher-loop Corrections,''
    JHEP \textbf{03} (2020), 031 
    [arXiv:1912.04581].
\bibitem{Maru:2023esr}
    N.~Maru and H.~Tanaka,
    ``Wilson-line Scalar Mass in Flux Compactification on an Orbifold $T^2/Z_2$,''
    arXiv:2303.01747 [hep-th].
\bibitem{Hirose:2024vvx}
    T.~Hirose, H.~Otsuka, K.~Tsumura and Y.~Uchida,
    ``Nambu-Goldstone Modes in Magnetized $T^{2n}$ Extra Dimensions,''
    arXiv:2403.16801 [hep-th].
\bibitem{xi}
    T.~Hirose and N.~Maru,
    ``Cancellation of One-loop Corrections to Scalar Masses in Yang-Mills Theory with Flux Compactification,''
    JHEP \textbf{08} (2019), 054
    [arXiv:1904.06028].
\bibitem{Maru} 
    T.~Hirose and N.~Maru,
    ``Cancellation of One-loop Corrections to Scalar Masses in Flux Compactification with Higher Dimensional Operators,''
    J. Phys. G \textbf{48} (2021) no.5, 055005
    [arXiv:2012.03494].
\bibitem{Maru2}
    T.~Hirose and N.~Maru,
    ``Nonvanishing finite scalar mass in flux compactification,''
    JHEP \textbf{06} (2021), 159
    [arXiv:2104.01779].
\bibitem{Akamatsu}
    K.~Akamatsu, T.~Hirose and N.~Maru,
    ``Gauge symmetry breaking in flux compactification with a Wilson-line scalar condensate,''
    Phys. Rev. D \textbf{106} (2022) no.3, 035035
    [arXiv:2205.09320].
\bibitem{KYS}
    K.~Kojima, Y.~Okubo and C.~S.~Takeda,
    ``Mass spectrum in a six-dimensional SU(n) gauge theory on a magnetized torus,''
    JHEP \textbf{08} (2023), 083
    [arXiv:2306.00644 [hep-th]].
\bibitem{Buchmuller:2019zhz}
    W.~Buchmuller, E.~Dudas and Y.~Tatsuta,
    ``Quantum corrections for D-brane models with broken supersymmetry,''
    JHEP \textbf{12} (2019), 022 
    [arXiv:1909.03007].
\bibitem{Franken:2022pug}
    V.~Franken,
    ``Tachyons and (non)vanishing scalar masses in six-dimensional gauge theories with flux compactification,''
    arXiv:2203.03307 [hep-th].
\bibitem{Buchmuller:2020nnl}
    W.~Buchmuller, E.~Dudas and Y.~Tatsuta,
    ``Tachyon condensation in magnetic compactifications,''
    JHEP \textbf{03} (2021), 070 
    [arXiv:2010.10891].
\bibitem{Haba:2002py}
    N.~Haba, M.~Harada, Y.~Hosotani and Y.~Kawamura,
    ``Dynamical rearrangement of gauge symmetry on the orbifold S1 / Z(2),''
    Nucl. Phys. B \textbf{657} (2003), 169-213
    [erratum: Nucl. Phys. B \textbf{669} (2003), 381-382]
    [arXiv:hep-ph/0212035].
\bibitem{Haba:2003ux}
    N.~Haba, Y.~Hosotani and Y.~Kawamura,
    ``Classification and dynamics of equivalence classes in SU(N) gauge theory on the orbifold S**1 / Z(2),''
    Prog. Theor. Phys. \textbf{111} (2004), 265-289
    [arXiv:hep-ph/0309088].
\bibitem{Kawamura:2022ecd}
    Y.~Kawamura, E.~Kodaira, K.~Kojima and T.~Yamashita,
    ``On representation matrices of boundary conditions in SU(n) gauge theories compactified on two-dimensional orbifolds,''
    JHEP \textbf{04} (2023), 113
    [arXiv:2211.00877].
\end{thebibliography}
\end{document}